\documentclass[11pt]{article}
\usepackage{jheppub}
\usepackage{epsfig} 
\usepackage{amssymb}
\usepackage{amsmath}
\allowdisplaybreaks[4]
\usepackage{epstopdf}
\usepackage{extarrows}

\usepackage{dsfont,bbm}
\usepackage{subfig}

\usepackage{shuffle}
\usepackage{booktabs}
\usepackage{array}

\graphicspath{{./figs/}}

\title{Deciphering the Maximal Transcendentality Principle via Bootstrap}

\author[a,b]{Yuanhong Guo,}
\emailAdd{guoyuanhong@itp.ac.cn}
\author[c]{Qingjun Jin,}
\emailAdd{qjin@gscaep.ac.cn}
\author[a,b]{Lei Wang,}
\emailAdd{wanglei@itp.ac.cn}
\author[a,b,d,e]{and Gang Yang}
\emailAdd{yangg@itp.ac.cn}
\affiliation[a]{CAS Key Laboratory of Theoretical Physics, Institute of Theoretical Physics, \\Chinese Academy of Sciences, Beijing 100190, China}
\affiliation[b]{School of Physical Sciences, University of Chinese Academy of Sciences, Beijing 100049, China}
\affiliation[c]{Graduate School of China Academy of Engineering Physics, No.~10 Xibeiwang East Road, Haidian District, Beijing, 100193, China}
\affiliation[d]{School of Fundamental Physics and Mathematical Sciences, Hangzhou Institute for Advanced Study, UCAS, Hangzhou 310024, China}
\affiliation[e]{International Centre for Theoretical Physics Asia-Pacific, Beijing/Hangzhou, China}

\abstract{
We prove the principle of maximal transcendentality for a class of form factors, including the general two-loop minimal form factors,  the two-loop three-point form factor of ${\rm tr}(F^2)$, and the two-loop four-point form factor of ${\rm tr}(F^3)$. Our proof is based on a recently developed bootstrap method using the representation of master integral expansions, together with some unitarity cuts that are universal in general gauge theories. The maximally transcendental parts of the two-loop four-gluon form factor of $\mathrm{tr}(F^3)$ are obtained for the first time in both planar $\mathcal{N}=4$ SYM and pure YM theories. This form factor can be understood as the Higgs-plus-four-gluon amplitudes involving a dimension-seven operator in the Higgs effective theory. In this case, we find that the maximally transcendental part of the $\mathcal{N}=4$ SYM result is different from that of pure YM, and the discrepancy is due to the gluino-loop contributions in $\mathcal{N}=4$ SYM.
In contrast, the scalar-loop contributions have no maximally transcendental parts. Thus, the maximal transcendentality principle still holds for the form factor results in $\mathcal{N}=4$ SYM and QCD, after a proper identification of the fundamental quarks and adjoint gluinos as $n_f \rightarrow 4N_c$. This seems to be the first example of the maximally transcendental principle that involves fermion-loop contributions. As another intriguing observation, we find that the four-point form factor of the half-BPS $\mathrm{tr}(\phi^3)$ operator is precisely a building block in the form factor of $\mathrm{tr}(F^3)$.
}

\begin{document}

\maketitle

\setcounter{footnote}{0}

%%%%%%%%%%%%%%%%%%%%
\section{Introduction}

It is known that there are a few direct connections between physical quantities in $\mathcal{N}=4$ SYM and QCD. At the tree level, gluon amplitudes are equivalent in the two theories. At one-loop level, through the supersymmetric decomposition \cite{Bern:1994cg}, one-loop $\mathcal{N}=4$ amplitudes are also useful building blocks for QCD amplitudes. 
An intriguing relation between $\mathcal{N}=4$ SYM and QCD also exists at high loop orders,  known as the principle of maximal transcendentality \cite{Kotikov:2002ab, Kotikov:2004er}. 
The \emph{``transcendentality"} here refers to the transcendentality degree which is a mathematical notion characterizing the ``complexity" of transcendental functions and numbers.
For example,  the transcendentality degree of algebraic numbers or rational functions is zero, the $\log(x)$ function and $\pi$ have transcendentality degree $1$, and the more general classical polylogrithm $\mathrm{Li}_n(x)$ and  the Riemann zeta value $\zeta_n$ has degree $n$,  recalling their definition:
\begin{equation}
\mathrm{Li}_n(x) = \sum_{k=1}^\infty {x^k \over k^n} = \int_0^x { \mathrm{Li}_{n-1}(t) \over t} d t \,, \qquad \mathrm{Li}_1(x) = - \log(1-x) \,, \qquad \zeta_n = \mathrm{Li}_n(x=1) \,.
\end{equation}
The maximal transcendentality principle (MTP) conjectures that for certain physical quantities, the maximally transcendental parts, \emph{i.e.}~the parts with the highest transcendentality degree, are equal in ${\cal N}=4$ SYM and QCD (up to certain identification of the fermions in the two theories).

The MTP was originally proposed in \cite{Kotikov:2002ab, Kotikov:2004er}, suggesting that the anomalous dimensions of twist-two operators in $\mathcal{N}=4$ SYM can be obtained from the maximally transcendental part of the corresponding QCD results \cite{Moch:2004pa}. 
A further observation beyond the anomalous dimensions was made for the form factors and Higgs amplitudes: the study of the two-loop three-point form factor of stress-tensor multiplet in $\mathcal{N}=4$ SYM in \cite{Brandhuber:2012vm} shows that it coincides with the maximally transcendental part of the two-loop Higgs plus three-gluon amplitudes in the heavy top-mass limit obtained in \cite{Gehrmann:2011aa}. This generalizes the scope of the MTP from anomalous dimensions to kinematics-dependent functions. 
More correspondence was also observed to Higgs-plus-three-parton amplitudes with high dimensional operators \cite{Brandhuber:2014ica, Brandhuber:2017bkg, Jin:2018fak, Brandhuber:2018xzk, Brandhuber:2018kqb, Jin:2019ile, Jin:2019opr, Jin:2020pwh} and with external quark states \cite{Jin:2019ile, Banerjee:2017faz}.
There is also other evidence of the correspondence for Wilson lines \cite{Li:2014afw, Li:2016ctv}.
The MTP was used to obtain the planar four-loop collinear anomalous dimension in $\mathcal{N}=4$ SYM \cite{Dixon:2017nat}, which was confirmed by \cite{Agarwal:2021zft}.
It is also worth pointing out that the MTP was found to be true for the non-planar cusp anomalous dimensions at four loops \cite{Henn:2019swt,Huber:2019fxe}, suggesting that it should apply beyond the planar (\emph{i.e.}~large $N_c$) limit.

So far the maximal transcendentality principle is still a conjecture and is known in the above-mentioned cases through explicit computations. (There are also known counterexamples such as gluon amplitudes, which we will discuss in the Discuss section.)
In this paper, we make a concrete step toward the understanding of the MTP by proving it for the known cases of form factors. 
We recall that an $n$-point form factor is defined as a matrix element between a local gauge-invariant operator and $n$ asymptotic on-shell states (see \cite{Yang:2019vag} for a recent review):
\begin{equation}
	\label{eq:FF-def}
	\mathcal{F}_{\mathcal{O}, n} = \int d^D x e^{-i q\cdot x} \langle \Phi(p_1) \ldots \Phi(p_n) | \mathcal{O}(x) |0 \rangle \,,
\end{equation}
where $p_i, i=1,..,n$ are on-shell momenta and $q=\sum_{i=1}^n p_i$ is the off-shell momentum associated to the local operator.
As mentioned, the form factors are also related to Higgs amplitudes in the Higgs effective theory where the top quark is integrated out \cite{Wilczek:1977zn, Shifman:1979eb, Dawson:1990zj, Djouadi:1991tka, Kniehl:1995tn}.
The central idea in our proof of the MTP for form factors is to apply a recently developed bootstrap method \cite{Guo:2021bym}.
In this method, one starts with a general ansatz of loop quantities in terms of a set of master integrals and then determines the integral coefficients via various physical constraints, such as infrared divergences and collinear limits. 
A further key idea is to apply a small set of unitarity cuts that are universal for general gauge theories as constraints, which are enough to fix the possibly remaining degrees of freedom.
It turns out that for a large class of form factors, these universal physical constraints are sufficient to determine the maximally transcendental parts uniquely, irrespective of which gauge theory is under consideration.
We will discuss these in detail for the minimal form factors and the three-point form factor of ${\rm tr}(F^2)$ up to two loops.

To test the MTP beyond the known examples, we also consider the two-loop four-gluon form factor of the $\mathrm{tr}(F^3)=\mathrm{tr}(F_{\mu}^{~\nu} F_{\nu}^{~\rho} F_{\rho}^{~\mu})$ operator and obtain its maximally transcendental parts for the first time. 
This is a more non-trivial four-point next-to-minimal form factor and has much richer structures. 
The form factor depends on seven independent Lorentz invariants: six Mandelstam variables $s_{ij}=(p_i+p_j)^2$ with $1\leqslant i<j \leqslant4$ and one parity-odd variable ${\rm tr}_5$. The general ansatz contains 221 master integrals, and the coefficients depend on five spinor factors.
We find that the physical constraints from the infrared divergences and the collinear limits, together with the cancellation of spurious poles, can fix a significant part of the ansatz. The remaining degrees of freedom can be fixed by a simple type of quadruple unitarity cuts.
A nice fact we find is that the difference of the two-loop form factors in different gauge theories can only depend on two free parameters.

We obtain the maximally transcendental parts of the form factor in both $\mathcal{N}=4$ SYM and pure YM theories. Interestingly, the results are different in the two theories.
The reason is that there are extra maximally transcendental contributions in ${\cal N}=4$ SYM from the diagrams involving the fermion (\emph{i.e.}~gluino) loop.
In contrast, the scalar-loop diagrams in ${\cal N}=4$ SYM have no maximally transcendental contribution.
This crucial fact implies that the results of $\mathcal{N}=4$ SYM and QCD are still identical, once one converts the fundamental quarks in QCD to be adjoint fermions, which can be achieved by a proper change of the color factors in the QCD form factor.
Thus, the MTP still holds in this case.

As another intriguing observation, we find that the four-point form factor of the half-BPS $\mathrm{tr}(\phi^3)$ operator \cite{Guo:2021bym} is identical to a part of the $\mathrm{tr}(F^3)$ form factor that carries the same spinor factors. This seems to be not trivial since the two form factors have different spinor structures.

This paper is organized as follows. 
In Section~\ref{sec:strategy}, we give a brief review of the general bootstrap strategy and then discuss the collinear limit for form factors in detail.
In Section~\ref{sec:miniFF}, we apply the bootstrap strategy to prove the MTP for two-loop minimal form factors, and some constraints on the lower transcendental parts are also considered.
In Section~\ref{sec:3ptFF} we consider the two-loop three-point form factors of $\mathrm{tr}(F^2)$ via the bootstrap method together with unitarity cuts.
In Section~\ref{sec:2loop4ptF3}, we compute the four-point form factors of $\mathrm{tr}(F^3)$ up to two loops and discuss the maximal transcendentality properties. 
A summary and discussion are given in Section~\ref{sec:discussion}.
Several appendices provide the integral conventions, some explicit results, as well as some technical details.
Appendix~\ref{app:UT} provides the definition of pure UT master integrals used in the paper.
Appendix~\ref{app:catani} discusses the Catani IR subtraction formula and its relation to the BDS subtraction.
Appendix~\ref{app:letterandcollinear} gives the definition of symbol letters as well as their collinear limit behavior for the four-point two-loop form factors.
Appendix~\ref{app:HigherOrder} discusses a technical point about the constraints from the $\mathcal{O}(\epsilon)$ order of form factors.
Appendix~\ref{app:BuildingBlocks} provides the building blocks obtained from the bootstrap of the four-point form factor.
Appendix~\ref{app:fullFF} gives the results of the finite remainder function of the four-point form factor.
Appendix~\ref{app:A4noMT} discusses the one-loop four-gluon amplitude, which provides a counterexample of the maximally transcendental principle for the amplitude case.

%%%%%%%%%%%%%%%%%%%%%%%%%%%%%%%
\section{Bootstrap strategy based on master integrals}
\label{sec:strategy}

It is known that an $l$-loop amplitude or form factor can be expanded in a set of integral basis as
\begin{equation}
	\label{eq:masterexpansion}
	\mathcal{F}^{(l)} = \sum_i \alpha_{i} \, I_{i}^{(l)} \,,
\end{equation}
where $I_{i}^{(l)}$ can be chosen as a set of master integrals obtained via the integration-by-part reduction \cite{Chetyrkin:1981qh, Tkachov:1981wb}. 
The number of IBP master integrals for a given quantity is also known to be finite \cite{Smirnov:2010hn}.
While the basis integrals are theory independent, the intrinsic physical information is contained in the coefficients $\alpha_i$, which will be the main target we investigate.

In the traditional Feynman diagram method, one usually starts with Feynman diagrams and then performs the integral reduction to get the coefficients. This typically requires complicated intermediate steps and the results are also often given in incomprehensible forms. 
Besides, the physical properties (mentioned below) are not manifest in such a computation but only provide consistency checks for the final results.

The bootstrap strategy takes a very different route: the final form of the result such as \eqref{eq:masterexpansion} is taken as the starting ansatz, and the physical consistency conditions are used at the very beginning of the computation, namely, they are used as constraints to solve the coefficients in the ansatz. 
In this way, the physical properties are manifest in each step, and this often leads to a result in a compact form.

In Section~\ref{sec:constraints} we first briefly discuss various physical properties that will be used as constraints in later computations, then we will provide some details about the collinear limit of form factors in Section~\ref{sec:collinearFF}.

%%%%%%%%%%%%%%%%%%%%%%
\subsection{Physical constraints}
\label{sec:constraints}

The constraints are from the general properties of physical quantities, 
including: (1) the loop quantity should reproduce the general infrared (IR) divergences, (2) it should satisfy the collinear factorization property, (3) the spurious poles must cancel in the full result, and (4) it should satisfy unitarity cuts or other possible constraints. 
Below we discuss them in more detail.

%%%%%%%%%%%%%%%%%
\paragraph{IR divergences.}

Amplitudes and form factors with massless external states have IR divergences, which have universal structures and are related to the number and types of external massless particles.
In the planar limit, for example, IR divergences are captured by the two-point Sudakov form factors \cite{Mueller:1979ih,Collins:1980ih,Sen:1981sd,Magnea:1990zb}, 
which are determined by two kinematics-independent nubmers:
the cusp anomalous dimension $\gamma_\text{cusp}$ \cite{Korchemsky:1985xj, Korchemsky:1988si}
and the collinear anomalous dimension  $\mathcal{G}_{\rm coll}$ (see \emph{e.g.}~\cite{Dixon:2017nat}).
For amplitudes or form factors with multiple external legs, IR divergences for general massless gauge theories can be conveniently taken into account by the Catani formula \cite{Catani:1998bh}. Since our main focus is on the maximally transcendental parts, it is convenient to use the Bern-Dixon-Smirnov (BDS) ansatz \cite{Bern:2005iz, Anastasiou:2003kj} which also captures the collinear behavior, as will be explained shortly below.
Some details of the Catani formula and its relation to the BDS form are given in Appendix~\ref{app:catani}.

%%%%%%%%%%%%%%%%%
\paragraph{Collinear limits.}

When two external legs are taken in the collinear limit, the form factors satisfy factorization formula as (see \emph{e.g.}~\cite{Kosower:1999xi}):
\begin{equation}
\label{eq:collinear-general}
	\mathcal{F}^{(L)}_n(1,\ldots, a^{h_a}, b^{h_b}, \ldots, n) \xrightarrow{ p_a || p_b } \sum_\ell \sum_\sigma \mathbf{Sp}_{-\sigma}^{(\ell)}(a^{h_a},b^{h_b}) \, \mathcal{F}^{(L-\ell)}_{n-1}(1,\ldots, (a+b)^\sigma, \ldots, n) \,.
\end{equation}
For example, in the linear limit $p_a \, || \, p_b \, || \, P=p_a+p_b$
\begin{equation}
	p_a \rightarrow z P , \qquad p_b \rightarrow (1-z)P \,,
\end{equation}
the one-loop splitting amplitude $\mathbf{Sp}^{(1)}$ can be given as \cite{Bern:1994zx, Kosower:1999rx, Bern:1999ry}
\begin{equation}
\label{eq:splittingAmp00}
	\mathbf{Sp}^{(1)}(P\rightarrow a \, b; z) = \mathbf{Sp}^{(0)}(P \rightarrow a\,b; z)\ r_1^{[1], \text{MT}}(P^2, z) + \textrm{(lower transendental part)}  \,,
\end{equation}
where the maximally transcendental part of the one-loop splitting function (denoted by the superscript `MT') is
\begin{equation}
\label{eq:splittingAmp}
	r_1^{[1], \text{MT}}(P^2, z)  = \frac{ e^{\epsilon \gamma_\text{E}}\Gamma (-\epsilon)^2 \Gamma (\epsilon +1)}{\Gamma (1-2 \epsilon )} (- P^2 )^{-\epsilon}\Big\{ 1-z^{-\epsilon}-\left({1-z}\right)^{-\epsilon} + \epsilon ^2 \big[ \log (z)\log (1-z) -\zeta_2 \big] + {\cal O}(\epsilon^3) \Big\} \,.
\end{equation}
We stress that \eqref{eq:splittingAmp} is universal for general gauge theories, and this formula will be used to bootstrap the one-loop three- and four-point form factors in Section~\ref{subsec:3ptff1loop} and Section~\ref{sec:solveFF4pt1loop}.

Beyond one-loop order, there is a convenient way to capture both the IR and collinear behavior 
by using the BDS ansatz \cite{Bern:2005iz} for $\mathcal{N}=4$ SYM (or the maximally transcendental parts in general gauge theories, see more discussion in Appendix~\ref{app:catani}). The loop correction at two loops can be given as 
\begin{equation}
	\label{eq:BDSansatz}
	\mathcal{I}^{(2)} = \frac{1}{2} \left( \mathcal{I}^{(1)}(\epsilon) \right)^2 + f^{(2)}(\epsilon) \mathcal{I}^{(1)}(2\epsilon) + \mathcal{R}^{(2)} + \mathcal{O}(\epsilon) \,,
\end{equation}
where
\begin{equation}
\label{eq:f2def}
	f^{(2)}(\epsilon)=-2 \zeta_{2}-2 \zeta_{3} \epsilon-2 \zeta_{4} \epsilon^{2} \,.
\end{equation}
The original two-loop BDS ansatz is proposed with only the first two terms in \eqref{eq:BDSansatz} \cite{Bern:2005iz, Anastasiou:2003kj}:
\begin{equation}
	\label{eq:BDSansatzOrigin}
	\mathcal{I}^{(2), {\rm BDS}} = \frac{1}{2} \left( \mathcal{I}^{(1)}(\epsilon) \right)^2 + f^{(2)}(\epsilon) \mathcal{I}^{(1)}(2\epsilon) \,,
\end{equation}
which were constructed in a way that they capture all the IR divergences and also have correct collinear behavior of amplitudes. This original ansatz is correct for the four- and five-point amplitudes in ${\cal N}=4$ SYM, but for higher-point amplitudes an extra finite remainder function is needed \cite{Drummond:2008aq, Bern:2008ap}, denoted as  $\mathcal{R}^{(2)}$ in  \eqref{eq:BDSansatz}.  
The same BDS-ansatz structure also generalizes to form factors \cite{Brandhuber:2012vm}.
Since the remainder function is free from both IR and collinear singularities, it has the important property that the $n$-point remainder reduces trivially to $(n-1)$-point remainder in the collinear limit as 
\begin{equation}
	\label{eq:remainderCL}
	\mathcal{R}_n^{(2)}  \ \xlongrightarrow[\mbox{}]{\mbox{$p_i \parallel p_{i+1}$}} \ \mathcal{R}_{n-1}^{(2)} \,.
\end{equation}
This will provide useful constraints for the two-loop three- and four-point form factors in Section~\ref{sec:3ptFF} and Section~\ref{sec:2loop4ptF3}.

%%%%%%%%%%%%%%%%%
\paragraph{Spurious pole cancellation.}

For form factors with non-trivial spinor structures (such as the four-point form factors considered later), the coefficients of master integrals can contain spurious poles (\emph{i.e.}~unphysical poles). 
The cancellation of spurious poles typically requires a combination of both the spinor factors and the master integrals, which can provide non-trivial constraints on the coefficients of master integrals. 
The details of the spurious poles as well as applying their cancellation as constraints will be given in Section~\ref{sec:2loop4ptF3} for the discussion of four-point form factors.

%%%%%%%%%%%%%%%%%
\paragraph{Lightlike limit of $q$.}

The form factor as defined in \eqref{eq:FF-def} contains an off-shell momentum $q=\sum_i p_i$ which is carried by the operator. An interesting limit to consider is the lightlike limit $q^2 \rightarrow 0$.\footnote{This limit of $q^2\rightarrow0$ has also been considered for the three-point form factor of $\mathrm{tr}(F^2)$ in \cite{Lin:2021lqo}.}
Since the form factors are equivalent Higgs-plus-gluons amplitudes where $q^2 = m_H^2$ \cite{Brandhuber:2012vm, Gehrmann:2011aa}, this lightlike limit of $q$ can be understood as the massless limit $m_H \rightarrow0$ of the Higgs particle, therefore it is reasonable to expect that
the form factor should have a smooth limit. 
This can provide useful constraints and will play a role in Section~\ref{sec:3ptFF}.

%%%%%%%%%%%%%%%%%
\paragraph{Unitarity cuts.}

When the above constraints are not enough to fix the full results, one can use another powerful tool -- the unitarity cut constraints \cite{Bern:1994zx, Bern:1994cg, Britto:2004nc}.
The unitarity cut method is a powerful method that in principle can determine the full result. 
Here in our application together with the bootstrap strategy, the nice point is that after using the above physical constraints, only a small number of simple unitarity cuts are needed to fix the remaining free parameters. 
Moreover, these cuts can often be chosen such that they are universal for general gauge theories, and this fact will play an important role in the proof of MTP.

At this point, it may be good to compare our strategy with the symbol-bootstrap method that has been used for computing the finite remainder of amplitudes (see \emph{e.g.}~\cite{Dixon:2011pw, Dixon:2013eka,Dixon:2014iba,Golden:2014pua,Drummond:2014ffa, Caron-Huot:2016owq,Dixon:2016nkn, Drummond:2018caf, Caron-Huot:2019vjl,Dixon:2020cnr, Zhang:2019vnm, He:2020vob, Golden:2021ggj}) and form factors \cite{Brandhuber:2012vm, Dixon:2020bbt, Dixon:2022rse} in ${\cal N}=4$ SYM.
Unlike the symbol bootstrap where one considers only the finite remainder functions,
the bootstrap method used here starts with an ansatz of the full form factor in terms of master integrals. 
This requires the knowledge of master integrals and in this sense, it contains more input information than the symbol bootstrap. 
However, since the master integrals are theory-independent, this strategy can be used for general observables in general theories. 
In particular, physical constraints that are not available in symbol bootstrap can be applied here, such as the IR divergences and the unitarity-cut constraints. 
As we will see, these new features make it possible to prove the maximal transcendentality principle that relates form factors in different theories.

%%%%%%%%%%%%%%%%%%%%%%%%%%%%%%%
\subsection{Collinear limit of form factors}
\label{sec:collinearFF}

In this subsection, we provide some details about the collinear limit for form factors. This is mainly used in the computation of the four-point form factor in Section~\ref{sec:2loop4ptF3}. In that cases, the kinematic variables are a bit complicated and the collinear limit must be taken properly. In the meanwhile, the introduced momentum twistor variables will also help to understand the structure of the symbol letter variables of the four-point form factors.

\begin{figure}[tb]
	\centering
	\includegraphics[scale=0.5]{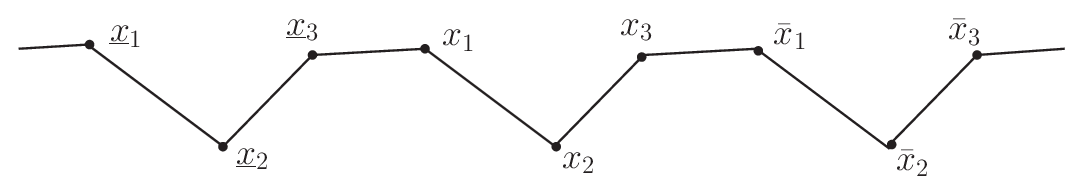}
	\caption{Dual periodic Wilson line configuration for the three-point form factor.}
	\label{fig:WL3pt_x}
\end{figure}

\begin{figure}[tb]
	\centering
	\includegraphics[scale=0.5]{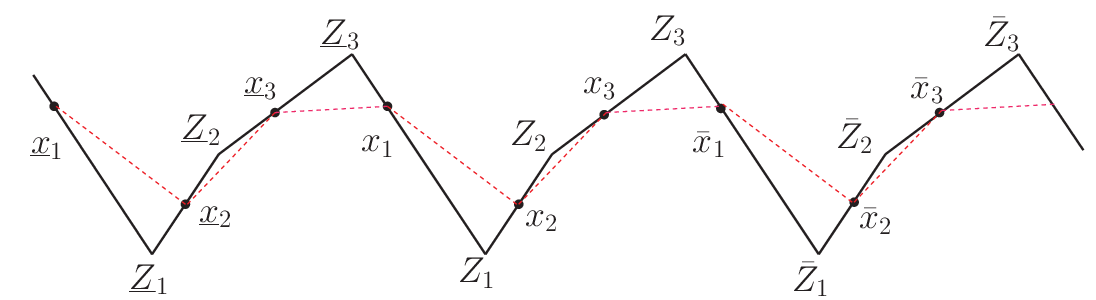}
	\caption{Momentum twistor space picture for the dual periodic Wilson line configuration for the three-point form factor. The red dashed line corresponds to the $x$-configuration in Figure~\ref{fig:WL3pt_x}.}
	\label{fig:WL3pt}
\end{figure}

We introduce the periodic Wilson line configuration in the momentum space \cite{Alday:2007hr, Maldacena:2010kp,Brandhuber:2010ad,Gao:2013dza}, as shown in Figure~\ref{fig:WL3pt_x} for the three-point form factor case. 
One has
\begin{equation}
	x_i - x_{i+1} = p_i = \lambda_i \widetilde\lambda_i \,, \qquad x_{\underline{i}} - x_i = x_i - x_{\bar{i}} = q \,.
\end{equation} 
As in the case of scattering amplitudes, each dual coordinate $x_i$ corresponds to a line in the (dual) twistor space which is represented by two momentum twistor variables $Z_{i-1}, Z_i$ \cite{Hodges:2009hk, Mason:2009qx}. The momentum twistor variables can be defined as 
\begin{equation}
	\label{eq:Z-def}
	Z_{i}^A = (\lambda_i^\alpha, \mu_i^\beta) \,, \qquad \mu_i^\beta = x_i^{\alpha\beta} \cdot \lambda_{i \alpha} =  x_{i+1}^{\alpha\beta} \cdot \lambda_{i \alpha}  \,.
\end{equation}
The periodic Wilson line configuration in momentum twistor space is shown in Figure~\ref{fig:WL3pt}.

%%%%%%%%%%%%%%%%
%%%%%%%%%%%%%%%%

%%%%%%%%%%%%%%%%
%%%%%%%%%%%%%%%%

The momentum twistor variables and dual spacetime coordinates can be related using the following formula
\begin{equation}
	\label{eq:REL}
	\langle i | \, x_{i j}\, x_{j k} \, | k \rangle = \frac{\langle Z_i Z_{j-1} Z_j  Z_k \rangle}{\langle j-1, j \rangle } \,,
\end{equation}
where $x_{ij} \equiv x_i - x_j$, $\langle i j \rangle \equiv \epsilon^{\alpha\beta} \lambda_{i \alpha} \lambda_{j \beta}$, and $\langle Z_i Z_j Z_k Z_l \rangle = \epsilon_{ABCD}Z_i^A Z_j^B Z_k^C Z_l^D$.
From \eqref{eq:REL}, one can obtain following useful relations
\begin{equation}
	\label{eq:REL-x2}
	x_{i j}^2 = {\langle Z_{i-1} Z_i Z_{j-1} Z_j \rangle \over \langle i-1, i \rangle \langle j-1, j \rangle} \,.
\end{equation}
In the following, we will use the abbrevation for the four-brackets $\langle Z_i Z_j Z_k Z_l \rangle = \langle i j k l \rangle$.

For the three-point case, using \eqref{eq:REL}-\eqref{eq:REL-x2}, one can rewrite the ratio variables in terms of twistor four-brackets and spinor products, such as
\begin{align}
	& u = \frac{s_{12}}{q^2} = {x_{13}^2 \over x_{1 \bar{1}}^2} = {\langle \underline{3} 1 2 3 \rangle \over \langle \underline{3} 1 3 \bar{1} \rangle} {\langle 3 \bar{1} \rangle \over \langle 2 3 \rangle} \,, \qquad\qquad
	1 - {1\over u} = {\langle \underline{3} \bar{1} 2 3 \rangle \over \langle \underline{3} 1 2 3 \rangle}
	{\langle \underline{3} 1 \rangle \over \langle \underline{3} \bar{1} \rangle} \,, \\
	& v = \frac{s_{23}}{q^2} = {x_{2\bar{1}}^2 \over x_{2 \bar{2}}^2} = {\langle 123 \bar{1} \rangle \over \langle 1 2  \bar{1} \bar{2} \rangle} {\langle \bar{1}\bar{2}\rangle \over \langle 3 \bar{1} \rangle} \,, \qquad\qquad 1 - {1\over v} = {\langle 1 \bar{2}  3\bar{1} \rangle \over \langle 1 2 3 \bar{1} \rangle}{\langle 12 \rangle \over \langle 1 \bar{2} \rangle} \,, \\
	& w = \frac{s_{13}}{q^2} = {x_{3\bar{2}}^2 \over x_{3 \bar{3}}^2} = {\langle 2 3 \bar{1}\bar{2} \rangle \over \langle 2 3 \bar{2}\bar{3} \rangle} {\langle\bar{2}\bar{3} \rangle\over\langle \bar{1} \bar{2} \rangle}\,,\qquad\qquad 1- {1\over w} = {\langle 2 \bar{3} \bar{1} \bar{2} \rangle \over \langle 2 3 \bar{1} \bar{2} \rangle}{\langle 23 \rangle \over \langle 2 \bar{3} \rangle} \,.
\end{align}
These variables are enough to provide the symbol letters for the remainder functions of three-point form factors of ${\rm tr}(F^2)$ \cite{Brandhuber:2012vm, Dixon:2020bbt}. In this case, the collinear limit is relatively trivial, for example, in the limit of $p_1 \parallel p_3$, one has $s_{13}\rightarrow0$, $q^2 \rightarrow s_{12}+s_{23}$ and
\begin{equation}
u \rightarrow u, \qquad v \rightarrow 1-u, \qquad w \rightarrow 0 \,. 
\end{equation}

%%%%%%%%%%%%%%%%
\begin{figure}[tb]
	\centering
	\includegraphics[scale=0.5]{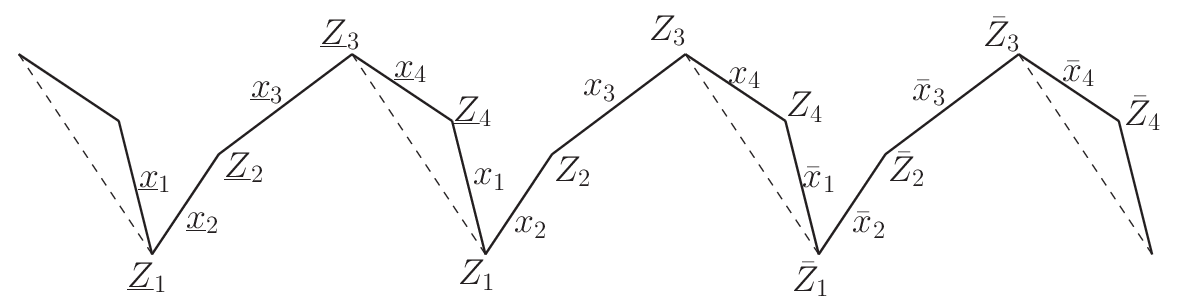}
	\caption{Dual periodic Wilson line configuration for the four-point form factor in momentum twistor space.}
	\label{fig:WL4pt}
\end{figure}
%%%%%%%%%%%%%%%%

The kinematics of the four-point form factor is more complicated. The periodic Wilson line configuration in momentum twistor space is shown in Figure~\ref{fig:WL4pt}. 
We define the ratio variables:
\begin{equation}\label{eq:defuij}
u_{ij} = \frac{s_{ij}}{q^2} \,, \qquad u_{ijk} = \frac{s_{ijk}}{q^2} \,,
\end{equation}
where $s_{ij \ldots k} = (p_i+p_j+\ldots+p_k)^2$.
They can be represented by momentum twistor as
\begin{align}
	& u_{i,i+1} = \frac{x_{i,i+2}^2}{x_{\underline{i},i}^2} = \frac{ \langle i-1,i, i+1, i+2 \rangle \langle \underline{i}-1, \underline{i} \rangle }{ \langle \underline{i}-1, \underline{i}, i-1, i \rangle \langle i+1, i+2 \rangle } \, , \\
	& u_{i,i+1,i+2} =  \frac{x_{i,i+3}^2}{x_{\underline{i}, i}^2} = \frac{\langle i-1, i, i+2, i+3 \rangle \langle \underline{i}-1, \underline{i} \rangle }{\langle \underline{i}-1, \underline{i}, i-1, i \rangle \langle i+2, i+3 \rangle } \, . \nonumber
\end{align}
Note that there are only five independent ratio variables, which can be chosen as five of $u_{ij}$:
\begin{equation}
\{ u_{12}, \, u_{23}, \, u_{34}, \, u_{14}, \, u_{13}, \, u_{24}\} \,, \qquad u_{12}+u_{23}+u_{34}+u_{14}+u_{13}+u_{24} = 1 \,.
\end{equation}

There are also other variables which necessarily appear in the two-loop functions, which we define as
\begin{align}
	x_{ijkl}^{\pm} = \frac{q^2+s_{ij}-s_{kl}\pm\sqrt{\Delta_{3,ijkl}}}{2s_{ij}}, \quad
	y_{ijkl}^{\pm} = \frac{\mathrm{tr}_{\pm}(ijkl)}{2s_{ij}s_{il}}, \quad
	z_{ijkl}^{\pm\pm} = 1+y_{ijkl}^{\pm}-x_{lijk}^{\pm} ,
\end{align}
where $\Delta_{3,ijkl} = {\rm Gram}(p_i+p_j, p_k+p_l)$ also appears in the one-loop three-massive triangle integral, 
and $\mathrm{tr}_{\pm}(ijkl) = s_{ij} s_{kl}-s_{ik} s_{jl}+s_{il} s_{jk} \pm \mathrm{tr}_5$, in which the parity-odd kinematics 
\begin{equation}\label{eq:tr5}
\mathrm{tr}_5 = 4i \epsilon_{\mu\nu\rho\sigma}p_1^{\mu}p_2^{\nu}p_3^{\rho}p_4^{\sigma} = \langle 1|2|3|4|1 ]-[1|2|3|4|1\rangle
\end{equation} 
is related to the Gram determinant $\Delta_5 = {\rm Gram}(p_1, p_2, p_3, p_4)$ by $\Delta_5 = \mathrm{tr}_5^2$.
These variables are used to define the symbol letters in the remainder function where we are defined in Appendix~\ref{app:letters}.

Since the $y^\pm$ variables are related to the parity-odd variable $\mathrm{tr}_5$, we discuss them in detail below.
Using the relations
\begin{equation}
	y_{ijkl}^+ = \frac{\mathrm{tr}_{+}(ijkl)}{2s_{ij}s_{il}} = \frac{\langle l | k | j ]}{\langle l | i | j ]} \,, \qquad 
	y_{ijkl}^- = \frac{\mathrm{tr}_{-}(ijkl)}{2 s_{ij} s_{il}} = \frac{\langle j | k | l ]}{\langle j | i | l ]} \, ,
\end{equation}
one has
\begin{align}
	& y_{1234}^+ = \frac{\langle 1234 \rangle}{\langle \underline{4}123 \rangle} \frac{\langle 1 \underline{4} \rangle}{\langle 1 4 \rangle} \,, \qquad\qquad\qquad\qquad\quad\ y_{ijkl}^- = \frac{u_{jk} u_{kl}}{u_{ij} u_{il}} \big(y_{ijkl}^+ \big)^{-1} \,, \nonumber \\		
	& y_{1324}^+ = \frac{\langle 13 \rangle \langle 24 \rangle}{\langle 14 \rangle \langle 23 \rangle} \Big(\frac{u_{123}-u_{12}}{u_{23}}-1 \Big)^{-1} \,, \qquad  y_{3124}^+ = y_{1324}^+ \big|_{p_1 \leftrightarrow p_3} \,, \nonumber \\
	& y_{1342}^+ = \frac{\langle 13 \rangle \langle 24 \rangle}{\langle 12 \rangle \langle 34 \rangle} \Big(\frac{u_{134}-u_{14}}{u_{34}}-1 \Big)^{-1} \,, \qquad  y_{3142}^+ = y_{1342}^+ \big|_{p_1 \leftrightarrow p_3} \,. \nonumber
\end{align}

Unlike the three-point case, the collinear limit for the four-point form factor should be carefully taken. 
Consider the collinear limit $p_4 \parallel p_3$, one can parametrize $Z_4$ in the following way:
\begin{equation}
	Z_4 = Z_3 + \delta {\langle \bar{1} \bar{2} 1 3 \rangle \over \langle \bar{1} \bar{2} 1 2 \rangle} Z_2 + \tau \delta {\langle \bar{2} 1 2 3 \rangle \over \langle \bar{1} \bar{2} 1 2 \rangle} \bar{Z}_1 + \eta {\langle \bar{1} 1 2 3 \rangle \over \langle \bar{1} \bar{2} 1 2 \rangle} \bar{Z}_2 \,,
\end{equation}
where the ratio of four brackets are introduced to balance the twistor weight. The collinear limit can be achieved by taking first $\eta \rightarrow 0$, followed by $\delta \rightarrow 0$. The parameter $\tau$ is finite which gives the momentum fraction shared by $p_4$ in the limit. 
Such a parametrization for the collinear limit was introduced for the amplitudes case in \cite{CaronHuot:2011ky}.
The same limit applies simultaneously to $\underline{Z}_4, \bar{Z}_4$ due to the periodicity condition.
And we can also define the collinear limit for spinors using \eqref{eq:Z-def} as
\begin{align}
	& \lambda_4 = \lambda_3 + \delta {\langle \bar{1} \bar{2} 1 3 \rangle \over \langle \bar{1} \bar{2} 1 2 \rangle} \lambda_2 + \tau \delta {\langle \bar{2} 1 2 3 \rangle \over \langle \bar{1} \bar{2} 1 2 \rangle} \bar{\lambda}_1 + \eta {\langle \bar{1} 1 2 3 \rangle \over \langle \bar{1} \bar{2} 1 2 \rangle} \bar{\lambda}_2 \,.
\end{align}
The explicit collinear limit for all letter variables are given in Appendix~\ref{app:letterCL}.

Finally, let us mention an alternative way to consider the collinear limit by expressing all variables in terms of spinor variables $\{\lambda_i, \tilde\lambda_j\}$. For example, the collinear limit $p_i \parallel p_j$ can be taking by using following parameterization of spinor variables
\begin{align}
	& \lambda_i \rightarrow \lambda_i \,, \qquad\qquad\quad\qquad \tilde{\lambda}_i \rightarrow (1-t) \tilde{\lambda}_i \,,  \\
	& \lambda_j \rightarrow  \lambda_i + \delta \frac{\langle il \rangle}{\langle kl \rangle} \lambda_k \,, \qquad \tilde{\lambda}_j \rightarrow t \tilde{\lambda}_i + \tilde{\delta} \frac{\left[il\right]}{\left[kl\right]} \tilde{\lambda}_k \,, \nonumber
\end{align}
where $\{\lambda_k, \tilde{\lambda}_k\}$ and $\{\lambda_l, \tilde{\lambda}_l\}$ are the two pairs of reference spinors and the indices $\{i,j,k,l\}$ are not equal to each other. Then the collinear limit can be achieved by taking $\delta, \tilde{\delta} \rightarrow 0$, where there is no need to distinguish the order of limits here. The formula gives the limit behavior as $p_i \rightarrow (1-t) p_i$ and $p_j \rightarrow t p_i$, thus the momentum conservation is satisfied by $p_i+p_j = \lambda_i \tilde{\lambda}_i + \lambda_j \tilde{\lambda}_j \rightarrow \lambda_i \tilde{\lambda}_i + \mathcal{O}(\delta,\tilde{\delta})$.

%%%%%%%%%%%%%%%%%%%%%%%%%%%%%%%
\section{Two-loop minimal form factors}
\label{sec:miniFF}

In this section, we consider a special class of form factors, the so-called minimal form factors:
\begin{equation}
	\mathcal{F}_{\mathcal{O}_n, \text{min}} = \int d^D x e^{-i q\cdot x} \langle p_1\, p_2 \ldots p_n | \mathcal{O}_n(x)|0 \rangle \,.
\end{equation}
There are two requirements in the definition of minimal form factors. 
Consider the minimal form factor $\mathcal{F}_{\mathcal{O}_{n}, {\rm min}}$ of a length-$n$ operator $\mathcal{O}_{n} = \mathrm{tr}(\mathcal{W}_1 ... \mathcal{W}_n)$.
First, the number of external on-shell states should be equal to the length of the operator.
Second, the minimal form factor at tree level is required to be non-zero, \emph{i.e.}~$\mathcal{F}_{\mathcal{O}_{n}, {\rm min}}^{(0)} \neq 0$, thus the external states should have same field configuration as $\{{\cal W}_i\}$.
For example, the form factor $\mathcal{F}_{\mathrm{tr}(F^3)}(1^{q},2^{\bar{q}},3^-)$ is not a minimal form factor, as it is zero at tree-level.
On the other hand,
the form factor of $\mathrm{tr}(\phi^n)$ and $n$ external on-shell scalar states, or the form factor of $\mathrm{tr}(F^n)$ with $n$ external gluon states, are minimal form factors.
Such form factors have been studied to two-loop order in $\mathcal{N}=4$ SYM and QCD \cite{Wilhelm:2014qua, Brandhuber:2014ica, Loebbert:2015ova, Brandhuber:2016fni, Loebbert:2016xkw, Brandhuber:2017bkg, Jin:2018fak, Brandhuber:2018xzk, Brandhuber:2018kqb, Jin:2019ile, Jin:2019opr, Jin:2020pwh,Lin:2020dyj}. These minimal form factors have played an important role for computing anomalous dimensions of high-length operators. The results of minimal form factors show that their maximally transcendental parts are given by same functions up to two-loop order, and lower transcendentality parts also present some universal structures \cite{Jin:2019ile, Jin:2019opr}. 

The goal of the section is to prove these universal structures.
We will show that by imposing only the IR constraint, it is enough to fix the maximally transcendental part of minimal form factors up to two loops. 
We also show that the IR divergences can be used to put strong constraints on the lower transcendental parts.
As a brief outline,
we will first consider the maximally transcendental parts up to two loops in Section~\ref{sec:MTminimal}.
Next, we will consider the constraints for the lower transcendental part in Section~\ref{sec:LTminimal}. 
Finally, we apply the bootstrap results together with some unitarity-cut arguments to explain the universality of the results in different theories in Section~\ref{sec:univer-mimi}.

%%%%%%%%%%%%
\paragraph{Setup.}

We define the $L$-loop $n$-point correction function $\mathcal{I}_{\text{min}, n}^{(L)}$ by factorizing out the tree-level minimal form factor as
\begin{equation}
	\mathcal{F}_{\mathcal{O}_n, \text{min}}^{(L)} = \mathcal{F}_{\mathcal{O}_n, \text{min}}^{(0)} \mathcal{I}_{\text{min}, n}^{(L)} \,.
\end{equation}
The loop correction function can be expanded in set of master integrals
\begin{equation}
	\mathcal{I}_{\text{min}, n}^{(L)} = \sum_i c_i(\epsilon) {I}_{i}^{(L)} \,,
\end{equation}
where $\{{I}_{i}^{(L)}\}$ is a set of master integrals and the coefficient $c_i(\epsilon)$ depends on $\epsilon = {(4-D)/2}$ and Mandelstam variables. An important remark follows: throughout this paper, we will always choose the set of master integrals, such that they have uniform transcendentality degree $2L$ at $L$ loops. All such masters which are used in this paper are collected in Appendix~\ref{app:UT}. They are often called pure uniformly transcendental (UT) integrals as defined in the canonical differential equations \cite{Henn:2013pwa}. Since the maximal transcendentality degree of $L$-loop form factors is $2L$, the coefficients $c_i(\epsilon)$ must be polynomials in $\epsilon$; here as in usual case, $\epsilon$ is assigned to carry transcendentality degree of $-1$.
For the convenience of later discussion, we reorganize the loop correction function in terms of different transcendentality degree as
\begin{equation}
\label{eq:minidegCalF}
	\mathcal{I}_{\text{min}, n}^{(L)} = \sum_{a=2L}^{-\infty} \epsilon^{2L-a} \mathcal{I}_{\text{min}, n}^{(L), \text{deg-}a} \, ,
\end{equation}
where $\mathcal{I}_{\text{min}, n}^{(L), \text{deg-}a}$ corresponds to the correction to the transcendentality degree-$a$ part and is a linear combination of pure UT master integrals with coefficients free of $\epsilon$.
Explicitly, at one and two loops one has
\begin{align}
	\mathcal{I}_{\text{min}, n}^{(1)} &= \mathcal{I}_{\text{min}, n}^{(1), \text{deg-2}} + \epsilon \, \mathcal{I}_{\text{min}, n}^{(1), \text{deg-1}} + (\textrm{lower degrees}) \,, \\
	\mathcal{I}_{\text{min}, n}^{(2)} &= \mathcal{I}_{\text{min}, n}^{(2), \text{deg-4}} + \epsilon \, \mathcal{I}_{\text{min}, n}^{(2), \text{deg-3}} + \epsilon^2 \, \mathcal{I}_{\text{min}, n}^{(2), \text{deg-2}} + (\textrm{lower degrees}) \,,
\end{align}
where $\mathcal{I}_{\text{min}, n}^{(L), \text{deg-a}}$ is free from $\epsilon$.
Our goal is to compute $\mathcal{I}_{\text{min}, n}^{(L), \text{deg-a}}$ in terms of master integral expansions.

%%%%%%%%%%%%%%%%%
\subsection{Maximally transcendental part}
\label{sec:MTminimal}

The maximally transcendental parts are given by $\mathcal{I}_{\text{min}, n}^{(1), \text{deg-2}}$ at one loop and $\mathcal{I}_{\text{min}, n}^{(2), \text{deg-4}}$ at two loops. For simplicity, we will focus on the planar form factors with all particles in adjoint representation. We will discuss other representations such quarks in QCD in Section~\ref{sec:univer-mimi}.

\subsubsection{One-loop case}
\label{sec:MTmini1loop}

We start with the simple one-loop case as a warm-up.
For the minimal form factor at the one-loop level, the interactions involve at most two external legs at one time, thus the only type of master integrals are one-loop bubble integrals as defined in \eqref{eq:bubbleintegral}:\footnote{We consider gauge theories with all fields being massless in the paper, thus there are no tadpole integrals.}
\begin{equation}
	I_{\text{Bub}}^{(1)}(i, j) = \frac{1}{\epsilon^2} - \frac{\log(-s_{ij})}{\epsilon} + \mathcal{O}(\epsilon^0) \,.
\end{equation}
The general ansatz of the planar form factor can be given by the summation of the bubble integrals as:
\begin{equation}
	\mathcal{I}_{\text{min}, n}^{(1), \text{deg-2}} =  \sum_{i=1}^n  c_{(i,i+1)} I_{\text{Bub}}^{(1)}(i, i+1) \,,
\end{equation}
where the external legs are circular as $I_{\text{Bub}}^{(1)}(n, n+1) = I_{\text{Bub}}^{(1)}(n, 1)$, and the similar convention is also adopted later.
The maximally transcendental part of the one-loop minimal form factor has universal IR divergences (which are determined by the maximally transcendental part of one-loop Sudakov form factor, see \emph{e.g.}~\cite{Bern:2005iz} for a review):
\begin{equation}
	\mathcal{I}_{\text{min}, n}^{(1), \text{deg-2}} \big|_\text{div.} = - \sum_{i=1}^n \bigg( \frac{1}{\epsilon^2} - \frac{\log(-s_{i,i+1})}{\epsilon} \bigg) \,.
\end{equation}
Matching the expression of bubble integrals, we find that
\begin{equation}
c_{(i,i+1)} = -1 \,.
\end{equation}
Therefore the maximally transcendental part of the one-loop minimal form factor is fixed uniquely as
\begin{equation}
\label{eq:miniMT1loop}
	\mathcal{I}_{\text{min}, n}^{(1), \text{deg-2}} = -\sum_{i=1}^n I_{\text{Bub}}^{(1)}(i, i+1) \, .
\end{equation}

%%%%%%%%%%%%%%%
\subsubsection{Two-loop case}
\label{sec:MTmini2loop}

Next, we consider the maximally transcendental part of the two-loop minimal form factor.
In this case, the interactions involve at most three external legs at one time. 
The two-loop master integrals for minimal form factors are shown in Figure~\ref{fig:master-mini-FF}. They are explicitly defined in Appendix~\ref{app:UT}.

%%%%%%%%%%%%%%%%%%%%%%%%%%%%%%%%%
\begin{figure}[t]
	\centering
	\subfloat[Sun]{\includegraphics[scale=0.5]{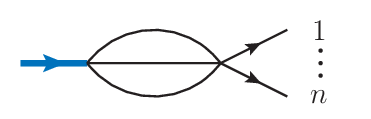}}
	\subfloat[dBub]{\includegraphics[scale=0.4]{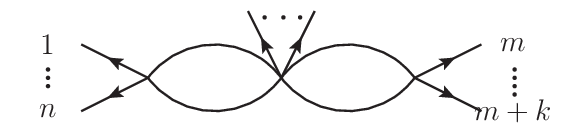}}	
	\subfloat[TBub0]{\includegraphics[scale=0.4]{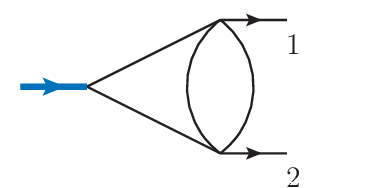}}
	
	\subfloat[TT2]{\includegraphics[scale=0.3]{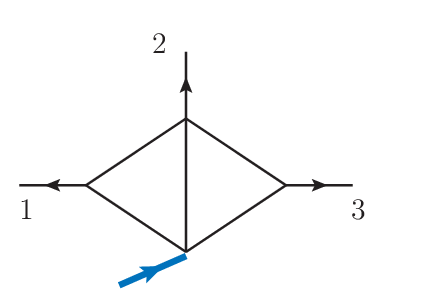}}
	\subfloat[TBub2]{\includegraphics[scale=0.4]{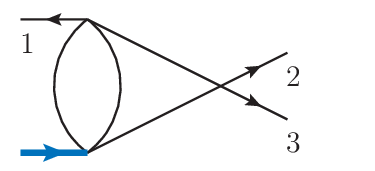}}
	\subfloat[BubBox]{\includegraphics[scale=0.4]{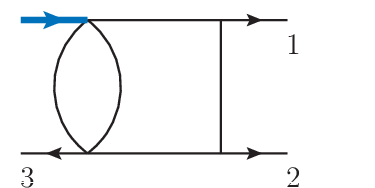}}
	\subfloat[NTBox1]{\includegraphics[scale=0.4]{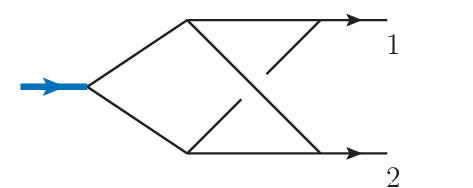}}
	\caption{Master integrals for the two-loop minimal form factors. \label{fig:master-mini-FF}}
\end{figure}
%%%%%%%%%%%%%%%%%%%%%%%%%%%%%%%%%

It is convenient to introduce the two-loop density function which involves up to four external legs $\{i,j,k,l\}$, and its most general form can be given as an expansion of master integrals:
\begin{align}
	\mathcal{I}_{2}^{(2)}(i, j) = &  c_1 I_{\text{Sun}}^{(2)}(i, j) + c_2 I_{\text{dBub}}^{(2)}(i, j; i, j) + c_3 I_{\text{TBub0}}^{(2)}(i, j) + c_4 I_{\text{NTBox1}}^{(2)}(i, j) \,, \nonumber\\
	\mathcal{I}_{3}^{(2)}(i, j, k) = & c_5 I_{\text{Sun}}^{(2)}(i, j, k) + c_6 I_{\text{dBub}}^{(2)}(i, j; i, j, k) + c_7 I_{\text{TT2}}^{(2)}(i, j, k) + c_8 I_{\text{TBub2}}^{(2)}(i, j, k) \nonumber \\
	& 
	+ c_9 I_{\text{BubBox}}^{(2)}(i, j, k) \,, \nonumber \\
	\mathcal{I}_{4}^{(2)}(i, j, k, l) = & c_{10} I_{\text{dBub}}^{(2)}(i, j; k, l) \,,
	\label{eq:MinAnsatz}
\end{align}
where $I_{\text{NTBox1}}^{(2)}(i,j)$ in $\mathcal{I}_{2}^{(2)}(i,j)$ are need to notice that they contribute to the leading-color part of Sudakov form factor($n=2$) although which are the non-planar master integrals. Without loss of generality, we take them into account, and which should be left in the case of higher points($n \geqslant 3$).

Then the two-loop ansatz for the planar minimal form factor can be given by the summation of the above functions as
\begin{equation}
	\mathcal{I}_{\text{min}, n}^{(2), \text{deg-4}} = \sum_{i=1}^n \bigg( \mathcal{I}_{2}^{(2)}(i, i+1) + \mathcal{I}_{3}^{(2)}(i, i+1, i+2) + \sum_{j=i+2}^{n+1} \mathcal{I}_{4}^{(2)}(i, i+1, j, j+1) \bigg) \,.
\end{equation}

Since we will use IR divergences to constrain the results, we can organize the two-loop results in terms of two types of building-blocks as
\begin{equation}
	\mathcal{I}_{\text{min}, n}^{(2), \text{deg-4}} = \sum_{\alpha} x_{\alpha} G_{\text{min}, \alpha}^{(2)} + \sum_{\beta} y_{\beta} \tilde{G}_{\text{min}, \beta}^{(2)} \,,
\end{equation}
and we will ask the first part $G_{\text{min},\alpha}^{(2)}$ to provide the correct IR divergences, and require the second part $\tilde{G}_{\text{min},\beta}^{(2)}$ are IR finite. Note $G_{\text{min},\alpha}^{(2)}$ and $\tilde{G}_{\text{min},\beta}^{(2)}$ are linear combination of master integrals.

%%%%%%%%%%%%%%%%%%%%%%%%%%%%%%%%%%%%%
\paragraph{Constructing building-blocks $G_{\text{min}, \alpha}^{(2)}$.}

We construct the building-blocks $G_{\text{min},\alpha}^{(2)}$ such that they provide the correct infrared divergences.
As reviewed in Section~\ref{sec:constraints}, the IR divergence can be captured by the BDS-ansatz function $\mathcal{I}_{\text{min}, n}^{(2), \text{BDS}}$, which is determined by the one-loop correction as
\begin{equation}
	\label{eq:BDS2loopmim}
	\mathcal{I}_{\text{min}, n}^{(2), \text{BDS}, \text{deg-4}} = \frac{1}{2}\left( \mathcal{I}_{\text{min}, n}^{(1), \text{deg-2}}(\epsilon) \right)^2 + f^{(2)}(\epsilon) \mathcal{I}_{\text{min}, n}^{(1), \text{deg-2}}(2\epsilon) \,.
\end{equation}
Note that we keep only the maximally transcendental part of degree $4$.
As reviewed in Appendix~\ref{app:catani}, the BDS ansatz and the Catani IR formula have the same maximally transcendental IR divergences, therefore the discussion with BDS ansatz here will not only apply to $\mathcal{N}=4$ SYM but also to general gauge theories.
Using the one-loop result \eqref{eq:miniMT1loop}, one can expand \eqref{eq:BDS2loopmim} as
\begin{align}
	\label{eq:nptInfraredStructure}
	\mathcal{I}_{\text{min}, n}^{(2), \text{BDS}, \text{deg-4}}
	= & \sum_{i = 1}^{n} \left( I_\text{Bub}^{(1)}(i, i+1)I_\text{Bub}^{(1)}(i+1, i+2) - f^{(2)}(\epsilon) I_\text{Bub}^{(1)}(i, i+1; 2\epsilon) \right) \nonumber \\
	& + \frac{1}{2} \sum_{j \neq i-1, i, i+1} \left( I_\text{Bub}^{(1)}(i, i+1)^2 + I_\text{Bub}^{(1)}(i, i+1) I_\text{Bub}^{(1)}(j, j+1) \right) \,.
\end{align}
We can divide the BDS-ansatz function into three parts, and introduce three $G_{\text{min}, \alpha}^{(2)}$ functions to capture their divergences as
\begin{align}
	G_{\text{min}, 1}^{(2)}(i, i+1, i+2) \big|_{\text{div.}} &=  I_{\text{Bub}}^{(1)}(i, i+1) I_{\text{Bub}}^{(1)}(i+1, i+2) \big|_{\text{div.}} \, , \nonumber\\
	G_{\text{min},2}^{(2)}(i, i+1) \big|_{\text{div.}}  &=  f^{(2)}(\epsilon) I_{\text{Bub}}^{(1)}(i, i+1; 2\epsilon) \big|_{\text{div.}} \, , \nonumber \\
	G_{\text{min},3}^{(2)}(i, j)  &=  I_\text{dBub}^{(2)}(i, i+1; j, j+1) \,, 
	\label{eq:MinFFIR}
\end{align}
where $ I_\text{dBub}^{(2)}(i, i+1; j, j+1) =  I_\text{Bub}^{(1)}(i, i+1) I_\text{Bub}^{(1)}(j, j+1)$ with  $j \neq i-1$ or $i+1$.

The choice of $G_{\text{min}, 1}^{(2)}$ and $G_{\text{min}, 2}^{(2)}$ is not unique. To pick up a simple solution, we require that:
(1) $G_{\text{min}, 1}^{(2)}(i, j, k)$ have flip symmetry by exchanging external legs $p_i$ and $p_k$; 
(2) $G_{\text{min},2}^{(2)}(i, j)$ only contain the master integrals with two external legs $p_i$ and $p_j$;
and (3) neither $G_{\text{min}, 1}^{(2)}(i, j, k)$ nor $G_{\text{min}, 2}^{(2)}(i, j)$ involve the non-planar master integrals $I_{\text{NTBox1}}^{(2)}$.
We can then obtain the following unique combinations
\begin{align}
	G_{\text{min}, 1}^{(2)}(i, j, k)  = & 4 I_\text{Sun}^{(2)}(i, j)+\frac{1}{2} I_{\text{TBub0}}^{(2)}(i, j) + \frac{1}{4} I_\text{dBub}^{(2)}(i, j; i, j) \nonumber \\
	& +I_\text{BubBox}^{(2)}(i, j, k) - \frac{1}{2} I_{\text{TT2}}^{(2)}(i, j, k) + \left( p_i \leftrightarrow p_k \right) \,, \nonumber \\
	G_{\text{min}, 2}^{(2)}(i, j)  = & - \left( \frac{1}{2} I_{\text{TBub0}}^{(2)}(i, j)+I_\text{Sun}^{(2)}(i, j) \right) \,.
\end{align}

%%%%%%%%%%%%%%%%%%%%%%%%%%
\paragraph{Constructing building-blocks $\tilde{G}_{\text{min}, \beta}^{(2)}$.}

Next, we consider all possible building-blocks $\tilde{G}_{\text{min}, \beta}^{(2)}$ which are infrared finite. 
By considering all possible combinations of master integrals, it is not hard to find that there are two independent blocks of $\tilde{G}_{\text{min}, \beta}^{(2), \text{deg-4}}$ which can be given as
\begin{align}
	\label{eq:tildeGmindeg4}
	\tilde{G}_{\text{min}, 1}^{(2), \text{deg-4}}(i, j) = & \frac{1}{2} I_{\text{NTBox1}}^{(2)}(i, j) +  I_{\text{dBub}}^{(2)}(i, j; i, j) + 11 I_{\text{Sun}}^{(2)}(i, j) + \frac{5}{2} I_{\text{TBub0}}^{(2)}(i, j) \nonumber \\
	= & -\zeta_4 + \mathcal{O}(\epsilon) \,,  \\
	\tilde{G}_{\text{min}, 2}^{(2), \text{deg-4}}(i, j, k) = & I_{\text{dBub}}^{(2)}(i, j; i, j) +4 I_{\text{Sun}}^{(2)}(i, j) - (i \leftrightarrow k) = -12 \zeta_3 \log\left( \frac{s_{ij}}{s_{jk}} \right) + \mathcal{O}(\epsilon) \nonumber \,.
\end{align}
It is worthwhile noticing that all the master integrals occurred in \eqref{eq:tildeGmindeg4} are range-two integrals that involve only two external legs.
There are some considerations when we define the above two functions:
\begin{itemize}
	\item[1)] $\tilde{G}_{\text{min}, 1}^{(2), \text{deg-4}}$ contains the non-planar integrals $I_{\text{NTBox1}}^{(2)}(i, j)$. For planar minimal form factor with $n \geqslant 3$, $I_{\text{NTBox1}}^{(2)}(i, j)$ will not contribute, thus the block $\tilde{G}_{\text{min}, 1}^{(2), \text{deg-4}}$ will not occur. In particular, the color factor of the non-planar topologies will vanish for the $n=3$ case with external particles in adjoint representation (such as three-gluon).
	
	\item[2)] $\tilde{G}_{\text{min}, 2}^{(2), \text{deg-4}}(i, j, k)$ is anti-symmetric by exchanging the external momenta $p_i$ and $p_k$. Such building blocks will vanish if their coefficients are free from kinematics invariants and the form factor has the symmetry of flipping $p_i\leftrightarrow p_k$ or cyclicly permuting external momentum, the later will then be explained.
\end{itemize}

%%%%%%%%%%%%%%%%%%
\paragraph{Full form factor.}

Now we can give the form factor results in terms of the above two types of building blocks.

For general $n \geqslant 3$ minimal form factors, the infrared structure \eqref{eq:nptInfraredStructure} require the following form
\begin{align}
	\label{eq:Min3ptFF}
	\mathcal{I}_{\text{min}, n}^{(2), \text{deg-4}} = & \sum_{j \neq i-1, i, i+1} \frac{1}{2} \left( G_{\text{min}, 3}^{(2)}(i, i) + G_{\text{min}, 3}^{(2)}(i, j) \right) 
	 + \sum_{i=1}^n \left( G_{\text{min}, 1}^{(2)}(i, i+1, i+2) - G_{\text{min}, 2}^{(2)}(i, i+1) \right) \nonumber \\
	& + \sum_{i=1}^{n} \left( y_{1,i} \tilde{G}_{\text{min}, 1}^{(2), \text{deg-4}}(i, i+1) +y_{2,i} \tilde{G}_{\text{min}, 2}^{(2), \text{deg-4}}(i, i+1, i+2) \right) \,,
\end{align}
where $y_{\beta,i}$ are parameters which can be fixed to be $0$ for the following cases. For planar minimal form factors, as mentioned $\tilde{G}_{\text{min}, 1}^{(2), \text{deg-4}}$ cannot occur since they contain non-planar master integrals, thus $y_{1,i}$ are zero. The maximally transcendental part of the minimal form factor has constant coefficients, moreover the form factors of operators $\mathrm{tr}(\phi^n)$ and $\mathrm{tr}((F_{\mu\nu})^n)$ have the symmetry of cycling external particles, which require all $y_{2,i}$ must be the same, in such case the summation of $\tilde{G}_{\text{min}, 2}(i, i+1, i+2)$ will vanish, thus one has also $y_{2,i}=0$. Actually, with some simple argument of unitarity cuts, one can show that this is also true for general minimal form factors; we will discuss this more in Section~\ref{sec:univer-mimi}.

For the $n=2$ Sudakov form factor, the master integrals with three external particles do not contribute, thus only building-blocks $G_{\text{min}, 2}^{(2), \text{deg-4}}(1,2)$, $G_{\text{min}, 3}^{(2), \text{deg-4}}(1,2)$, and $\tilde{G}_{\text{min}, 1}^{(2), \text{deg-4}}(1, 2)$
can appear, the last one contains the non-planar master integrals $I_{\text{NTBox1}}^{(2)}(i, j)$ which are necessary. By matching the infrared structure \eqref{eq:nptInfraredStructure} at $n=2$:
\begin{equation}
	\label{eq:2ptInfraredStructure}
	\mathcal{I}_{\text{min}, 2}^{(2), \text{BDS}, \text{deg-4}} = 2 \left( I_\text{Bub}^{(1)}(i, i+1)^2 - f^{(2)}(\epsilon) I_\text{Bub}^{(1)}(i, i+1; 2\epsilon) \right) \,,
\end{equation}
we obtain
\begin{align}
	\mathcal{I}_{\text{min}, 2}^{(2), \text{deg-4}} = & 2 G_{\text{min}, 3}^{(2)}(1, 2) - 2 G_{\text{min}, 2}^{(2)}(1, 2) + y_0 \tilde{G}_{\text{min}, 2}^{(2), \text{deg-4}}(1, 2) \\
	= & 2 ( 1+y_0 ) I_\text{dBub}^{(2)}(1, 2; 1, 2) + \left( 1 +\frac{5y_0}{2} \right) I_{\text{TBub0}}^{(2)}(1, 2) \nonumber \\
	& + (2+11y_0) I_\text{Sun}^{(2)}(1, 2) + \frac{y_0}{2} I_\text{NTBox1}^{(2)}(1, 2) \,, \nonumber
\end{align}
with the single unfixed parameter $y_0$, which in principle can be fixed by a unitarity cut. Later in Section~\ref{sec:3ptFF}, we will show the parameter $y_0$ can also be fixed by considering the collinear and the $q^2\rightarrow0$ limits of two-loop three-point form factor, which gives $y_0 = 2$.

%%%%%%%%%%%%%%%%%%%
\paragraph{Remainder functions.}

The above discussion is about the coefficients of the master integrals and provides the maximally transcendental part to all orders in $\epsilon$. Below we focus on the finite order and consider the funite remainder function $\mathcal{R}_{\text{min}, n}^{(2), \text{deg-4}}$, which can be compared with the known results in literature.

We first compute the building blocks that compose the remainder:
\begin{align}
	G_{\text{min}, 1}^{(2)}(i, j,k) - I_{\text{Bub}}^{(1)}(i, j) I_{\text{Bub}}^{(1)}(j, k) = & T_4\left( \frac{s_{ij}}{s_{ijk}}, \frac{s_{jk}}{s_{ijk}}, \frac{s_{ik}}{s_{ijk}} \right) + \left( i \leftrightarrow k \right) + \mathcal{O}(\epsilon) \,, \nonumber \\
	G_{\text{min}, 2}^{(2)}(i, j) - f^{(2)}(\epsilon) I_{\text{Bub}}^{(1)}(i, j; 2\epsilon) = & -3 \zeta_4 + \mathcal{O}(\epsilon) \,,
\end{align}
where $T_4(u,v,w)$ is defined as
\begin{align}
	T_4(u, v, w) = & G(1-u,1-u,1,0, v) \\
	& - {\rm Li}_4(1-u)-{\rm Li}_4(u)+{\rm Li}_4\left(\frac{u - 1}{u}\right)
	-  \log \left( \frac{1-u}{w }\right) \left[  {\rm Li}_3\left(\frac{u - 1}{u}\right) - {\rm Li}_3\left(1-u\right) \right] \nonumber \\
	&- \log \left(u\right) \left[{\rm Li}_3\left(\frac{v}{1-u}\right)+{\rm Li}_3\left(-\frac{w}{v}\right) + {\rm Li}_3\left(\frac{v-1}{v}\right) -\frac{1}{3}  \log ^3\left(v\right) -\frac{1}{3} \log ^3\left(1-u\right)  \right] \nonumber \\
	&- {\rm Li}_2\left(\frac{u-1}{u}\right) {\rm Li}_2\left(\frac{v}{1-u}\right)+  {\rm Li}_2\left(u\right) \left[
	\log \left(\frac{1-u}{ w }\right) \log \left(v \right) +\frac{1}{2} \log ^2\left( \frac{ 1-u }{ w }\right) \right]  \nonumber \\
	&+  \frac{1}{24} \log ^4\left(u\right)
	-\frac{1}{8} \log ^2\left(u\right) \log ^2\left(v\right)  - \frac{1}{2} \log ^2\left(1-u\right) \log \left(u\right) \log \left( \frac{w}{v}\right) \nonumber \\
	& -\frac{1}{2} \log \left(1-u\right) \log ^2\left(u\right) \log \left(v\right) - \frac{1}{6} \log ^3\left(u\right) \log \left(w\right) + \zeta_3 \log (u) -\zeta_4 \nonumber \\
	& -\zeta_2 \Big[ \log \left(u\right) \log \left(\frac{1-v}{ v} \right)
	+ \frac{1}{2}\log ^2\left( \frac{ 1-u }{ w }\right) - \frac{1}{2}\log ^2\left(u\right)   \Big] \,. \nonumber
\end{align}
Therefore, we can define the density function $\mathcal{R}_{3}^{(2), \text{deg-4}}(i,j,k)$ as
\begin{equation}
	\mathcal{R}_{3}^{(2), \text{deg-4}}(1,2,3) = T_4(u, v, w) + T_4(v, u, w) + 3 \zeta_4 \,.
\end{equation}
The result is consistent with the known results \cite{Brandhuber:2014ica, Lin:2020dyj}.
The planar minimal form factors with $\mathrm{tr}(\phi^n)$ are
\begin{equation}
	\mathcal{R}_{\text{min}, n}^{(2), \text{deg-4}} = \sum_{i=1}^{n} \mathcal{R}_{3}^{(2), \text{deg-4}}(i,i+1,i+2) \, .
\end{equation}

Finally, we consider the  form factor results for the special $n=2, 3$ cases. The $n=2$ Sudakov form factor is \cite{vanNeerven:1985ja}
\begin{equation}
	\mathcal{R}_{\text{min}, 2}^{(2), \text{deg-4}} = 6 \zeta_4 - y_0 \zeta_4 = 4 \zeta_4 \,,
\end{equation}
where $y_0=2$ is used.
The $n=3$ minimal form factor of $\mathrm{tr}(\phi^3)$ has a compact form as
\begin{align}
	\mathcal{R}_{\text{min}, 3}^{(2), \text{deg-4}} = & {\cal R}_{3}^{(2), \text{deg-4}}(1,2,3)+{\cal R}_{3}^{(2), \text{deg-4}}(2,3,1)+{\cal R}_{3}^{(2), \text{deg-4}}(3,1,2) \\
	= & \frac{3}{4} \text{Li}_4\left(-\frac{u v}{w}\right)-\frac{3}{2} \log (w) \text{Li}_3\left(-\frac{u}{v}\right)-\frac{3}{2} \text{Li}_4(u) \nonumber \\
	& + \frac{1}{32} \log ^2(u) \left( \log ^2(u)+\log ^2(v)+\log ^2(w)-4 \log (v) \log (w) \right) \nonumber \\
	& +\frac{1}{8} \zeta_2 \left( 5 \log ^2(u) - 2 \log (v) \log (w) \right)+\frac{1}{2} \zeta_3 \log (u)+\frac{7}{16} \zeta_4 + \left( \text{full perm.}(u,v,w) \right) \,. \nonumber
\end{align}

%%%%%%%%%%%%%%%%%%%%%%%%%%%%%
\subsection{Lower transcendentality parts}
\label{sec:LTminimal}

The above discussion can be generalized to lower transcendentality parts. 
We will show that a similar procedure will explain the universal building blocks of transcendentality degree-$3$ and degree-$2$ functions for QCD minimal form factors observed in \cite{Jin:2019opr}.

We will focus on the two-loop case. To apply the infrared structure, we consider again the BDS function $\mathcal{I}_{\text{min}, n}^{(2), \text{BDS}}$:\footnote{Here we use BDS ansatz for the $\mathcal{N}=4$ case. In the case of QCD, one can use Catani subtraction, as explained in Appendix~\ref{app:catani}.}
\begin{equation}
	\mathcal{I}_{\text{min}, n}^{(2), \text{BDS}, \text{deg-a}} =\sum_{b=a-2}^{2} \frac{\mathcal{I}_{\text{min}, n}^{(1), b}(\epsilon) \mathcal{I}_{\text{min}, n}^{(1), \text{deg-(a-b)}}(\epsilon)}{2} + f^{(2)}(\epsilon) \mathcal{I}_{\text{min}, n}^{(1), \text{deg-a}}(2\epsilon) \,,
\end{equation}
where full $\mathcal{I}_{\text{min}, n}^{(2), \text{BDS}}$ is decomposed similarly as \eqref{eq:minidegCalF} as
\begin{equation}
	\mathcal{I}_{\text{min}, n}^{(2), \text{BDS}} = \sum_{a=2L}^{-\infty} \epsilon^{2L-a} \mathcal{I}_{\text{min}, n}^{(2), \text{BDS}, \text{deg-a}} \,.
\end{equation}
We also assume the lower transcendentality part of one-loop form factor are known as
\begin{equation}
	\mathcal{I}_{\text{min}, n}^{(1), \text{deg-a}} = \sum_{i = 1}^{n} c_{a,i}^{(1)} I_{\text{Bub}}^{(1)}(i, i + 1) + \frac{1}{\epsilon} Z_i^{(1)} \delta_{a,1} \,,
\end{equation}
where $Z_i^{(1)}$ represents the possible UV renormalization constant which contributes only in degree-$1$ part.
One can see that the BDS function $\mathcal{I}_{\text{min}, n}^{(2), \text{BDS}, \text{deg-a}}$ consists of bubbles integrals similar to \eqref{eq:nptInfraredStructure}. Therefore, the degree-$a$ part of the form factor can be expressed as
\begin{align}
	\label{eq:nptLowerTransFF}
	\mathcal{I}_{\text{min}, n}^{(2), \text{deg-a}} = & \sum_{j \neq i-1, i, i+1} \frac{1}{2} \left[ \left(c_{ai}^{(1)} \right)^2 G_{\text{min}, 3}^{(2)}(i, i) + c_{ai}^{(1)} c_{aj}^{(1)} G_{\text{min}, 3}^{(2)}(i, j) \right] \nonumber \\
	& + \sum_{i=1}^n \left( c_{ai}^{(1)} c_{a,i+1}^{(1)} G_{\text{min}, 1}^{(2)}(i, i+1, i+2) + c_{ai}^{(1)} G_{\text{min}, 2}^{(2)}(i, i+1) \right) 
	+ (\textrm{UV-part}) \nonumber \\
	& + \sum_{\beta} y_{\beta} \tilde{G}_{\text{min}, \beta}^{(2), \text{deg-a}} \,,
\end{align}
where the first two lines capture the full divergences and the UV-part contains terms depending on $Z_i^{(1)}$, we will not discuss the latter in detail here. 
The remaining problem is to construct the concrete form of the building-blocks $\tilde{G}_{\text{min}, \beta}^{(2), \text{deg-a}}$, which can be determined by requiring that $\tilde{G}_{\text{min}, \beta}^{(2), \text{deg-a}}$ starting at the order of $\mathcal{O}(\epsilon^{a-2L})$ because of the factor $\epsilon^{2L-a}$.

Additionally, the first and second summations in LHS of \eqref{eq:nptLowerTransFF} will not contribute to the finite remainder. These terms in the complete form factor result will be multiplied by a factor $\epsilon^{2L-a}$ where $a<4$, thus they will only contribute to the $\mathcal{O}(\epsilon)$ part of the remainder.
Thus the lower transcendental part of the finite remainder will only depend on the $\tilde{G}_{\text{min}, \beta}^{(2), \text{deg-a}}$ part and the UV part, the latter is trivially determined by the one-loop result. In other words, only some special function blocks will appear in the remainder, we will give them as follows.

For transcendentality degree-$3$, there are
\begin{align}
	\tilde{G}_{\text{min},1}^{(2), \text{deg-3}}(i,j) =  & \tilde{G}_{\text{min},1}^{(2), \text{deg-4}}(i,j) = \mathcal{O}(\epsilon^0) \,, \nonumber\\
	\tilde{G}_{\text{min},2}^{(2), \text{deg-3}}(i,j) = & I_{\text{dBub}}^{(2)}(i, j; i, j) +4 I_{\text{Sun}}^{(2)}(i,j) = \frac{6\zeta_3}{\epsilon} + \mathcal{O}(\epsilon^0) \,, \nonumber \\
	\tilde{G}_{\text{min},3}^{(2), \text{deg-3}}(i,j,k) = & I_{\text{Sun}}^{(2)}(i,j) +\frac{1}{2} I_{\text{TBub}0}^{(2)}(i,j)  - ( i \leftrightarrow k) = -\frac{ \zeta_2 }{\epsilon} \log \left( \frac{s_{ij}}{s_{jk}} \right) + \mathcal{O}(\epsilon^0) \,, \nonumber \\
	\tilde{G}_{\text{min},4}^{(2), \text{deg-3}}(i,j,k) = & I_{\text{TBub}2}^{(2)}(i,j,k) + I_{\text{TT}2}^{(2)}(i,j,k) - I_{\text{BubBox}}^{(2)}(k,j,i) - I_{\text{Sun}}^{(2)}(i,j) \nonumber \\
	= & \frac{1}{\epsilon} T_3\left( \frac{s_{ij}}{s_{ijk}}, \frac{s_{jk}}{s_{ijk}}, \frac{s_{ik}}{s_{ijk}} \right) + \mathcal{O}(\epsilon^0) \,, 
	\label{eq:buidlingblock1}
\end{align}
where $T_3(u, v, w)$ occurs in \cite{Jin:2019opr}, 
\begin{align}
T_3(u,v,w) = & \Big[ -{\rm Li}_3 \left(-{u\over w} \right) + \log(u) {\rm Li}_2\left({v \over 1-u} \right) - {1\over2} \log(u) \log(1-u) \log\left({w^2\over 1-u}\right) \nonumber\\
& + {1\over2} {\rm Li}_3\left(-{uv \over w}\right) + {1\over2} \log(u)\log(v)\log(w) + {1\over12}\log^3(w) + (u\leftrightarrow v) \Big] \nonumber\\
& +  {\rm Li}_3(1-v) - {\rm Li}_3(u) + {1\over2} \log^2(v) \log\left({1-v\over u}\right) - \zeta_2 \log\left( {u v \over w} \right) \,.
	\label{eq:T3}
\end{align}
For transcendentality degree-$2$, there are
\begin{align}
	\tilde{G}_{\text{min}, 1}^{(2), \text{deg-2}}(i, j) = & \tilde{G}_{\text{min},1}^{(2), \text{deg-4}}(i, j) = \mathcal{O}(\epsilon^{-1}) \,, \nonumber\\
	\tilde{G}_{\text{min}, 2}^{(2), \text{deg-2}}(i, j) = & \tilde{G}_{\text{min}, 1}^{(2), \text{deg-3}}(i, j) = \mathcal{O}(\epsilon^{-1}) \,, \nonumber \\
	\tilde{G}_{\text{min}, 3}^{(2), \text{deg-4}}(i, j) = & \tilde{G}_{\text{min}, 3}^{(2), \text{deg-3}}(i, j, k) = \mathcal{O}(\epsilon^{-1}) \,, \nonumber \\
	\tilde{G}_{\text{min}, 4}^{(2), \text{deg-2}}(i,j) = & 2I_{\text{Sun}}^{(2)}(i,j) + I_{\text{TBub0}}^{(2)}(i,j) = \frac{\zeta_2}{\epsilon^2} + \mathcal{O}(\epsilon^{-1}) \, , \nonumber \\
	\tilde{G}_{\text{min}, 5}^{(2), \text{deg-2}}(i, j, k) = & 2 I_{\text{Sun}}^{(2)}(i, j) + I_{\text{TBub2}}^{(2)}(k, j, i) = -\frac{ 1 }{\epsilon^2} T'_2 \left( \frac{s_{ij}}{s_{ijk}} \right) + \mathcal{O}(\epsilon^{-1}) \,, \nonumber \\
	\tilde{G}_{\text{min}, 6}^{(2), \text{deg-2}}(i, j, k) = & I_{\text{TT2}}^{(2)}(i, j, k) = \frac{1}{\epsilon^2} T_2 \left( \frac{s_{ij}}{s_{ijk}}, \frac{s_{jk}}{s_{ijk}} \right) + \mathcal{O}(\epsilon^{-1}) \,. 
	\label{eq:buidlingblock2}
\end{align}
where $T'_2(u)$ and $T_2(u, v)$ occur in reference~\cite{Jin:2019opr}, 
\begin{align}
	T'_2(u) = & \mathrm{Li}_2(1-u) + \frac{1}{2}\log^2(u) \, , \nonumber\\
	T_2(u, v) = & \mathrm{Li}_2(1-u) + \mathrm{Li}_2(1-v) + \log(u) \log(v) -\zeta_2 \, .
	\label{eq:T2}
\end{align}

We will not discuss the building blocks for transcendentality degree-$1$ and degree-$0$ because there are too many possibilities of building blocks for these parts; however, they consist only of $\log(-s_{ij})$ and rational functions that depend on kinematics.

%%%%%%%%%%%%%%%%%%%%%%%%%%%%
\subsection{Universal transcendentality structures}
\label{sec:univer-mimi}

In previous subsections, we impose the constraints that form factors should have the correct universal IR divergences. 
We will show that the remaining degrees of freedom can be classified by functions that are free from IR divergences, such as $\tilde{G}_{\text{min}, \beta}^{(2)}$ in \eqref{eq:tildeGmindeg4}.
In this subsection, we will further determine these remaining degrees of freedom by applying unitarity cuts.
We will show that the minimal form factors in $\mathcal{N}=4$ SYM and QCD theories will have the same maximally transcendental parts, 
and moreover, they also contain universal lower transcendentality building blocks up to simple logarithm functions and Riemann zeta numbers.\footnote{Our discussion for the lower transcendental part is for the bare form factors. The UV renormalization will only modify the log and rational functions up to two-loop order.}

%%%%%%%%%%%%%%%
\paragraph{One-loop case.}

Let us consider first the one-loop case. 
We show in Section~\ref{sec:MTmini1loop} that by matching the universal IR divergence, it is enough to fix the maximally transcendental part. 
In this way, we arrive at the conclusion that maximally transcendental one-loop corrections $\mathcal{I}_{\text{min}, n}^{(1), \text{deg-2}}$ are universal in different theories and for different operators (up to a simple change of color factors, see discussion below). 
We stress again that this is due to the fact that the maximally transcendent part of one-loop IR divergence is the same for general gauge theories. 

%%%%%%%%%%%%%%%%%%%%%%%%%%%
\begin{figure}[t]
	\centering
	\includegraphics[scale=0.5]{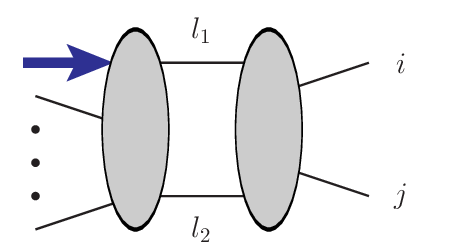}
	\caption{Unitarity cut for one-loop minimal form factors in gauge theories.}
	\label{fig:samecutsmin1loop}
\end{figure}
%%%%%%%%%%%%%%%%%%%%%%%%%%%

Now we will use a different argument based on unitarity cuts which will not only constrain the maximally transcendental but also lower transcendental parts.
Considering the double cut in Figure~\ref{fig:samecutsmin1loop}, which is enough to determine the full coefficient of $I_{\text{Bub}}^{(1)}(i,j)$. The cut integrand is
\begin{equation}
	\label{eq:min1loopcut}
	\mathcal{F}_{\mathcal{O}_{n}, n}^{(0)}(..,-l_1,-l_2,..)  \mathcal{A}_{4}^{(0)}(l_2,l_1,i,j) \,,
\end{equation}
where the tree-level form factor $\mathcal{F}_{\mathcal{O}_{n}, n}^{(0)}(..,-l_1,-l_2,..)$ are non-zero only if the types of internal particles $l_1$ and $l_2$ are the same as external particles $i$ and $j$, namely, it must be a minimal form factor.
To be more concrete, we consider minimal form factors in $\mathcal{N}=4$ SYM and QCD with operators that contain only gluon and quark (or gluino) fields. Then it should be clear that the kinematic parts of the cut integrands \eqref{eq:min1loopcut} are the same for $\mathcal{N}=4$ SYM and QCD, and the difference only appears in the color factors that involve fermions: since the fermion is in adjoint representation in $\mathcal{N}=4$ SYM while being fundamental in QCD.
In this case, one can identify the two results if one converts the quark representation in QCD results from the fundamental to the adjoint.
For the one-loop correction function $\mathcal{I}_{\text{min}, n}^{(1)}$, this can be achieved by a simple replacement for the quadratic Casimir as $C_F \rightarrow C_A$.

%%%%%%%%%%%%%%%%%%%%%%%%%%%
\begin{figure}[t]
	\centering
	\subfloat[]{\includegraphics[scale=0.5]{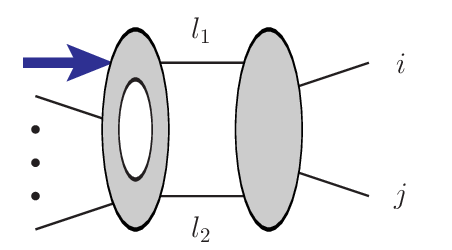}}
	\subfloat[]{\includegraphics[scale=0.5]{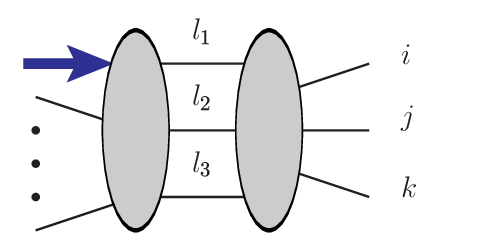}}
	\caption{Unitarity cuts which are same for two-loop minimal form factors in gauge theories. }
	\label{fig:samecutsmin}
\end{figure}
%%%%%%%%%%%%%%%%%%%%%%%%%%%

%%%%%%%%%%%%%%%%%
\paragraph{Two-loop case.}

In previous two subsections, we find that after the IR constraints, the remaining degrees of freedom for the two-loop minimal form factors are 
related to functions $\tilde{G}_{\text{min}, \beta}^{(2), \text{deg-a}}$ in \eqref{eq:tildeGmindeg4}, \eqref{eq:buidlingblock1}, and \eqref{eq:buidlingblock2}.
For the maximally transcendental part, since the blocks $\tilde{G}_{\text{min}, 1}^{(2), \text{deg-4}}$ contain non-planar integrals, they will not contribute if we focus on the planar form factors.
We have also argued that if the form factor has cyclic symmetry, the terms $\tilde{G}_{\text{min}, 2}^{(2), \text{deg-4}}$ also vanish.
Below we would like to apply unitarity cuts to prove the universality of these functions in general gauge theories, including also the lower transcendental parts.

We first consider the maximally transcendental parts and focus on the planar form factor for simplicity.
To fix the coefficients of $\tilde{G}_{\text{min}, 2}^{(2), \text{deg-4}}$, one can consider the double-cut (a) in Figure~\ref{fig:samecutsmin} which can determine the coefficients of $I_{\text{dBub}}^{(2)}(i, j; i, j)$.\footnote{The other block $\tilde{G}_{\text{min}, 1}^{(2), \text{deg-4}}$ contains also $I_{\text{dBub}}^{(2)}(i, j; i, j)$. By doing the full-color double-cut Figure~\ref{fig:samecutsmin}(a), one can also fix the non-planar corrections. Our following arguments also apply to such cases.}
The cut integrand is
\begin{align}
	& \mathcal{F}_{\mathcal{O}_{n}, n}^{(1)}(\ldots,-l_1,-l_2,\ldots) \mathcal{A}_{4}^{(0)}(l_2,l_1,i,j) \\
	= & \mathcal{I}_{\text{min}, n}^{(1), \text{deg-2}}(\ldots,-l_1,-l_2,\ldots) \times \left[ \mathcal{F}_{\mathcal{O}_{n}, {\rm min}}^{(0)}(\ldots,-l_1,-l_2,\ldots)  \mathcal{A}_{4}^{(0)}(l_2,l_1,i,j) \right] + (\textrm{lower trans.}) \,.\nonumber
\end{align}
We stress that only the first term in the second line contributes to maximally transcendental part, in which we have used the universality of one-loop results. In particular, the universal one-loop correction function $\mathcal{I}_{\text{min}, n}^{(1), \text{deg-2}}(\ldots,-l_1,-l_2,\ldots)$ contributes only if the internal particles $l_1$ and $l_2$ are the same types from external particles $i$ and $j$. In addition, the terms in the square brackets provide the same one-loop cut integrand \eqref{eq:min1loopcut}. 
These show that the coefficients of double-bubble masters $I_{\text{dBub}}^{(2)}(i, j; i, j)$ are the same for general operators in general gauge theories. 
Since we know that $\tilde{G}_{\text{min}, 2}^{(2), \text{deg-4}}$ blocks are zero for form factors with cyclic symmetry, they also vanish in form factors of general operators.
For two-loop minimal form factors in $\mathcal{N}=4$ SYM and QCD, their maximally transcendental parts are equivalent by a similar change of color factors for fermions, as discussed in the one-loop case.

Next we use unitarity-cut argument to derive constraints on the lower transcendental parts. 
Consider the triple-cut (b) in Figure~\ref{fig:samecutsmin} with the cut integrand given as
\begin{equation}
	\label{eq:min2looptriplecut}
	\mathcal{F}_{\mathcal{O}_{n}, \text{min}}^{(0)}(\ldots, -l_1, -l_2, -l_3, \ldots)  \mathcal{A}_{6}^{(0)}(l_3,l_2,l_1,i,j,k) \,,
\end{equation}
which are able to detect all range-3 master integrals $I_{\text{Sun}}^{(2)}(i,j,k)$, $I_{\text{BubBox}}^{(2)}(i,j,k)$, $I_{\text{TBub2}}^{(2)}(i,j,k)$, and $I_{\text{TT2}}^{(2)}(i,j,k)$. Since the tree form factor in the cut must be the minimal form factor (otherwise it is zero), the internal particles for the cut legs $\{l_1, l_2, l_3\}$ are the same types as external particles $\{i, j, k\}$. Similar to the one-loop case, this implies that this cut computation is the same also for $\mathcal{N}=4$ SYM and QCD,\footnote{Strictly speaking, the lower transcendentality part of this cut can be different in ${\cal N}=4$ SYM and QCD because of different regularization-scheme choices. But it can be shown that $T_3$, $T_2$, and $T'_2$ terms are not affected by choices of schemes.} if we change the color factors for fermions accordingly like $C_F \rightarrow C_A$. Thus the coefficients of all range-3 master integrals in a bare form factor will be the same for the two theories,
and the results only differ by the range-2 master integrals, which depend only on logarithm functions and Riemann zeta numbers. In particular, in the finite remainder, the function blocks $T_3(u,v,w)$, $T_2(u,v)$ and $T'_2(u)$ (\eqref{eq:T3} and \eqref{eq:T2}) for the lower transcendental parts must be the same for $\mathcal{N}=4$ SYM and QCD, as observed in \cite{Jin:2019ile, Jin:2019opr}.

%%%%%%%%%%%%%%%%%%%%%%%%%%%%%%%
\section{Two-loop three-point form factor of $\mathrm{tr}(F^2)$}
\label{sec:3ptFF}

In this section, we consider three-point form factor of $O_2 = \mathrm{tr}(F^2)$:
\begin{equation}
	\mathcal{F}_{\mathcal{O}_2, 3} = \int d^D x e^{-i q\cdot x} \langle p_1 \, p_2 \, p_3 | \mathcal{O}_2(x)|0\rangle \,.
\end{equation}
This is a next-to-minimal form factor, and whose collinear limits will provide new important constraints. Moreover, this form factor is equivalent  to the Higgs-plus-three-parton amplitudes in the heavy top mass limit  by integrating out the heavy top quark \cite{Wilczek:1977zn, Shifman:1979eb, Dawson:1990zj, Djouadi:1991tka, Kniehl:1995tn, Chetyrkin:1997sg, Chetyrkin:1997un}. The full two-loop QCD corrections were obtained in \cite{Gehrmann:2011aa}, and the two-loop result in $\mathcal{N}=4$ SYM was obtained in \cite{Brandhuber:2012vm}.
It turns out that the maximally transcendental parts of the results in QCD and ${\cal N}=4$ SYM satisfy \cite{Brandhuber:2012vm, Jin:2019ile}:
\begin{equation}
	\label{eq:MTP3pt}
	\mathcal{F}^{(L), \mathcal{N}=4}_{\mathcal{L}\sim{\mathrm{tr}(F^2)}}(1,2,3) = \mathcal{F}^{(L), \text{QCD}}_{\mathrm{tr}(F^2), \text{M.T.}}(1^g,2^g,3^g) 
	= \mathcal{F}^{(L), \text{QCD}}_{\mathrm{tr}(F^2), \text{M.T.}}(1^q, 2^{\bar{q}}, 3^g) \Big|_{C_F \rightarrow C_A} ,
\end{equation}
for $L=1,2$.
One main goal of this section is to provide a proof for the relation \eqref{eq:MTP3pt}.
Here in ${\cal N}=4$ SYM theory, ${\cal L}$ is the chiral Lagrangian which contains ${\rm tr}(F^2)$ as a component, see \emph{e.g.}~\cite{Eden:2011yp, Brandhuber:2011tv}.\footnote{The chiral Lagrangian belongs also to the larger stress-tensor supermultiplet which is half-BPS. We mention that the same maximally transcendental function was also found in the two-loop three-point form factor of the non-BPS Konishi operator in $\mathcal{N}=4$ SYM \cite{Banerjee:2016kri}.} For our discussion of the maximally transcendental part, it is enough to focus on ${\rm tr}(F^2)$ since other components in the supermultiplet have only contribution of lower transcendentality.
Since the field strength operator can be decomposed as self-dual and anti-self-dual parts as
\begin{equation}
{\rm tr}(F_{\mu\nu}F^{\mu\nu}) = {1\over2} \big[ {\rm tr}(F_{\alpha\beta}F^{\alpha\beta}) + {\rm tr}({\bar F}_{\dot\alpha \dot\beta} {\bar F}^{\dot\alpha \dot\beta}) \big] \,,
\end{equation}
for simplicity (and without loss of generality), in the following discussion we will take the operator as the self-dual part ${\cal O}_2 = {\rm tr}(F_{\alpha\beta}F^{\alpha\beta})$. The minimal two-point tree-level form factors is ${\cal F}_{{\cal O}_2}^{(0)}(1^-,2^-) = \langle 12 \rangle^2$.

It is worthwhile pointing out that the same three-point form factor in ${\cal N}=4$ SYM has been computed up to eight loops \cite{Dixon:2020bbt, Dixon:2022rse} in the planar limit using the symbol bootstrap method where the form factor OPE played a crucial role \cite{Sever:2020jjx, Sever:2021nsq, Sever:2021xga}. The form factor results with full-color dependence have also been constructed up to four loops \cite{Lin:2021kht, Lin:2021qol, Lin:2021lqo} using color-kinematics duality.
It may be expected the maximally transcendental principle also applies to these higher loop results.

As a brief outline,
in Section~\ref{sec:ansatzN=4}, we will show that the constraints from IR together with collinear limits can fix the form factor in $\mathcal{N}=4$ up to the two-loop order. In Section~\ref{sec:MTP3pt}, we consider further the form factors in QCD, and together with the use of unitarity cuts for the $\mathcal{F}^{(l), \text{QCD}}_{\mathrm{tr}(F^2)}(1^q,2^{\bar{q}},3^g)$ case, then the relations \eqref{eq:MTP3pt} can be proven.

%%%%%%%%%%%%%%%%%%%%%%%%%%%%%%%
\subsection{Bootstrapping the $\mathcal{N}=4$ form factor}
\label{sec:ansatzN=4}

As a warm-up, we consider the form factors of stress-tensor supermultiplet in $\mathcal{N}=4$ super-Yang-Mills theory, which are uniformly transcendental with weight $2L$ at $L$ loops. The loop correction $\mathcal{I}_{\mathcal{O}_2, 3}^{(L)}$ can be defined by factorizing out the tree-level form factor from the loop-level as
\begin{equation}
	\mathcal{F}_{\mathcal{O}_2, 3}^{(L)}(1,2,3) = \mathcal{F}_{\mathcal{O}_2, 3}^{(0)} \mathcal{I}_{\mathcal{O}_2, 3}^{(L)}(1,2,3) \, .
\end{equation}
and $\mathcal{I}_{\mathcal{O}_2, 3}^{(L)}$ are functions depending on three Mandelstam variables \{$s_{12}$, $s_{23}$, $s_{13}$\}.

%%%%%%%%%%%%%%%
\subsubsection{One-loop case}
\label{subsec:3ptff1loop}

At one loop, $\mathcal{I}_{\mathcal{O}_2, 3}^{(1)}$ can be expanded in terms of 7 master integrals as
\begin{align}
	\label{eq:3pt1loopAnsatz}
	\mathcal{I}_{\mathcal{O}_2, 3}^{(1)} = & c_1 I^{(1)}_{\text{Bub}}(1, 2) + c_2 I^{(1)}_{\text{Bub}}(2, 3) + c_3 I^{(1)}_{\text{Bub}}(1, 3) + c_4 I^{(1)}_{\text{Bub}}(1, 2, 3) \nonumber \\
	& + c_5 I^{(1)}_{\text{Box}}(1, 2, 3) + c_6 I^{(1)}_{\text{Box}}(2, 3, 1) + c_7 I^{(1)}_{\text{Box}}(3, 1, 2) \, .
\end{align}

First, the infrared structure for one-loop takes an uniform form as
\begin{align}
	\label{eq:3pt1loopIR}
	\mathcal{I}_{\mathcal{O}_2, 3}^{(1)}  \Big|_{\text{IR}} = -\frac{3}{\epsilon^2} +\frac{\log(-s_{12}) +\log(-s_{23}) +\log(-s_{13})}{\epsilon} \,.
\end{align}
By requiring that the infrared part of \eqref{eq:3pt1loopAnsatz} to match \eqref{eq:3pt1loopIR}, one finds the coefficients should satisfy 
\begin{align}
	& c_4 = -{3\over2} (c_1 + c_2 + c_3 + 3) \,, \qquad c_5 = -{1\over4}( c_1 + c_2 - c_3 + 1) \,, \nonumber\\
	& c_6 = -{1\over4}( -c_1 + c_2 + c_3 + 1) \,, \qquad c_7 = -{1\over4}( c_1 - c_2 + c_3 + 1) \,.
\end{align}

Next we consider the constraints of the collinear limits. 
In the linear limit $p_3 \, || \, p_1 \, || \, (p_1+p_3)=p_1'$, one has
\begin{equation}
	s_{13} \rightarrow \delta q^2, \qquad s_{12} \rightarrow z q^2, \qquad s_{23} \rightarrow (1- z - \delta) q^2 \,,
\end{equation}
where $q^2 = (p_1+p_2+p_3)^2 = (p_1'+p_2)^2$, $0 \leqslant z \leqslant 1$ is a finite number and  $\delta \ll 1$. As discussed in Section~\ref{sec:constraints}, using \eqref{eq:collinear-general}, the three-point form factor in the limit satisfies 
\begin{equation}
	\mathcal{F}^{(1)}_{\mathcal{O}_2, 3}(3, 1,2) \xrightarrow{ p_1 || p_3 } \sum_\sigma \Big[ \mathbf{Sp}_{-\sigma}^{(0)}(3^{h_3},1^{h_1}) \, \mathcal{F}_{\mathcal{O}_2, 2}^{(1)}(1'^\sigma,2) + \mathbf{Sp}_{-\sigma}^{(1)}(3^{h_3},1^{h_1}) \, \mathcal{F}^{(0)}_{\mathcal{O}_2, 2}(1'^\sigma,2) \Big] \,.
\end{equation}
Since only $\mathcal{F}_{\mathcal{O}_2, 2}(1'^-,2^-)$ is non-zero, after subtracting the tree factors, one obtains
\begin{align}
	\label{eq:3pt1loopCL1}
	\mathcal{I}_{\mathcal{O}_2, 3}^{(1)}  \xrightarrow{ p_1 || p_3 } \, \mathcal{I}_{\mathcal{O}_2, 2}^{(1)}(p_1', p_2) + r_1^{[1], \text{MT}}(s_{13}, z) \,,
\end{align}
where $\mathcal{I}_{\mathcal{O}_2, 2}^{(1)}=- 2 I_{\text{Bub}}^{(1)}(p_1', p_2)$ is the known minimal two-point form factor, and $r_1^{[1], \text{MT}}$ is the splitting function given in \eqref{eq:splittingAmp}.
Explicitly, the form factor in the collinear limit is given by
\begin{align}
	\label{eq:3pt1loopCL2}
	\mathcal{I}_{\mathcal{O}_2, 3}^{(1)} \, \xrightarrow{ p_1 \parallel p_3 } & -\frac{3}{\epsilon^2} +
	\frac{\log (\delta) + 3 \log (-q^2)+\log (1-z)+\log (z)}{\epsilon} \\
	& - \log (\delta) \left( \log(-q^2) + \log (1-z) + \log (z) \right) - \frac{1}{2} \log ^2(\delta) \nonumber \\
	& +\log (-q^2) (-\log (1-z)-\log (z)) - \frac{3}{2} \log ^2(-q^2) \nonumber\\
	& +\frac{1}{2} \left(- \log ^2(1-z)- \log ^2(z)+2 \log (z) \log (1-z)+\zeta_2 \right) \,. \nonumber
\end{align}
By using the ansatz expression \eqref{eq:3pt1loopAnsatz} for $\mathcal{I}_{\mathcal{O}_2, 3}^{(1)}$, and requiring its collinear limit to match with \eqref{eq:3pt1loopCL2}, the coefficients in \eqref{eq:3pt1loopAnsatz} can be fixed to be 
\begin{align}
c_2 = c_1 \,, \quad c_3 = 1 \,, \quad c_4 = -c_1-2 \,, \quad c_5 = -\frac{c_1}{2} \,, \quad c_6 = -\frac{1}{2} \,,  \quad c_7 = -\frac{1}{2} \,,
\end{align}
leaving one parameter $c_1$ unfixed. 
Note that we only use one collinear limit, the other collinear limits $p_3 \parallel p_2$ and  $p_2 \parallel p_1$ can be further considered, which will give $c_1=1$. 

To summarize, the infrared structure and the collinear limits are enough to fix the one-loop form factor uniquely as
\begin{equation}
	\mathcal{I}_{{\cal O}_2, 3}^{(1)} = I_{\text{Bub}}^{(1)}(1,2) - I_{\text{Bub}}^{(1)}(1,2,3) - \frac{1}{2}I_{\text{Box}}^{(1)}(1,2,3) +(\text{cyclic perm.(1,2,3)}) \,.
\end{equation}

%%%%%%%%%%%%%%%
\subsubsection{Two-loop case}

\begin{figure}[tb]
	\centering
	\includegraphics[scale=0.5]{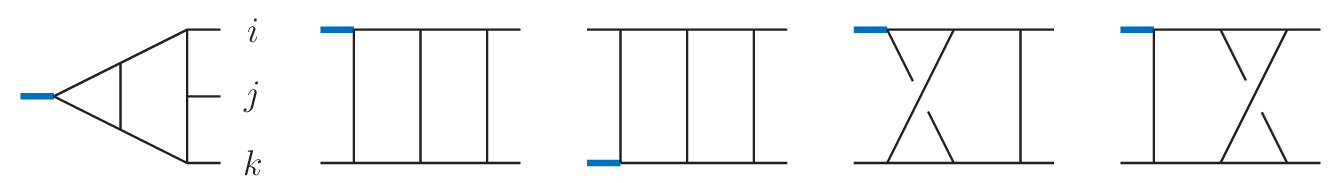}
	\caption{Integral topologies of maximal number of propagators for the two-loop three-point form factor.}
	\label{fig:maxTopology3pt}
\end{figure}

At two loops, there are in general 89 master integrals, the integral topologies with maximal number of propagators, which are enough to cover all the master integrals we need, are shown in Figure~\ref{fig:maxTopology3pt}. Then ansatz for the maximally transcendental part can be generally written as
\begin{align}
	\label{eq:3pt2loopAnsatz}
	\mathcal{I}_{\mathcal{O}_2, 3}^{(2)} = & c_1 I_{\text{Sun}}^{(2)}(1,2) +c_2 I_{\text{Sun}}^{(2)}(1,2,3) +c_3 I_{\text{TBub0}}^{(2)}(1,2) +c_4 I_{\text{dBub}}^{(2)}(1,2;1,2) \\
	& + c_5 I_{\text{dBub}}^{(2)}(1,2;1,2,3) + c_6 I_{\text{dBub}}^{(2)}(1,2,3;1,2,3) + c_{7} I_{\text{TBub1}}^{(2)}(1,2,3) \nonumber \\
	& +c_{8} I_{\text{TBub2}}^{(2)}(1,2,3) + c_{9} I_{\text{TT0}}^{(2)}(1,2,3)+c_{10} I_{\text{TT1}}^{(2)}(1,2,3) + c_{11} I_{\text{TT1a}}^{(2)}(1,2,3) \nonumber \\
	& + c_{12} I_{\text{TT2}}^{(2)}(1,2,3) + c_{13} I_{\text{TBox0}}^{(2)}(1,2,3) + c_{14} I_{\text{BoxBub}}^{(2)}(1,2,3)+c_{15} I_{\text{BubBox0}}^{(2)}(1,2,3) \nonumber \\
	& +c_{16} I_{\text{BubBox}}^{(2)}(1,2,3) + c_{17} I_{\text{dBox1a}}^{(2)}(1,2,3)+c_{18} I_{\text{dBox1b}}^{(2)}(1,2,3) +c_{19} I_{\text{NTBox1}}^{(2)}(1,2) \nonumber \\
	& +c_{20} I_{\text{NTBox2}}^{(2)}(1,2,3) + c_{21} I_{\text{NTBox3a}}^{(2)}(1,2,3) + c_{22} I_{\text{NTBox3b}}^{(2)}(1,2,3)+c_{23} I_{\text{NdBox1a}}^{(2)}(1,2,3) \nonumber \\
	& +c_{24} I_{\text{NdBox1b}}^{(2)}(1,2,3) + c_{25} I_{\text{NdBox2a}}^{(2)}(1,2,3)+c_{26} I_{\text{NdBox2b}}^{(2)}(1,2,3) + \left( \text{full perm.(1,2,3)} \right) \nonumber \,,
\end{align}
where the parameters $c_i$ in ansatz can be solved by imposing constraints. We comment that some master integrals have symmetry of the external momentum, thus there are only 89 independent master integrals and parameters.

%%%%%%%%%%%%%%%%
%%%%%%%%%%%%%%%%

As we emphasized, all these master integrals are pure UT integrals with uniform transcendentality degree-$4$ and the explicit definitions are given in Appendix~\ref{app:UT}. Their expressions were obtained in \cite{Gehrmann:2000zt, Gehrmann:2001ck} as two-dimensional harmonic polylogarithms which depend on Mandelstam variables \{$s_{12}$, $s_{23}$, $s_{13}$\}, plugging them into our ansatz \eqref{eq:3pt2loopAnsatz}, we can obtain the form factor expression in powers of $\epsilon$, which starts from $\epsilon^{-4}$.
We can apply the constraints to solve for the coefficients of the ansatz in \eqref{eq:3pt2loopAnsatz}.

First, we consider the form factor $\mathcal{F}_{\mathcal{O}_2, 3}^{(2)}$ has the symmetry of cycling external momentum, $\mathcal{F}_{\mathcal{O}_2, 3}^{(2)}(p_1,p_2,p_3)=\mathcal{F}_{\mathcal{O}_2, 3}^{(2)}(p_2,p_3,p_1)$. The symmetry will reduce the number of parameters from 89 to 24.

To apply further physical constraints, we consider the BDS ansatz function
\begin{equation}
	\mathcal{I}_{\mathcal{O}_2, 3, \text{BDS}}^{(2)} = \frac{1}{2} \left(\mathcal{I}_{\mathcal{O}_2, 3}^{(1)}(\epsilon)\right)^2+f^{(2)}(\epsilon) \mathcal{I}_{\mathcal{O}_2, 3}^{(1)}(2\epsilon) \,.
\end{equation}
As reviewed in Section~\ref{sec:constraints}, the BDS ansatz (using only one-loop data) provides the divergent part of $1/\epsilon^m, m=4,3,2,1$. By matching with our ansatz \eqref{eq:3pt2loopAnsatz}, the number of parameters is reduced to 13.

Furthermore, the finite remainder obtained by subtracting the BDS ansatz part has the collinear limits 
\begin{equation}
	\label{eq:3ptremainderCL}
	\mathcal{R}_{\mathcal{O}_2, 3}^{(2)} = \mathcal{I}_{\mathcal{O}_2, 3}^{(2)}-\mathcal{I}_{\mathcal{O}_2, 3, \text{BDS}}^{(2)} \ \xlongrightarrow{\mbox{$p_i \parallel p_j$}} \mathcal{R}_{\mathcal{O}_2, 2}^{(2)} \, ,
\end{equation}
where $p_i$ and $p_j$ can be any pair of momentum, and the Sudakov form factor result $\mathcal{R}_{\mathcal{O}_2, 2}^{(2)} = (6 - y_0) \zeta_4$ is given in Section~\ref{sec:miniFF}, here we take $y_0$ as an input known parameter.
The limit can be calculated with the series of two-dimensional harmonic polylogarithms, or by evaluating the master integrals with a very small $s_{ij} = \delta q^2$ with $\delta \ll 1$ in high-precision. After applying all three collinear limits, there is only one parameter unfixed. 

After this step, we can organize the form factor result in the following form
\begin{equation}
    \mathcal{I}_{\mathcal{O}_2, 3}^{(2)} = G_1^{(2)} + y_0 G_2^{(2)} + c' \tilde{G}^{(2)} \,,
\end{equation}
where $c'$ is the unfixed parameter, and each function on the RHS can be expanded by master integrals as
\begin{align}
	G_1^{(2)} = & -I_{\text{dBub}}^{(2)}(1,2;1,2,3) +\frac{2}{3} I_{\text{dBub}}^{(2)}(1,2,3;1,2,3) -3 I_{\text{BubBox}}^{(2)}(1,2,3) \\
	& +\frac{1}{2} I_{\text{BubBox0}}^{(2)}(1,2,3) +\frac{1}{2} I_{\text{dBox1a}}^{(2)}(1,2,3) +\frac{1}{2} I_{\text{dBox1b}}^{(2)}(1,2,3) +\frac{1}{8} I_{\text{NdBox1a}}^{(2)}(1,2,3) \nonumber \\
	& +\frac{1}{8} I_{\text{NTBox2}}^{(2)}(1,2,3) -\frac{1}{4} I_{\text{NTBox3b}}^{(2)}(1,2,3) -\frac{11}{6} I_{\text{Sun}}^{(2)}(1,2,3) -\frac{11}{4} I_{\text{Sun}}^{(2)}(1,2) \nonumber \\
	& +\frac{7}{4} I_{\text{TBub2}}^{(2)}(1,2,3) +\frac{1}{2} I_{\text{TT0}}^{(2)}(1,2,3) -\frac{1}{4} I_{\text{TT1}}^{(2)}(1,2,3) +\frac{5}{4} I_{\text{TT2}}^{(2)}(1,2,3) + (\text{full permute}) \,, \nonumber \\
	G_2^{(2)} = & \frac{3}{4} I_{\text{Sun}}^{(2)}(1,2) +\frac{13}{3} I_{\text{Sun}}^{(2)}(1,2,3) +\frac{3}{2} I_{\text{TBub1}}^{(2)}(1,2,3) -\frac{1}{4} I_{\text{TT1}}^{(2)}(1,2,3) \\
	& -\frac{1}{2} I_{\text{TT1a}}^{(2)}(1,2,3) -\frac{5}{4} I_{\text{TT2}}(1,2,3) +\frac{1}{3} I_{\text{dBub}}^{(2)}(1,2,3;1,2,3) \nonumber \\
	& -\frac{1}{4} I_{\text{NTBox3a}}^{(2)}(1,2,3) +\frac{3}{4} I_{\text{NTBox3b}}^{(2)}(1,2,3) + (\text{cyclic perm.(1,2,3)}) \,, \nonumber \\
	\tilde{G}^{(2)} = & I_{\text{dBub}}^{(2)}(1,2;1,2) - I_{\text{dBub}}^{(2)}(1,2,3;1,2,3)  +\frac{35}{4} I_{\text{Sun}}^{(2)}(1,2) -13 I_{\text{Sun}}^{(2)}(1,2,3) \\
    & +\frac{5}{2} I_{\text{TBub0}}^{(2)}(1,2) -\frac{9}{2} I_{\text{TBub1}}^{(2)}(1,2,3) +\frac{3}{4} I_{\text{TT1}}^{(2)}(1,2,3) +\frac{3}{2} I_{\text{TT1a}}^{(2)}(1,2,3) +\frac{15}{4} I_{\text{TT2}}^{(2)}(1,2,3) \nonumber \\
    & +\frac{1}{2} I_{\text{NTBox1}}^{(2)}(1,2) +\frac{3}{4} I_{\text{NTBox3a}}^{(2)}(1,2,3) -\frac{9}{4} I_{\text{NTBox3b}}^{(2)}(1,2,3) + (\text{cyclic perm.(1,2,3)}) \,, \nonumber
\end{align}
where the building-block $\tilde{G}^{(2)}$ is finite and vanishes in the collinear limits, thus it is unconstrained by infrared structure and collinear limit. The function $G_2^{(2)}$ is infrared finite and has collinear limit behavior as $G_2^{(2)} \rightarrow - \zeta_4 + \mathcal{O}(\epsilon)$ for any $p_i \parallel p_j$.

We note that the building-block $\tilde{G}^{(2)}$ contains the nonplanar master integrals $I_{\text{NTBox1}}^{(2)}(i,j)$, which cannot occur in two-loop three-point cases since its color factor is zero when the particles are adjoint representation. This gives $c' = 0$. Then the parameters in ansatz are all fixed.
We summarize the above process of applying constraints in Table~\ref{tab:solvingAnsatz3ptphi2}.

%%%%%%%%%%%%%%%  TABLE   %%%%%%%%%%%%%%%%%%%%%
\begin{table}[t]
	\centering
	\vskip .1 cm 
	\begin{tabular}{| c | c |} 
		\hline
		Constraints				&  Remaining parameters afterwards   \\ \hline \hline
		Starting ansatz   		&  89  \\ \hline
		Symmetry(cycles)   		&  24  \\ \hline
		IR   					&  11  \\ \hline
		Collinear limit 		&  1  \\ \hline
		Color factor structure	&  0  \\ \hline
	\end{tabular}
	\caption{Solving ansatz via various constraints for two-loop three-point $\mathrm{tr}(\phi^2)$.
		\label{tab:solvingAnsatz3ptphi2}}
\end{table}
%%%%%%%%%%%%%%%%%%%%%%%%%%%%%%%%%%%%%%%%%%

Finally, we comment on the input parameter $y_0$, which is introduced by the Sudakov form factor. 
Interestingly, supposing we do not know its value, which can be fixed by requiring that the form factor $\mathcal{F}_{\mathcal{O}_2, 3}^{(2)}$ has a smooth limit behavior when $q^2 \rightarrow 0$.
Let us explain how this works. It is not hard to see that for each of many master integrals, the $\epsilon$-expansion series are divergent when taking $q^2 \rightarrow 0$. For example, $I_{\text{Sun}}^{(2)}(1,2,3)$ contains the logarithm terms as
\begin{equation}
	I_{\text{Sun}}^{(2)}(1,2,3) \propto (-q^2)^{-2\epsilon} = \sum_{k=0}^{\infty} \frac{\log^k(-q^2)}{k!} \epsilon^k \,.
\end{equation}
Some of harmonic polylogarithms contained in other master integrals also diverge logarithmically in the limit. 
Therefore, the smooth limit behavior of $q^2 \rightarrow 0$ provides non-trivial constraints on the form factor: these logarithmic divergences from different master integrals should cancel with each other in the full form factor result. 
One can study the above limit by taking the series expansion analytically,  or an alternative way is to apply Cauchy's convergence test which equivalent to the smooth limit: for any positive number $\eta$, there should always exist $\delta>0$, such that the following inequality holds for $0<|x_1|<\delta$ and $0<|x_2|<\delta$:
\begin{equation}
	\left| \left(\mathcal{I}_{\mathcal{O}_2, 3}^{(2)}\right) \Big|_{q^2 = x_1} - \left(\mathcal{I}_{\mathcal{O}_2, 3}^{(2)}\right) \Big|_{q^2 = x_2} \right| < \eta \,.
\end{equation}
This is easy to be implemented numerically, and in this way, one obtains constraints on the master coefficients. We mention that the limit of $q^2 = s_{12}+s_{23}+s_{13} \rightarrow 0$ implies that some $s_{ij}$ must be positive, so it is not possible to stay in the Euclidean region (with all $s_{ij}<0$) in the limit, and proper analytical continuation is required.
By applying this constraint, one can find that $y_0 = 2$, which is consistent with the known result.

%%%%%%%%%%%%%%%%%%%%%%%%%%%%%%%%%
\subsection{Understanding maximal transcendentality principle}
\label{sec:MTP3pt}

Now we consider the form factors in QCD. Unlike the $\mathcal{N}=4$ case where there is only one \emph{supersymmetric} form factor, in QCD one needs to distinguish three types of external particles, which are $(1^-, 2^-, 3^-)$, $(1^-, 2^-, 3^+)$ and $(1^q, 2^{\bar{q}}, 3^-)$ \cite{Gehrmann:2011aa}. 
We will only focus on their maximally transcendental parts and prove the relations of maximal transcendentality \eqref{eq:MTP3pt}.
We first consider the bootstrap constraints and then apply also the unitarity-cut arguments.

%%%%%%%%%%%%%%%%%%
\paragraph{Bootstrap constraints.}

The three types of external particles have different physical constraints:
%%%%%%%%%%%

\begin{itemize}
\item 
The configuration $(1^-, 2^-, 3^-)$ has full $S_3$ permutational symmetries as the $\mathcal{N}=4$ SYM form factor, and thus the constraints one can use in this case are identical to the case of $\mathcal{N}=4$ SYM.  
\item
The case $(1^-, 2^-, 3^+)$ in QCD has only the sub-symmetry of $S_3$ by exchanging $p_1$ and $p_2$. The collinear limits are also different, and one can only take $p_2 \parallel p_3$ or $p_1 \parallel p_3$. (Recall that we focus on the self-dual operator ${\cal O}_2 = {\rm tr}(F_{\alpha\beta}F^{\alpha\beta})$, and the collinear $p_1 \parallel p_2$ is zero in the limit since ${\cal F}_{{\cal O}_2}^{(0)}(1'^\pm,3^+) = 0$.)
\item
For the $(1^-, 2^-, 3^-)$ and $(1^-, 2^-, 3^+)$ cases, an analysis of color factor for Feynman diagrams shows that only the pure gluon configurations contribute the leading $N_c^2$ color parts, and they have no $N_c$-subleading corrections. In particular, the fifth topology in Figure~\ref{fig:maxTopology3pt} has zero color factors which means that the non-planar master integrals $I_{\text{NTBox1}}^{(2)}$, $I_{\text{NTBox2a}}^{(2)}$ and $I_{\text{NTBox2b}}^{(2)}$ will not contribute.
\item
The  $(1^q, 2^{\bar{q}}, 3^-)$ case has least constraints: there is no symmetry property to use and the only collinear limit constraint is $p_1 \parallel p_2$.\footnote{The form factor ${\cal F}^{(0)}(1^q, 2^{\bar{q}}, 3^-) = {\langle 2 3\rangle^2 / \langle12\rangle}$ is non-zero by taking the limit $p_1 \parallel p_3$. However, to apply the factorization property \eqref{eq:collinear-general}, the collinear limit requires a physical pole, which does not exist in the above limit.}
\item
Finally, one can apply the condition that form factors should have a smooth limit when $q^2 \rightarrow 0$. This will be automatically satisfied for the first two configurations but provide non-trivial constraints for the $(1^q, 2^{\bar{q}}, 3^-)$ case.

\end{itemize}
%%%%%%%%%%%
Since the bootstrap procedures are similar to the $\mathcal{N}=4$ case in the previous subsection, we will not go into details but only
summarize the main steps in Table~\ref{tab:solvingAnsatz3ptphi2F2-1loop} and Table~\ref{tab:solvingAnsatz3ptphi2F2},
for one and two loops respectively.

%%%%%%%%%%%%%%%  TABLE  %%%%%%%%%%%%%%%%%%%%%
\begin{table}[t]
	\centering
	\vskip .1 cm 
	\begin{tabular}{| p{5cm}<{\centering} | p{2.5cm}<{\centering} | p{2.5cm}<{\centering} | p{2.5cm}<{\centering} |} 
		\hline
		External particles	&	$(1^-, 2^-, 3^-)$ & $(1^-, 2^-, 3^+)$ & $(1^q, 2^{\bar{q}}, 3^-)$ \\ \hline\hline
		Constraints			&	\multicolumn{3}{c|}{Remaining parameters} \\ \hline 
		Starting ansatz   	&	7 & 7 & 7 \\ \hline
		IR   				&	4 & 4 & 4 \\ \hline
		Collinear limit 	&	0 & 0 & 1 \\ \hline
	\end{tabular}
	\caption{Bootstrap for one-loop form factors of $\mathrm{tr}(F^2)$.
		\label{tab:solvingAnsatz3ptphi2F2-1loop}}
\end{table}
%%%%%%%%%%%%%%%%%%%%%%%%%%%%%%%%%%%%%%%%%%

%%%%%%%%%%%%%%%  TABLE  %%%%%%%%%%%%%%%%%%%%%
\begin{table}[t]
	\centering
	\vskip .1 cm 
	\begin{tabular}{| p{5cm}<{\centering} | p{2.5cm}<{\centering} | p{2.5cm}<{\centering} | p{2.5cm}<{\centering} |} 
		\hline
		External particles	&	$(1^-, 2^-, 3^-)$ & $(1^-, 2^-, 3^+)$ & $(1^q, 2^{\bar{q}}, 3^-)$ \\ \hline\hline
		Constraints			&	\multicolumn{3}{c|}{Remaining parameters} \\ \hline 
		Starting ansatz   	&	89 & 89 & 89 \\ \hline
		Symmetry			&	24 & 53 & 89 \\ \hline
		IR   				&	11 & 21 & 48 \\ \hline
		Collinear limit 	&	1 & 5 & 21 \\ \hline
		Color factor		&	0 & 2 & 21 \\ \hline
		Smooth light-like limit of $q$
							&	0 & 0 & 11 \\ \hline
	\end{tabular}
	\caption{Bootstrap for two-loop  form factors of $\mathrm{tr}(F^2)$.
		\label{tab:solvingAnsatz3ptphi2F2}}
\end{table}
%%%%%%%%%%%%%%%%%%%%%%%%%%%%%%%%%%%%%%%%%%

We can see that all parameters in the cases $(1^-, 2^-, 3^-)$ and $(1^-, 2^-, 3^+)$ can be fixed with the constraints. We would like to stress that the bootstrap constraints are theory-independent, therefore the maximally transcendental parts of form factors $\mathcal{F}_{\mathrm{tr}(F^2), 3}^{(L)}(1^-,2^-,3^\pm)$ for $L=1,2$ should be the same for general gauge theories.
On the other hand, for the $(1^q, 2^{\bar{q}}, 3^-)$ case, there is one parameter left that is not fixed by IR and collinear constraints at one loop, 
and there are 11 parameters left at two loops as shown in Table~\ref{tab:solvingAnsatz3ptphi2F2}.
Below we show that with the unitarity-cut method, its maximally transcendental part can be also proved to be universal.

\paragraph{Unitarity-cut for the $(1^q, 2^{\bar{q}}, 3^-)$ case.}

Below we would like to show that the QCD result is equivalent to the $\mathcal{N}=4$ result for $(1^q, 2^{\bar{q}}, 3^-)$ case by converting QCD quarks from the fundamental to the adjoint representation.
Note that due to supersymmetry, the $\mathcal{N}=4$ SYM result is the same for all possible choices of external states, this is then enough to prove \eqref{eq:MTP3pt}.

\begin{figure}[t]
	\centering
	\includegraphics[scale=0.55]{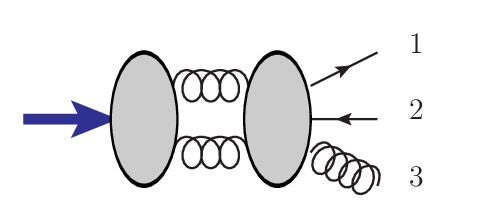}
	\caption{A unitarity cut for the one-loop form factor ${\cal F}^{(1)}(1^{q}, 2^{\bar{q}}, 3^-)$.}
	\label{fig:cut1loop3point}
\end{figure}

To determine this remaining parameter at one loop, one can use the unitarity cut shown in Figure~\ref{fig:cut1loop3point}. Actually, there is no need to perform this computation, since this cut is the same for $\mathcal{N}=4$ SYM and QCD if one changes the quadratic Casimir $C_F \rightarrow C_A$ in QCD. Thus we prove the relations in \eqref{eq:MTP3pt} at one loop.

%%%%%%%%%%%%%%%%%%%%%%%%%%%
%%%%%%%%%%%%%%%%%%%%%%%%%%%

%%%%%%%%%%%%%%%%%%%%%%%%%%%
\begin{figure}[t]
	\centering
	\includegraphics[scale=0.5]{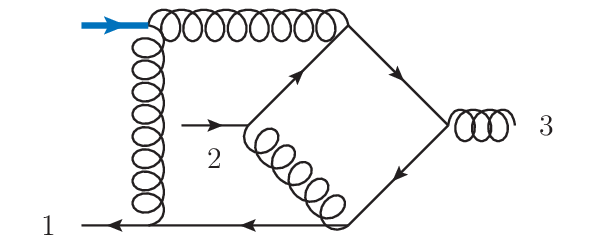}
	\caption{The QCD Feynman diagram that contributes to the fifth topology in Figure~\ref{fig:maxTopology3pt}.}
	\label{fig:colorF2}
\end{figure}
%%%%%%%%%%%%%%%%%%%%%%%%%%%

The two-loop case is less trivial.
We first recall that the fifth topology in Figure~\ref{fig:maxTopology3pt} has zero color factors for adjoint particles, therefore, when we convert quarks to be adjoint, the non-planar master integrals $I_{\text{NTBox1}}^{(2)}$, $I_{\text{NTBox2a}}^{(2)}$ and $I_{\text{NTBox2b}}^{(2)}$ can not contribute.
The related Feynman diagram contribution in QCD is shown in Figure~\ref{fig:colorF2}. Its color factor can be computed as
\begin{equation}
	t_F^2 (C_A-C_F) (C_A-2C_F) (T^{a_3})_{i_1}^{~\bar{j}_2} \,,
\end{equation}
which indeed vanishes when taking $C_F \rightarrow C_A$.
In this way, one can eliminate 6 parameters that are related to these non-planar master integrals. 

%%%%%%%%%%%%%%%%%%%%%%%%%%%
\begin{figure}[t]
	\centering
	\subfloat[]{\includegraphics[scale=0.55]{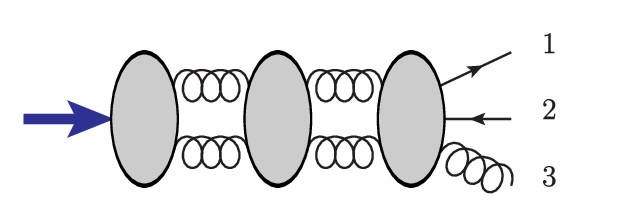}}
	\subfloat[]{\includegraphics[scale=0.5]{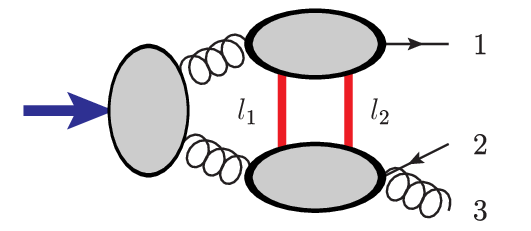}}
	\subfloat[]{\includegraphics[scale=0.5]{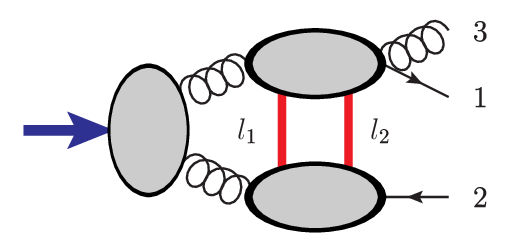}}
	\caption{Unitarity cuts for the two-loop form factor $\mathcal{F}^{(2)}(1^{q}, 2^{\bar{q}}, 3^-)$.}
	\label{fig:cut2loop3point-MT}
\end{figure}
%%%%%%%%%%%%%%%%%%%%%%%%%%%

For the remaining 5 parameters, we will apply the unitarity cuts in Figures~\ref{fig:cut2loop3point-MT} to fix the related five master integrals as follows: 
\begin{align}
\textrm{cut-(a)}: & \quad I_{\text{dBub}}^{(2)}(1,2,3;1,2,3) \,, \nonumber\\
\textrm{cut-(b)}: & \quad I_{\text{TT1}}^{(2)}(3,1,2) \,, \quad I_{\text{dBox1a}}^{(2)}(2,3,1) \,, \nonumber\\
\textrm{cut-(c)}: & \quad I_{\text{TT1}}^{(2)}(1,2,3) \,, \quad I_{\text{dBox1a}}^{(2)}(3,1,2) \,. \nonumber
\end{align}
The cut-(a) is universal for general gauge theory since the cut legs can only be gluons.
On the other hand, the cut-(b) and (c) are more complicated, since the cut legs $\{l_1,l_2\}$ can have different particle configurations. 
For example, the cut-(b) integrand in $\mathcal{N}=4$ SYM is:
\begin{align}
\label{eq:cutintF2A4A4}
	\int \prod_{i=1}^2 d^4\eta_{l_i} \mathcal{F}_{\mathcal{O}_{2},2}^{(0), \text{MHV}}(-l_3^g,-l_4^g) \mathcal{A}_4^{(0), \text{MHV}}(1^\psi,l_2, l_1,l_3^g) \mathcal{A}_4^{(0), \text{MHV}} (2^{\bar{\psi}},3^g,l_4^g,-l_1,-l_2) \,,
\end{align}
where the fermionic integration corresponds to sum $l_i$ over all possible super-states in the $\mathcal{N}=4$ on-shell superfield \cite{Nair:1988bq}:
\begin{equation}
	\Phi(l,\eta)=g_+(l) +  \eta^A \, \bar\psi_A(l) + {\eta^A\eta^B \over 2!} \, \phi_{AB}(l) + { \epsilon_{ABCD} \eta^A\eta^B\eta^C \over 3!} \, \psi^D(l) + \eta^1\eta^2\eta^3\eta^4 \, g_-(l) \,.
	\label{eq:onshellN=4superspace}
\end{equation}
More explicitly, \eqref{eq:cutintF2A4A4} can be expanded as
\begin{align}
	& \mathcal{F}_{\mathcal{O}_{2}, 2}^{(0)}(-l_3^g, -l_4^g) \mathcal{A}_{4}^{(0)}(1^\psi, l_2^{{\bar\psi}/g}, l_1^{g/{\bar\psi}}, l_3^g) \mathcal{A}_{5}^{(0)}(2^{\bar\psi}, 3^g, l_4^g, -l_1^{g/\psi}, -l_2^{\psi/g})  \nonumber\\
	& +\mathcal{F}_{\mathcal{O}_{2}, 2}^{(0)}(-l_3^g, -l_4^g) \mathcal{A}_{4}^{(0)}(1^\psi, l_2^{\psi/\phi}, l_1^{\phi/\psi}, l_3^g) \mathcal{A}_{5}^{(0)}(2^{\bar\psi}, 3^g, l_4^g, -l_1^{\phi/\bar\psi}, -l_2^{\bar\psi/\phi}) \,. 
	\label{eq:cutB3pt}
\end{align}
The configuration in the first line contains only gluon and gluino states, thus they will map to the QCD result by converting color factors of fermions accordingly. 
The second line in \eqref{eq:cutB3pt}, however, involves the scalar particles which are special for the $\mathcal{N}=4$ SYM theory.
To have the maximally transcendental principle, it is crucial that the scalar configuration have not any leading transcendental contribution.
An explicit calculation shows that the configuration involving scalars indeed do not contribute to the maximally transcendental part. 
The case of cut-(c) is similar.
Therefore, all the configurations that contribute to maximally transcendental parts are the same for $\mathcal{N}=4$ SYM and QCD, and if one changes the quadratic Casimir $C_F \rightarrow C_A$ in QCD, the QCD form factor result is the same as that of $\mathcal{N}=4$ SYM in the case $(1^q, 2^{\bar{q}}, 3^-)$.

%%%%%%%%%%%%%%%%%%%%%%%%%%%%%%%
\section{Two-loop four-point form factor of $\mathrm{tr}(F^3)$}
\label{sec:2loop4ptF3}

In this section, we consider further two-loop four-point form factor of length-three operators:
\begin{align}
	\mathcal{F}_{\mathcal{O}_3,4} = \int d^D x e^{-i q\cdot x} \langle p_1 \, p_2 \, p_3 \, p_4 |\mathcal{O}_3(x)|0\rangle \,. \nonumber
\end{align}
The case of $\mathrm{tr}(\phi^3)$ form factor has been obtained by bootstrapping recently in \cite{Guo:2021bym}. In this paper we will compute a similar form factor which contains a length-three operator $\mathrm{tr}(F^3)$ and four external on-shell gluon states, defined concretely as
\begin{align}
	\mathcal{F}_{\mathrm{tr}(F^3),4} := & \mathcal{F}_{\mathrm{tr}(F^3),4}(1^-, 2^-, 3^-, 4^+;q) \\
	= & \int d^D x e^{-i q\cdot x} \langle g_-(p_1) g_-(p_2) g_-(p_3) g_+(p_4) | \mathrm{tr}(F^3)(x)|0\rangle \,. \nonumber
\end{align}
As in the previous three-point form factor, this form factor can be understood as the Higgs-plus-four-gluon scattering amplitudes in the Higgs EFT with a dimension-six operator.
Unlike $\mathrm{tr}(\phi^3)$, the operator $\mathrm{tr}(F^3)$ is non-BPS in $\mathcal{N}=4$ SYM, thus the form factor also receives contribution from lower transcendental parts in $\mathcal{N}=4$ SYM.
In this paper, we will focus on the maximal transcendental part.
We will first consider the form factors in $\mathcal{N}=4$ SYM and pure YM, and we apply bootstrap strategy to obtain the maximally transcendental parts up to two loops in Section~\ref{sec:ansatzTrF3}-\ref{sec:solvingAnsatz}. Then we will discuss the correspondence between QCD and $\mathcal{N}=4$ results in Section~\ref{sec:MTPtrF3}.
We finally discuss the connection between form factors of $\mathrm{tr}(F^3)$ and $\mathrm{tr}(\phi^3)$ in Section~\ref{sec:phi3andF3}.

%%%%%%%%%%%
\subsection{Ansatz of the form factor up to two loops}
\label{sec:ansatzTrF3}

We define the maximally transcendental part of the loop corrections $\mathcal{I}_{\mathrm{tr}(F^{3}), 4}^{(L), \text{M.T.}}$ as
\begin{equation}
	\mathcal{F}_{\mathrm{tr}(F^{3}), 4}^{(L), \text{M.T.}} = \mathcal{F}_{\mathrm{tr}(F^{3}), 4}^{(0)} \, \mathcal{I}_{\mathrm{tr}(F^{3}), 4}^{(L), \text{M.T.}} \,,
\end{equation}
in which the tree-level result takes the simple form as 
\begin{equation}
\label{eq:treeF3}
	\mathcal{F}_{\mathrm{tr}(F^{3}), 4}^{(0)} = \frac{ \langle12\rangle \langle23\rangle \langle31\rangle^2 }{ \langle34\rangle \langle41\rangle } \, .
\end{equation}
Our goal is to compute the loop corrections in terms of master integral expansion:
\begin{equation}
\mathcal{I}_{\mathrm{tr}(F^{3}), 4}^{(L), \text{M.T.}} = \sum_i x_i I_i^{(L)} \,.
\end{equation}
Since we choose $I_i^{(L)}$ as pure UT integrals of transcendentality $2L$, the coefficients $x_i$ are independent of dimensional parameter $\epsilon$.

Unlike the discussion in previous sections, there is a major complication for the four-point form factor studied here, namely, the master coefficients will have non-trivial dependence on kinematics factors as $x_i = \sum_a c_{i,a} B_a$, where $B_a$ are kinematic factors. We will classify the possible $B_a$ factors.

First, since the tree factor has been factorized out, the $B_a$ factors must carry no helicity weight, in other words, they can only be functions of cross ratios of spinor products, or Mandelstam $s_{ij}$ variables.
Inspired by result of the four-point form factor of $\mathrm{tr}(\phi^3)$ \cite{Guo:2021bym}, we can assume that the factor $B_a$ can be expressed with only the cross ratios of angle brackets, which take the following form
\begin{equation}
	\frac{\langle i j \rangle \langle k l \rangle}{\langle i k \rangle \langle j l \rangle} \,.
\end{equation}
The appropriate basis of cross ratios can be selected as
\begin{equation}
	\label{eq:DefB1B2}
	B_1 = \frac{\langle12\rangle \langle34\rangle}{\langle13\rangle \langle24\rangle} \,, \qquad B_2 = \frac{\langle14\rangle \langle23\rangle}{\langle13\rangle \langle24\rangle} \,,
\end{equation}
which appears in the result of $\mathrm{tr}(\phi^3)$ \cite{Guo:2021bym}.

Next, we require the $B_a$ factors contain only poles that appear in the numerator of the tree-level form factor \eqref{eq:treeF3}, and one has
\begin{equation}
\label{eq:PoleFromTree}
B_a \sim \left\{ \frac{1}{\langle 12 \rangle}\,,  \frac{1}{\langle 23 \rangle}\,, \frac{1}{\langle 13 \rangle } \,, \frac{1}{\langle 13 \rangle^2 } \right\} \,.
\end{equation}
Besides $B_1$ and $B_2$, the set of $B_a$ factors which contain these poles can be given as
\begin{equation}
	B_3 = B_1 B_2 = \frac{\langle12\rangle \langle34\rangle \langle14\rangle \langle23\rangle }{\langle13\rangle^2 \langle24\rangle^2} \, , \quad 
	B_4 = \frac{B_1}{B_2} = \frac{\langle12\rangle \langle34\rangle}{\langle14\rangle \langle23\rangle} \, , \quad 
	B_5 = \frac{B_2}{B_1} = \frac{\langle14\rangle \langle23\rangle}{\langle12\rangle \langle34\rangle} \,.
\end{equation}
We will assume that all master coefficients is a linear combination fo $B_a$, where $a =1, ..,5$.
We would like to comment that the poles \eqref{eq:PoleFromTree} will disappear when timing together with the tree form factors in the full form factor. On the other hand, since $B_a$ are cross ratios, new poles are introduced:
\begin{equation}
\label{eq:SpuriousPoleFromTree}
\textrm{spurious poles} : \quad \left\{ \frac{1}{\langle 34 \rangle} \,,  \frac{1}{\langle 14 \rangle} \,, \frac{1}{\langle 24 \rangle^2} \right\} ,
\end{equation}
which can not be canceled by tree form factor.
These poles are spurious poles in the sense that they must cancel within the loop correction function $\mathcal{I}_{\mathrm{tr}(F^{3}), 4}^{(L), \text{M.T.}}$.
As we will see, this requirement will provide important constraints on the ansatz results.

It is convenient to reorganize  the loop corrections as
\begin{equation}
\label{eq:ansatzFullF3}
	\mathcal{I}_{\mathrm{tr}(F^{3}), 4}^{(L), \text{M.T.}} = \sum_{a=1}^5 B_a \mathcal{G}_a^{(L)} \, ,
\end{equation}
where $\mathcal{G}_a^{(L)}$ are the expansion of a set of pure UT master integrals with only numerical coefficients. In addition, they are not all independent because of the symmetry as follows
\begin{equation}
\label{eq:Gsymmetry}
	\mathcal{G}_1^{(L)} = \mathcal{G}_2^{(L)} |_{(p_1\leftrightarrow p_3)} \, , \qquad \mathcal{G}_3^{(L)} = \mathcal{G}_3^{(L)} |_{(p_1\leftrightarrow p_3)} \, , \qquad \mathcal{G}_4^{(L)} = \mathcal{G}_5^{(L)} |_{(p_1\leftrightarrow p_3)} \,,
\end{equation}
which is determined by the symmetry of full form factor by ${p_1\leftrightarrow p_3}$ and the symmetry properties of $B_a$.
Therefore, one only needs to focus on $\mathcal{G}_1^{(L)}$, $\mathcal{G}_3^{(L)}$ and $\mathcal{G}_5^{(L)}$.
Below we discuss the ansatz in terms of master integrals in more detail.

For the one-loop form factor, there are 12 master integrals
\begin{equation}
	\label{eq:ansatz1loop}
	\mathcal{G}_a^{(1)} = c_{a, 1}^{(1)} I_{\text{Bub}}^{(1)}(1, 2) + c_{a, 2}^{(1)} I_{\text{Bub}}^{(1)}(1, 2, 3) + c_{a, 3}^{(1)} I_{\text{Box}}^{(1)}(1, 2, 3) + \text{cyclic perm.}(1,2,3,4) \,.
\end{equation}
Using the symmetry properties \eqref{eq:Gsymmetry}, one can find there are in total 32 = 12+12+8 (for $\mathcal{G}_1^{(1)}$, $\mathcal{G}_5^{(1)}$ and $\mathcal{G}_3^{(1)}$ respectively) independent parameters for the one-loop correction.

%%%%%%%%%%%%%%%%
\begin{figure}[tb]
	\centering
	\includegraphics[scale=0.5]{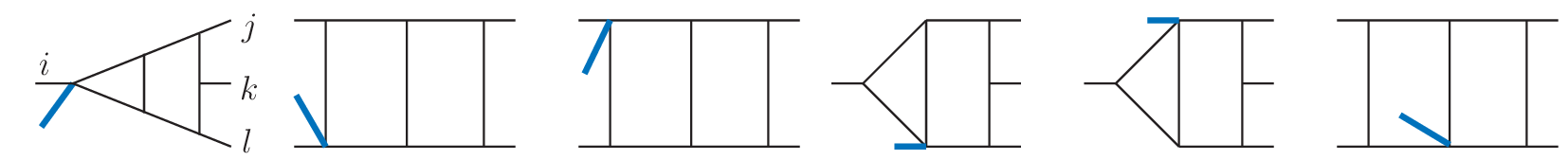}
	\caption{Integral topologies of maximal number of propagators for the planar two-loop form factor.}
	\label{fig:maxTopology}
\end{figure}
%%%%%%%%%%%%%%%%

For the two-loop form factor, the integral topologies with the maximal number of propagators (which are enough to cover all the master integrals we need) are shown in Figure~\ref{fig:maxTopology}.
Since the operator has length-three, the massive $q$-leg (denoted by blue color) should be connected to a four-vertex. 
To obtain the full form factor, one needs to consider all possible insertions of the $q$-leg, since the operator is a color-singlet.
The most general ansatz contains 221 pure UT master integrals: 
\begin{equation}
	\label{eq:ansatz2loop}
	\mathcal{G}_a^{(2)} = \sum_{i=1}^{221} c_{a,i}^{(2)} I_{i}^{(2)} \,.
\end{equation}
The definition of the involved master integrals are given in Appendix~\ref{app:UT}, and they have been computed in \cite{Abreu:2020jxa, Canko:2020ylt}, see also \cite{Chicherin:2021dyp, Abreu:2021smk}.
The main goal is to solve the coefficients $c_{a, i}^{(2)}$. 
Using again the symmetry properties \eqref{eq:Gsymmetry}, 
one finds there are 221 free parameters from $\mathcal{G}_1^{(2)}$ and $\mathcal{G}_5^{(2)}$ respectively, and 118 from $\mathcal{G}_3^{(L)}$. In total, our two-loop ansatz contains 560 free parameters.

%%%%%%%%%%%%%%%%%%%%%
\subsection{Solving one-loop result}
\label{sec:solveFF4pt1loop}

We start with the one-loop ansatz given by \eqref{eq:ansatzFullF3} and \eqref{eq:ansatz1loop}. The bootstrap method can be applied using similar steps as in Section~\ref{subsec:3ptff1loop}. 

First, the infrared structure is
\begin{equation}
	\mathcal{I}_{\mathrm{tr}(F^3),4}^{(1)} \Big|_{\text{IR}} = \left( B_1+B_2 \right) \sum_{i=1}^4 \left( -\frac{1}{\epsilon^2} + \frac{\log(-s_{i,i+1})}{\epsilon} \right) \, .
\end{equation}
where $B_1+B_2=1$ is used. To match with the ansatz, one finds that
\begin{equation}
	\mathcal{G}_1^{(1)} \Big|_\text{IR} = \sum_{i=1}^n \left( -\frac{1}{\epsilon^2}+\frac{\log(-s_{i,i+1})}{\epsilon} \right)  \, , \qquad \mathcal{G}_3^{(1)} \Big|_\text{IR} = \mathcal{G}_5^{(1)} \Big|_\text{IR} = 0 \, .
\end{equation}
This can fix 21 degrees of freedom, leaving 11 free parameters.

Next, we consider the constraints of collinear limit $p_3 \parallel p_4 \parallel p_3' = p_3+p_4$ (the other collinear limit $p_1 \parallel p_4 \parallel p_1' = p_1+p_4$ is related by symmetry). 
The Mandelstam variables in the limit can be taken as 
\begin{equation}
	\label{eq:colllinearofs}
	s_{34} \rightarrow \delta q^2 \,, \quad s_{i3} \rightarrow z(1-\delta) s_{i3}'\,, \quad s_{i4} \rightarrow (1-z)(1-\delta) s_{i3}' \,, 
\end{equation} 
where $\delta\ll1$, $q^2 = (p_1+p_2+p_3+p_4)^2 = (p_1+p_2+p_3')^2$ and $s_{i3}' = (p_i+p_3')^2$ with $i = 1, 2$. Meanwhile the limits of $B_a$ factors are
\begin{equation}
	\label{eq:colllinearofB}
	\{B_1, B_3, B_4\} \rightarrow 0 \,, \qquad B_2 \rightarrow 1 \,, \qquad B_5 \rightarrow \infty \,, 
\end{equation} 
since $B_{1,3,4}  \propto \langle 34 \rangle$, $B_5 \propto 1/\langle 34 \rangle$, and $B_2 \rightarrow1$ because $B_1 + B_2 = 1$. The parameterization \eqref{eq:colllinearofs} can be used directly for the one-loop master integrals since they are free from $\mathrm{tr}_5$.
Similar to the discussion of the three-point form factor in Section~\ref{subsec:3ptff1loop},
the four-point form factor in the collinear limit satisfies
\begin{equation}
	\mathcal{I}^{(1), \text{M.T.}}_{\mathrm{tr}(F^3), 4} \xrightarrow{ p_3 || p_4 } \mathcal{I}_{\mathrm{tr}(F^3), 3}^{(1), \text{M.T.}}(1, 2, 3') +r_1^{[1], \text{MT}}(s_{34}, z) \,.
\end{equation}
where $\mathcal{I}_{\mathrm{tr}(F^3), 3}^{(1), \text{M.T.}}$ is the three-point minimal form factor, and $r_1^{[1], \text{MT}}$ is the splitting function given in \eqref{eq:splittingAmp}.
Together with \eqref{eq:ansatzFullF3} and \eqref{eq:colllinearofB}, one can find the following constraints on $\mathcal{G}_i$:
\begin{equation}
	\label{eq:4pt1loopCL}
	\mathcal{G}_2^{(1)} \rightarrow \mathcal{I}_{\mathrm{tr}(F^3), 3}^{(1), \text{M.T.}}(1, 2, 3') +r_1^{[1], \text{MT}}(s_{34}, z) \,, \qquad \mathcal{G}_5^{(1)} \rightarrow 0 \,.
\end{equation}
We point out that $\mathcal{G}_{a}^{(1)}$ can only have logarithmic divergence in the collinear limit, thus the terms $B_a \mathcal{G}_{a}^{(1)}$ for $a=1,3,4$ will vanish directly, leaving only constraints on $\mathcal{G}_2^{(1)}$ and $\mathcal{G}_5^{(1)}$.\footnote{Note that in principle $B_5 \mathcal{G}_5^{(1)}$ is allowed to have lower transcendental remnants in the collinear limit, which is irrelevant here since we focus only on maximally transcendental part; see also the discussion at the end of this subsection.} 
By using the constraints \eqref{eq:4pt1loopCL}, one can solve for 8 parameters and  the number of free parameters is reduced to 3.

Third, we consider the constraint from the cancellation of spurious poles \eqref{eq:SpuriousPoleFromTree}. The form factor should be finite, when $\langle 14 \rangle$, $\langle 24 \rangle$, and $\langle 34 \rangle$ approach 0, namely
\begin{equation}
	\mathcal{F}_{\mathrm{tr}(F^3), 4}^{(1)} \ \xlongrightarrow{\mbox{$\langle ij \rangle \rightarrow 0$}} \text{finite} \,, \qquad \langle ij \rangle \in \left\{ \langle 14 \rangle, \langle 24 \rangle, \langle 34 \rangle \right\} \,.
\end{equation}
Only $B_5$ has spurious pole $\langle 34 \rangle$ and this provides a constraint on $\mathcal{G}_5^{(1)}$, which is equivalent to the constraint of spurious pole $\langle 14 \rangle$ on $\mathcal{G}_4^{(1)}$ through the symmetry of exchanging external momentum $p_1$ and $p_3$. And $B_1$, $B_2$ and $B_3$ all contain spurious pole $\langle 24 \rangle$, and one should consider them together to cancel the spurious pole. It is convenient to rearrange the sum of the three terms as
\begin{equation}
	\label{eq:1loopnewform}
	\sum_{a=1}^3 B_a \mathcal{G}_a^{(1)} = \frac{1}{2} \left( \mathcal{G}_1^{(1)} + \mathcal{G}_2^{(1)} + 2\mathcal{G}_3^{(1)} \right) + \frac{B_1-B_2}{2} \left( \mathcal{G}_1^{(1)} - \mathcal{G}_2^{(1)} \right) - \left( \frac{B_1-B_2}{2} \right)^2 \mathcal{G}_3^{(1)} \, ,
\end{equation}
and the spurious pole $\langle 24 \rangle$ only appears in $(B_1-B_2)/2$, which constrains the second and third terms. 
To summarize, the spurious pole cancellation impose the constraints:
\begin{equation}
	\label{eq:4pt1loopSP}
	\mathcal{G}_5^{(1)} \, \xlongrightarrow{\mbox{$\langle 34 \rangle \rightarrow 0$}} 0 \,, \qquad \mathcal{G}_1^{(1)}-\mathcal{G}_2^{(1)} \xlongrightarrow{\mbox{$ \langle 24 \rangle \rightarrow 0$}} 0 \,, \qquad \ \mathcal{G}_3^{(1)}  \xlongrightarrow{\mbox{$ \langle 24 \rangle \rightarrow 0$}} 0 \,,
\end{equation}
in the formula, we will not consider the cancellation in higher order of $\langle ij \rangle$, the reason will be discussed later.
We point out that the limit $\langle ij \rangle \rightarrow 0$ can be naviely treated as $s_{ij} \rightarrow 0$, because $\mathrm{tr}_5$ doesn't appear in the one-loop master integrals we used, whose limit behavior needs special treatment.
After this constraint, only 1 parameter left.

Now the result that contains a free parameter $x_0$ and satisfies the above constraints can be summarized as follows
\begin{align}
	\label{eq:F3oneloop}
	& \mathcal{G}_1^{(1)} = G^{(1)} - \tilde{G}^{(1)}(4,1,2) \,, \qquad  \mathcal{G}_2^{(1)} = \mathcal{G}_1^{(1)} |_{\left( p_1 \leftrightarrow p_3 \right)} \,, \qquad \mathcal{G}_4^{(1)} = \mathcal{G}_5^{(1)} = 0 \,, \\
	& \mathcal{G}_3^{(1)} = x_0 \left( \tilde{G}^{(1)}(4,1,2)+\tilde{G}^{(1)}(2,3,4) \right) \,, \nonumber
\end{align}
where the building-blocks $G^{(1)}$, $\tilde{G}^{(1)}(i,j,k)$ are
\begin{align}
	& G^{(1)} = -I_{\text{Bub}}^{(1)}(1,2) - I_{\text{Bub}}^{(1)}(2,3) - I_{\text{Bub}}^{(1)}(3,4,1) - \frac{1}{2}I_{\text{Box}}^{(1)}(3,4,1) \, , \\
	& \tilde{G}^{(1)}(i,j,k) = I_{\text{Bub}}^{(1)}(i,j) + I_{\text{Bub}}^{(1)}(j,k) - I_{\text{Bub}}^{(1)}(i,j,k) - \frac{1}{2}I_{\text{Box}}^{(1)}(i,j,k) \,. \nonumber
\end{align}
In particular, the building-block $\tilde{G}^{(1)}(i,j,k)$ is not only IR finite but is also trivial in the collinear limits as
\begin{equation}
	\tilde{G}^{(1)}(i,j,k) \xrightarrow{ \langle ik \rangle \rightarrow 0 } \mathcal{O}(\langle ik \rangle) \,.
\end{equation}
Thus it is not constrained by the above limit.

\begin{figure}[t]
	\centering
	\includegraphics[scale=0.5]{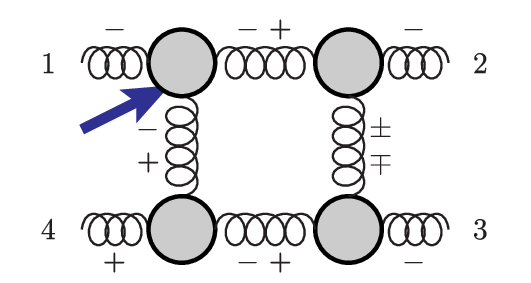}
	\caption{A simple quadruple cut to determine the coefficient of $I_{\text{Box}}^{(1)}(2,3,4)$.}
	\label{fig:fourcut}
\end{figure}

Finally, to fix the remaining single parameter, we use the unitarity method. We note that $x_0$ appears in the coefficient $-(B_2 + x_0 B_3)/2$ of $I_{\text{Box}}^{(1)}(2,3,4)$ in \eqref{eq:F3oneloop}. By performing a simple quadruple-cut as Figure~\ref{fig:fourcut}:
\begin{equation}
	\sum \mathcal{F}_{3}^{(0)} \mathcal{A}_{3}^{(0)} \mathcal{A}_{3}^{(0)} \mathcal{A}_{3}^{(0)} \longrightarrow - \mathcal{F}_4^{(0)} \frac{B_1}{2} I_{\text{Box}}^{(1)}(2,3,4) \big|_{\text{cut integrand}} \,,
\end{equation}
one obtains the box coefficient as $-(B_2-B_3)/2$ which fixes the parameter as $x_0 = -1$.

%%%%%%%%%%%%%%%%%%%%
%%%%%%%%%%%%%%%%%%%%

This completes the construction of the one-loop form factor. We summarize the above steps in Table~\ref{tab:solvingAnsatz1loop4ptF3}.
We point out that in the above construct, the same constraints apply to general gauge theories. In particular, the unitarity cut in Figure~\ref{fig:fourcut} only involves gluon states, thus it gives the same coefficients for any gauge theory containing a Yang-Mills sector. 
Therefore, the result we obtain applies to the form factor in general gauge theories, as long as one starts with the ansatz \eqref{eq:ansatzFullF3}.
As a cross-check, we have also performed an independent full unitarity-cut computation and found the same result as above using bootstrap.

%%%%%%%%%%%%%%%  TABLE   %%%%%%%%%%%%%%%%%%%%%
\begin{table}[t]
	\centering
	\vskip .1 cm 
	\begin{tabular}{| c | c |} 
		\hline
		Constraints								&	Parameters left  \\ \hline \hline
		Symmetry of $(p_1\leftrightarrow p_3)$	&	32  \\ \hline 
		IR   									&	11  \\ \hline
		Collinear limit							&	3  \\ \hline
		Spurious pole							&	1  \\ \hline
		Unitarity								&	0  \\ \hline
	\end{tabular} 
	\caption{Solving ansatz via constraints for the one-loop four-point form factor of $\mathrm{tr}(F^3)$.
		\label{tab:solvingAnsatz1loop4ptF3}
	}
\end{table}
%%%%%%%%%%%%%%%%%%%%%%%%%%%%%%%%%%%%%%%%%%

Finally, we make a comment about the term $(B_1-B_2)^2 \mathcal{G}_3^{(1)}/4$ in \eqref{eq:1loopnewform}. 
Since the  spurious pole in $(B_1-B_2)^2 \sim 1/\langle 24 \rangle^2$ is second-order, one may wonder if it could provide stronger constraint on $\mathcal{G}_3^{(1)}$ than \eqref{eq:4pt1loopSP}.
Indeed, if one considers the limit of $\mathcal{G}_3^{(1)}$ in the final result:
\begin{equation}
\label{eq:doublespuriouslimit}
	\mathcal{G}_3^{(1)} \ \xlongrightarrow{\mbox{$ \langle 24 \rangle \rightarrow 0$}} s_{24}\left( {1\over s_{12}}{\log \left( \frac{s_{14}}{s_{12}+s_{14}} \right)} +(p_1 \leftrightarrow p_3, p_2 \leftrightarrow p_4)\right) + \mathcal{O}(s_{24}^2) \, ,
\end{equation}
one find it is only of $\langle 24 \rangle$ order, 
which means that the term $(B_1-B_2)^2 \mathcal{G}_3^{(1)} \rightarrow \infty$ when $\langle 24 \rangle \rightarrow 0$. 
This seems to contradict the spurious pole cancellation condition.
However, one can see from the RHS of \eqref{eq:doublespuriouslimit} that, the divergent term is of transcendental degree one, thus its cancellation should involve the lower transcendental part of the one-loop form factor.
This will indeed impose new constraints on the lower transcendental part of the form factor, but it is in no contradiction with our above computation on the leading transcendental part.\footnote{This could also happen for the collinear limit. For example, in the collinear limit $p_3 \parallel p_4$, $B_5$ is divergent, and $B_5 \mathcal{G}_5^{(L)}$ is in principle allowed to be non-zero but only contributes to lower transcendental parts; though we find that $B_5 \mathcal{G}_5^{(L)}$ vanish in the limit up to two loops.}
It is a general feature that certain limit of a transcendental function may generate functions of lower transcendentality, for example, $\mathrm{Li}_2(1-\delta)$ has series expansion for $\delta \ll 1$
\begin{equation}
	\mathrm{Li}_2(1-\delta) = \zeta_2 + \sum_{n=1}^N  \left( \frac{\delta^n }{n} \log (\delta) - \frac{\delta^n }{n^2} \right) + {\cal O}(\delta^{N+1}) \, ,
\end{equation}
where $\zeta_2$ has degree 2, but the other terms (truncated to certain order ${\cal O}(\delta^N)$) have lower transcendental degrees. 
Such property can be used to impose constraints for lower transcendental parts.

%%%%%%%%%%%%%%%%%%%%%%%%%%
\subsection{Solving two-loop result}
\label{sec:solvingAnsatz}

We consider two-loop case in this subsection. We start with the ansatz  \eqref{eq:ansatzFullF3} and \eqref{eq:ansatz2loop}. 

To simplify the discussion of each step, we will first apply the constraints at the level of \emph{symbol} and then at the level of full functions.
The \emph{symbol} $\mathcal{S}$ of a function $T^{(k)}$ of transcendentality $k$ is represented in a tensor product form as \cite{Goncharov:2010jf}
\begin{equation}
\mathcal{S} (T^{(k)}) = \sum_{i_1,\ldots, i_k} R_{i_1} \otimes \cdots \otimes R_{i_k} \,,
\end{equation}
where $R_i$ are rational functions of kinematic variables. (Rational function has transcendentality degree 0, and by definition, its symbol is zero.)
It can be understood as a mathematical tool to simplify transcendental functions into tensor products of function arguments, for simple examples:
$\mathcal{S}(\log (x)) = x, \mathcal{S}(\mathrm{Li}_2(x)) = -(1-x) \otimes  x$.
For the problem we consider, the symbol of all two-loop masters have been given in \cite{Abreu:2020jxa}. After plugging them into our ansatz \eqref{eq:ansatz2loop}, we obtain an $\epsilon$-expansion expression of the form factor
\begin{equation}
\label{eq:ansatzSymbol}
	\mathcal{S} \left(\mathcal{F}_{\mathrm{tr}(F^3), 4}^{(2), \text{M.T.}} \right) = 
	\sum_{k\geq0} \epsilon^{k-4} \sum_I \alpha_I(c) \otimes_{i=1}^k w_{I_i} \,,
\end{equation}
where are $w_I$ are rational function of Mandelstam variables and are classified as symbol \emph{letters}. 
There are 46 independent letters and their definitions are given in Appendix~\ref{app:letters}.
Since the loop correction is uniformly transcendental, the tensor degree at given order in $\epsilon$-expansion is fixed, \emph{e.g.}~the finite order has degree $k=4$.
Moreover, the coefficient $\alpha_I(c)$ are linear combinations of $c_{a,i}$ in \eqref{eq:ansatz2loop}.

%%%%%%%%%%%%%%%%%%%%%%%%%%%%%%%%%%%%%%
\subsubsection{Constraints of IR, collinear and spurious poles}
\label{sec:bootstrapFF4pt2loop}

To impose the constraint of IR divergences and collinear limit, as reviewed in Section~\ref{sec:constraints}, one can consider the BDS function
\begin{equation}
\label{eq:badBDS}
	\tilde{\mathcal{I}}_{\mathrm{tr}(F^3), 4, \text{BDS}}^{(2), \text{M.T.}} = \frac{1}{2} \left(\mathcal{I}_{\mathrm{tr}(F^3), 4}^{(1), \text{M.T.}}(\epsilon)\right)^2+f(\epsilon) \mathcal{I}_{\mathrm{tr}(F^3), 4}^{(1), \text{M.T.}}(2\epsilon) \,.
\end{equation}
The one-loop result has been obtained in the previous subsection.
One complication is that the four-point form factor has multiple spinor factors $B_i$,
thus the one-loop square will introduce terms with new spinor factors such as $B_1^2$. 
Here a nice solution to avoid this complication is to introduce another BDS function which is linear in $B_a$:
\begin{equation}
\label{eq:goodBDS}
	\mathcal{I}_{\mathrm{tr}(F^3), 4, \text{BDS}}^{(2), \text{M.T.}} = \sum_{a=1}^2 \frac{1}{2} \mathcal{G}_a^{(1)} \left( B_a  \mathcal{G}_a^{(1)} + B_3 \mathcal{G}_3^{(1)} \right) + f(\epsilon) \sum_{a=1}^2 B_a\mathcal{G}_a^{(1)}(2\epsilon) \, .
\end{equation}
Similar form was used for the four-point form factor of $\mathrm{tr}(\phi^3)$ \cite{Guo:2021bym}.
One can prove that the new BDS function \eqref{eq:goodBDS} has same infrared part and same collinear limit behavior as \eqref{eq:badBDS} by computing the difference between the two functions:
\begin{equation}
	\tilde{\mathcal{I}}_{\mathrm{tr}(F^3), 4, \text{BDS}}^{(2), \text{M.T.}} - \mathcal{I}_{\mathrm{tr}(F^3), 4, \text{BDS}}^{(2), \text{M.T.}} = -B_3 \left( \mathcal{G}_1^{(1)} - \mathcal{G}_2^{(1)} - B_1 \mathcal{G}_3^{(1)} \right) \left( \mathcal{G}_1^{(1)} - \mathcal{G}_2^{(1)} + B_2 \mathcal{G}_3^{(1)} \right) \,.
	\label{eq:diffBDS4pt}
\end{equation}
Since $\mathcal{G}_1^{(1)} \big|_{\text{div.}} = \mathcal{G}_2^{(1)} \big|_{\text{div.}}$ and $\mathcal{G}_3^{(1)} \big|_{\text{div.}} = 0$, the difference is IR finite. Moreover, the above formula will vanish in the collinear limits $p_1 \parallel p_4$ or $p_3 \parallel p_4$, because the factor $B_3$ vanishes in the limit.\footnote{$\mathcal{G}_{a}^{(1)}$ have only logarithmic divergence, thus the difference \eqref{eq:diffBDS4pt} still goes to zero in the collinear limits.}

%%%%%%%%%%%%%%%%%%%%%
\paragraph{Constraints at the symbol level.}

Using the BDS function \eqref{eq:goodBDS} and one-loop result, one obtains the two-loop divergent terms at orders $1/\epsilon^m, m=4,3,2,1$. By matching their symbol with our ansatz \eqref{eq:ansatzSymbol}:
\begin{equation}
	\mathcal{S} \Big(\mathcal{I}_{\mathrm{tr}(F^3), 4, \text{BDS}}^{(2), \text{M.T.}} \Big) \Big|_\text{div.} = {\cal S} \Big( \mathcal{I}_{\mathrm{tr}(F^3), 4}^{(2), \text{M.T.}} \Big) \Big|_\text{div.} \,,
\end{equation}
we can solve for 353 parameters and the remaining degree of freedom is 207.

Next, the collinear limits for the finite remainder function $\mathcal{R}_{\mathrm{tr}(F^3), 4}^{(2), \text{M.T.}}$ should match the two-loop three-point remainder for the maximum transcendence part as
\begin{equation}
	\label{eq:4ptremainderCL}
	\mathcal{R}_{\mathrm{tr}(F^3), 4}^{(2), \text{M.T.}} = 
	 \Big(\mathcal{I}_{\mathrm{tr}(F^3), 4}^{(2), \text{M.T.}} - \mathcal{I}_{\mathrm{tr}(F^3), 4, \text{BDS}}^{(2), \text{M.T.}} \Big)_\text{fin.}
	  \ \xlongrightarrow[\mbox{or {$p_4 \parallel p_1$}}]{\mbox{$p_4 \parallel p_3$}} \ \mathcal{R}_{\mathrm{tr}(F^3), 3}^{(2), \text{M.T.}} \, .
\end{equation}
The formula should hold at Symbol level, same as one-loop case, we have
\begin{equation}
	\label{eq:4pt2loopCL}
	\mathcal{S} \left(\mathcal{G}_2^{(2)} \right) \ \xlongrightarrow{\mbox{$p_4 \parallel p_3$}} \ \mathcal{S} \left({\cal R}_{\mathrm{tr}(F^3), 3}^{(2), \text{M.T.}}\right) \, , \qquad \mathcal{S} \left(\mathcal{G}_5^{(2)}\right) \ \xlongrightarrow{\mbox{$p_4 \parallel p_3$}} 0 \, .
\end{equation}
The collinear limits of the symbol letters are given explicitly in Appendix~\ref{app:letterCL}. After this step, the number of free parameters reduces to 119.

Furthermore, the spurious poles should be eliminated similar to the one-loop formula \eqref{eq:4pt1loopSP}, and this provides the following constraints on the two-loop correction functions:
\begin{equation}
	\label{eq:4pt2loopSP}
	\mathcal{G}_5^{(2)} \, \xlongrightarrow{\mbox{$\langle 34 \rangle \rightarrow 0$}} 0 \,, \qquad \mathcal{G}_1^{(2)}-\mathcal{G}_2^{(2)} \xlongrightarrow{\mbox{$ \langle 24 \rangle \rightarrow 0$}} 0 \,, \qquad \ \mathcal{G}_3^{(2)}  \xlongrightarrow{\mbox{$ \langle 24 \rangle \rightarrow 0$}} 0 \,.
\end{equation}
As a technical point, we mention that the variable $\mathrm{tr}_5$, defined in \eqref{eq:tr5}, occurs in the two-loop master integrals. Its limit can be taken in the following way:
\begin{equation}
	\mathrm{tr}_5 = s_{14}s_{23} - s_{12}s_{34} + s_{24} s_{13} - 2 \langle 24 \rangle \left[ 41 \right] \langle 13 \rangle \left[ 32 \right] \ \xlongrightarrow{\mbox{$ \langle 24 \rangle \rightarrow 0$}} s_{14} s_{23} - s_{12} s_{34} \,,
\end{equation}
where a proper definition on the LHS is used.
Since the spurious pole must cancel to all order in the $\epsilon$ expansion of the form factor, we also consider its cancellation at $\mathcal{O}(\epsilon)$ order. We find that the $\epsilon^1$-order also provide useful new constraints. They will constrain the coefficients of $I_{\text{TP}}^{(2)}$ and $I_{\text{BPb}}^{(2)}$ and $I_{\text{dBox2c}}^{(2)}$, which are all $\mu$-term master integrals. (Recall that the integral $I_{\text{dBox2c}}^{(2)}$ has contribution only starting from $\epsilon$-order.)
We also find that once the cancellation of spurious poles is satisfied at $\epsilon^1$-order, it will hold for any order of $\epsilon$; further details are given in Appendix~\ref{app:HigherOrder}. 
The condition of spurious-pole cancellations can solve for 66 parameters, and the remaining freedom of degree is 53.

%%%%%%%%%%%%%%%%%%%%%%
\paragraph{Constraints at the function level.}

Since the Symbol does not concern the terms that contain transcendental numbers such as $\pi, \zeta_n$, possible constraints may not be captured by using the symbol alone. Thus we need to consider the full functional form of the master integrals, which have been computed in \cite{Canko:2020ylt}.
Since we only need to fix the coefficients, it is convenient to do numerical computation with high enough precision. Details of performing numerics will be discussed in Appendix~\ref{app:fullFF}, and here we focus on the solution to the constraints.
By repeating the above steps at the function level, the remaining degrees of freedom can be reduced to
40 (by IR), 24 (by collinear limits), and 20 (by the cancellation of spurious poles).\footnote{We mention that the numerical collinear limit can be taken with the parameterization \eqref{eq:colllinearofs}.}

We summarize the constraints and corresponding fixed parameters in Table~\ref{tab:solvingAnsatz2loop4ptF3}.
We point out that two of 20 degrees of freedom only change results at $\mathcal{O}(\epsilon)$ order which will be explained in the next subsection.
All remaining parameters can be fixed by simple unitarity cuts as discussed later in Section~\ref{sec:unitarity4pt}.

%%%%%%%%%%%%%%%  TABLE   %%%%%%%%%%%%%%%%%%%%%
\begin{table}[t]
	\centering
	\vskip .1 cm 
	\begin{tabular}{| l | c |} 
		\hline
		Constraints								&  Parameters left	\\ \hline \hline
		Starting ansatz							&  1105				\\ \hline
		Symmetry of $(p_1\leftrightarrow p_3)$	&  560				\\ \hline
		IR (Symbol)    							&  207				\\ \hline
		Collinear limit (Symbol)				&  119				\\ \hline
		Spurious pole (Symbol)					&  53				\\ \hline
		IR (Function)							&  40				\\ \hline
		Collinear limit (Funcion)				&  24				\\ \hline
		Spurious pole (Funcion)					&  20				\\ \hline
		Simple unitarity cuts	   				&  0				\\ \hline
	\end{tabular} 
	\caption{Solving for parameters via constraints.
		\label{tab:solvingAnsatz2loop4ptF3}
	}
\end{table}
%%%%%%%%%%%%%%%%%%%%%%%%%%%%%%%%%%%%%%%%%%

%%%%%%%%%%%%%%%%%%%%%%%%%%%%%%%%%
\subsubsection{Building blocks for the remaining parameters}
\label{sec:buildingblocksFF4pt}

Before applying further unitarity-cut constraints, it is instructive to first analyze the remaining degrees of freedom.

The terms depending on the remaining free parameters can be organized into three groups:
\begin{align}
	\label{eq:BuildingBlock}
	\left( B_1+B_2 \right) \tilde{G}_{1,\alpha}^{(2)} \,,  \qquad & \alpha = 1, \ldots, 8 \,, \\ 
	B_3 \tilde{G}_{2,\beta}^{(2)} \,, \qquad & \beta = 1, \ldots, 7\,, \nonumber \\
	B_4 \tilde{G}_{3,\gamma}^{(2)} + \left( p_1\leftrightarrow p_3 \right) \,, \qquad & \gamma = 1, \ldots, 5\,,\nonumber
\end{align}
where $\tilde{G}_{1,\alpha}^{(2)}$ are
\begin{align}
	& \tilde{G}_{1,1}^{(2)} = I_{\text{dBox2c}}^{(2)}(1,2,3,4) + I_{\text{dBox2c}}^{(2)}(3,2,1,4) \,, \qquad\quad\qquad\quad \tilde{G}_{1,2}^{(2)} = \tilde{G}_{1,1}^{(2)} |_{\left( p_2 \leftrightarrow p_4 \right)} \,, \nonumber \\
	& \tilde{G}_{1,3} = I_{\text{BPb}}^{(2)}(1,2,3,4)- I_{\text{BPb}}^{(2)}(4,3,2,1) + \left(p_1 \leftrightarrow p_3 \right) \,, \qquad \  \tilde{G}_{1,4}^{(2)} = \tilde{G}_{1,3}^{(2)} |_{\left( p_2 \leftrightarrow p_4 \right)}\,, \nonumber \\
	& \tilde{G}_{1,5} = I_{\text{TP}}^{(2)}(1,2,3,4) + I_{\text{TP}}^{(2)}(3,2,1,4) \, , \nonumber\\ 
	& \tilde{G}_{1,6}^{(2)} = \tilde{G}_{1,5} |_{\left( p_i \rightarrow p_{i+1} \right)} \,,  \qquad\quad
	 \tilde{G}_{1,7}^{(2)} = \tilde{G}_{1,5} |_{\left( p_i \rightarrow p_{i+2} \right)} \,,  \qquad\quad
	 \tilde{G}_{1,8}^{(2)} = \tilde{G}_{1,5} |_{\left( p_i \rightarrow p_{i+3} \right)} \,, 
\end{align}
and expressions for $\tilde{G}_{2,\beta}^{(2)}$ and $\tilde{G}_{3,\gamma}^{(2)}$ are a little lengthy and we give them in Appendix~\ref{app:BuildingBlocks}. 
The eight functions $\tilde{G}_{1,\alpha}^{(2)}$ in the first group functions are a little special, because they are combinations of three special class of UT integrals:
\begin{equation}
\label{eq:3mutermintegrals}
I_{\text{TP}}^{(2)} \,,  \qquad I_{\text{BPb}}^{(2)}\,, \qquad I_{\text{dBox2c}}^{(2)} \,,
\end{equation} 
of which the numerators are proportional to $\mu_{ij} = l_i^{-2\epsilon} \cdot l_j^{-2\epsilon}$, see Appendix~\ref{app:UT}.
It will be necessary to apply $D$-dimensional unitarity cuts to determine their coefficients. 
On the other hand, the other two groups $\tilde{G}_{2,\beta}^{(2)}$ and $\tilde{G}_{3,\gamma}^{(2)}$ are free of these $\mu$-term integrals.
We also point out that $\tilde{G}_{1,1}^{(2)}$ and $\tilde{G}_{1,2}^{(2)}$ contain only $I_{\text{dBox2c}}^{(2)}$ which is of $\mathcal{O}(\epsilon)$ order, thus they are irrelevant if one is only interested in getting the $\epsilon^0$ order results of the form factor.

All the above functions are free from infrared divergences, and they also satisfy the following collinear behavior:
\begin{equation}
	\tilde{G}_{1,\alpha}^{(2)} \xlongrightarrow{\mbox{$ p_i \parallel p_j $}} 0 \,, \qquad \tilde{G}_{2,\beta}^{(2)} \xlongrightarrow{\mbox{$ \langle 24 \rangle \rightarrow 0 $}} 0 \,, \qquad \tilde{G}_{3,\gamma}^{(2)} \xlongrightarrow{\mbox{$ \langle 14 \rangle \rightarrow 0 $}} 0 \,,
\end{equation}
where $p_i \parallel p_j$ means the collinear limits of any pair of momentum.

We can understand why they are not constrained in the above procedure as follows:
\begin{itemize}
	\item $\tilde{G}_{1,\alpha}^{(2)}$ free from all spurious poles because $B_1+B_2 = 1$.
	\item $\tilde{G}_{2,\beta}^{(2)}$ free from the collinear limits $p_3 \parallel p_4$ and $p_1 \parallel p_4$ because they both give $B_3 \rightarrow 0$.
	\item $\tilde{G}_{3,\gamma}^{(2)}$ vanish in the colliear limit $p_1 \parallel p_4$(meanwhile $B_4$ turns to infinity), which is covered by the requirement of spurious pole $\langle 14 \rangle \rightarrow 0$.
\end{itemize}

By analyzing the $\tilde{G}^{(2)}$ functions in \eqref{eq:BuildingBlock}, we find that the remaining 20 parameters can be fix by the coefficients of the master integrals 
\begin{equation}
	\label{eq:2loop4pointmaster}
	I_\text{BubBox}^{(2)} \,, \quad I_\text{TBox0}^{(2)} \,, \quad I_\text{dBox1a}^{(2)} \,, \quad I_\text{BPb}^{(2)} \,, \quad I_\text{dBox2c}^{(2)} \,, \quad I_\text{TP}^{(2)} \,.
\end{equation}
Each above master integral has 8 different external-particle orders $(i,j,k,l)$, thus there are 48 coefficients but only 20 of them are independent.

We would like to stress that the constraints that we apply here, including the infrared structures, the collinear limits, and the spurious-pole cancellations, are the same for general gauge theories. Therefore, form factors of different gauge theories are different only by the value for the remaining 20 parameters.

%%%%%%%%%%%%%%%%%%%%%%%%%%%%%%%%%%%%%%%%%%%
\subsubsection{Pure-YM result via $D$-dimensional unitarity cut}
\label{sec:unitarity4pt}

In this subsection, we apply unitarity cuts to fix the remaining free parameters. 
We will use the $D$-dimensional unitarity-IBP method to determine the form factor in pure YM theory.
Form factors in other theories will be discussed in the next subsection.

%%%%%%%%%%%%%%%%%%%%%%%%%%%
\begin{figure}[t]
	\centering
	\includegraphics[scale=0.6]{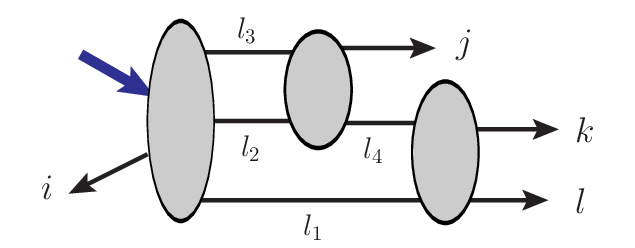}
	\caption{Unitarity cuts that can determine the remaining degrees of freedom.}
	\label{fig:unitaritycut}
\end{figure}
%%%%%%%%%%%%%%%%%%%%%%%%%%%

The remaining degrees of freedom are the coefficients of the master integrals \eqref{eq:2loop4pointmaster}. To determine them, we find it is sufficient to consider only one type of unitarity cuts shown in Figure~\ref{fig:unitaritycut}. 
Since \eqref{eq:2loop4pointmaster} involve $\mu$-term integrals $\{I_{\text{TP}}^{(2)}, I_{\text{BPb}}^{(2)}, I_{\text{dBox2c}}^{(2)}\}$ whose numerators depends on $\mu_{ij} = l_i^{-2\epsilon} \cdot l_j^{-2\epsilon}$. To determine their coefficients, it is not enough to consider four-dimensional cuts. 
We will employ the $D$-dimensional unitarity cut based on the strategy of unitarity-IBP that has been applied to multi-loop computation of form factors (and Higgs amplitudes) in \cite{Jin:2018fak, Jin:2019ile, Jin:2019opr, Jin:2020pwh} and pure gluon amplitudes in \cite{Boels:2018nrr, Jin:2019nya}.\footnote{Similar strategy has also been used in the numerical unitarity approach \cite{Abreu:2017xsl, Abreu:2017hqn}, 
and idea of applying cuts to simplify IBP has been used in \emph{e.g.}~\cite{Kosower:2011ty, Larsen:2015ped, Ita:2015tya, Georgoudis:2016wff}.}
The strategy can be outlined as follows:
\begin{equation}
\mathcal{F}_4^{(2)}\Bigr|_{\text{cut}} 
= \sum_{\rm hel.} {\cal F}_4^{(0)} {\cal A}^{(0)}_4 {\cal A}^{(0)}_4  
\xrightarrow[\text{gauge-invariant basis}]{\text{projection at}}  \sum_{\alpha} {f}^{\alpha}\mathbf{B}_{\alpha}
\xrightarrow{\ \text{cut-IBP}\ } \sum_{\text{cut permitted } I_i} c_i I^{(2)}_i \ .
\end{equation}
Below we explain each step in more detail.

First,  the product of the tree amplitudes under $D$-dimensional cuts provides the cut integrands. We compute the tree form factor by Feynman rules and the perform the helicity sum by using 
\begin{align}
\label{eq:polarsum}
\varepsilon^\mu(l_i) \circ \varepsilon^\nu(l_i) \equiv \sum_{\rm hel.}
\varepsilon^\mu(l_i) \varepsilon^\nu(l_i)
=\eta^{\mu\nu}-\frac{q^\mu_i l_i^\nu+q_i^\nu l_i^\mu}{q_i\cdot l_i},
\quad i=1,..,4 \,,
\end{align}
where $q_i^\mu$ are arbitrary light-like reference momenta.
In this way, we obtain the cut integrands which are tensor integrals containing inner products of polarization vectors and loop momenta.

Next, we perform integrand reduction and obtain scalar integrals by projecting the integrand to a set of gauge-invariant basis, as introduced in \emph{e.g.}~\cite{Boels:2018nrr}. 
Following \cite{Boels:2018nrr}, the basis of $n$-gluon form factors and scattering amplitudes can be constructed using gauge invariant building blocks $\mathbf{A}_i^{jk}$ and $\mathbf{D}_{ij}$:
\begin{equation}
	\mathbf{A}_i^{jk} = \delta^{\varepsilon_i p_i}_{p_j p_k} \,, \qquad \mathbf{D}_{ij} = \varepsilon^{\perp}_i \cdot \varepsilon^{\perp}_j=\frac{\delta^{\varepsilon_i p_1 p_2 p_3 p_4}_{\varepsilon_j p_1 p_2 p_3 p_4}}{\delta^{p_1 p_2 p_3 p_4}_{p_1 p_2 p_3 p_4}} \,, 
\end{equation}
in which $\varepsilon_i^{\perp}$ is the component of $\varepsilon_i$ which is perpendicular to all $p_j$, and
\begin{equation}
	\delta^{a_1 \cdots a_n}_{b_1 \cdots b_n} = \det \left( a_i\cdot b_j \right)_{n\times n} \, .
\end{equation}
For the four-gluon form factor under consideration, one can choose the following set of $\mathbf{A}$ and $\mathbf{D}$:
\begin{align}
	& \mathbf{A}_1^{23} \,, \ \mathbf{A}_1^{24} \,, \ \mathbf{A}_2^{34} \,, \ \mathbf{A}_2^{31} \,, \ \mathbf{A}_3^{41} \,, \ \mathbf{A}_3^{42} \,, \ \mathbf{A}_4^{12} \,, \ \mathbf{A}_4^{13} \,, \ \nonumber \\
	& \mathbf{D}_{12} \,, \ \mathbf{D}_{13} \,, \ \mathbf{D}_{14} \,, \ \mathbf{D}_{23} \,, \ \mathbf{D}_{24} \,, \ \mathbf{D}_{34} \,.
\end{align}
Each gauge invariant basis should contain all four polarization vectors $\varepsilon_i$ and also depend on them linearly. In total, there are 43 elements in the set of gauge invariant basis, including: 
16 $(\mathbf{A})^4$-type, 24 $(\mathbf{A})^2\mathbf{D}$-type, and 3 $(\mathbf{D})^2$-type ones. 
Since 43 is not a small number, it would be non-trivial to project the integrand on this set of basis.
Fortunately, there is an important simplification for our problem. The $\mathbf{D}$-terms vanish in four-dimension, so $(\mathbf{A})^2\mathbf{D}$ and $(\mathbf{D})^2$-type basis can be neglected in the HV scheme \cite{tHooft:1972tcz}, which is enough for our consideration of the maximally transcendental part.\footnote{Note that all internal momenta, as well as the helicity sum for the cut gluon states, are in $D$ dimensions.} Therefore we are left with a simpler gauge-invariant basis with only 16 $\mathbf{A}^4$-type elements, which will be denoted by $\mathbf{B}_{\alpha}$.

We would like to project the cut integrand in this set of basis as
\begin{equation}
\mathcal{F}_4^{(2)}\Bigr|_{\text{cut}} =  \sum_{\alpha} {f}^{\alpha}(l_i, p_j)\mathbf{B}_{\alpha}\,,
\end{equation}
where only the $\mathbf{B}_{\beta}$ term contains polarization vectors. 
This can computed by performing the following contraction with the gauge invariant basis as (see also \cite{Jin:2019opr})
\begin{equation}
	\label{contract-gib}
f^\alpha= \left( \mathcal{F}_{\text{cut}} \circ \mathbf{B}_{\alpha} \right) \mathbf{G}^{\alpha\beta} \, , 
\end{equation}
in which $\mathbf{G}_{\alpha\beta} = \mathbf{B}_{\alpha} \circ \mathbf{B}_{\beta}$, and $\mathbf{G}^{\alpha\beta}$ is the inverse of $\mathbf{G}_{\alpha\beta}$.
The `$\circ$' product is defined as in \eqref{eq:polarsum}, and here the helicity sum is for the four external polarization vectors $\varepsilon_i$. 
A technical challenge here is that the $16\times 16$ matrix $\mathbf{G}^{\alpha\beta}$ is still quite complicated, and it is the main obstacle to performing an analytical evaluation; therefore in this step, we have carried out the gauge-invariant basis contraction numerically.

The basis coefficients $f^\alpha$ are functions of Lorentz product of momenta, and thus it can be directly reduced using IBP reduction \cite{Chetyrkin:1981qh, Tkachov:1981wb}, with \emph{e.g.}~public codes \cite{Smirnov:2019qkx, Klappert:2020nbg}. 
In this way, we obtain the coefficients $c_i^\alpha$ of master integrals $I_i$ that contain four cut propagators as shown in Figure~\ref{fig:unitaritycut}.

So far we have not specified the helicities of external gluons. The polarization vectors can be set to the $\pm$ helicities by the following rules for the basis: 
\begin{equation}
	\mathbf{A}_i^{jk}\bigr|_{\epsilon_i\rightarrow \epsilon_i^+}\rightarrow [i|j|k|i] \, , \qquad \mathbf{A}_i^{jk}\bigr|_{\epsilon_i\rightarrow \epsilon_i^-}\rightarrow \langle i|j|k|i\rangle \, .
\end{equation}
After being divided by the tree-level form factor $\mathcal{F}_4^{(0)}$, $\frac{\mathbf{B}_{\alpha}}{\mathcal{F}_4^{(0)}}\bigr|_{\text{helicity}}$ can be rewritten in terms of $s_{ij}$ and $\mathrm{tr}_5$. For example
\begin{equation}
	{\mathbf{A}_1^{23} \mathbf{A}_2^{34} \mathbf{A}_3^{41} \mathbf{A}_4^{12} / \mathcal{F}_4^{(0)}} \bigr|_{(+---)} = -\frac{s_{12}s_{13}s_{14}}{2} \left( s_{14}s_{23}-s_{13}s_{24}+s_{12}s_{34} - \mathrm{tr}_5 \right).
\end{equation}
We can now reconstructed the analytical expression of the maximal transcendentality part of the master integral coefficients.
The coefficients are generally non-trivial rational functions of kinematics and dimensional regularization parameter $\epsilon$. For the four-gluon form factor at hand, there are a lot of kinematic variables, and in general, it can be difficult to reconstruct their complete form.
Since the focus of this paper is the maximal transcendentality part, the job is significantly simplified: using the simple ansatz \eqref{eq:ansatzFullF3}, the kinematic structure is known in advance, and there are 5 numerical parameters to determine which can be fixed by using only 5 numerical points.
We find the numerical results match perfectly with the ansatz form \eqref{eq:ansatzFullF3}, and the coefficients are also all small rational numbers. In this way, we fix the coefficient of all coefficients for the master integrals \eqref{eq:2loop4pointmaster}. 
The final solutions of master coefficients are provided in the auxiliary files.

We mention that besides the master integrals \eqref{eq:2loop4pointmaster}, the cut will also determine the following other master integrals: 
\begin{align}
& I_\text{BPa}^{(2)} \,, \ I_\text{BubBox2}^{(2)} \,, \ I_\text{dBox1b}^{(2)} \,, \ I_\text{dBox2a}^{(2)} \,, \ I_\text{dBox2b}^{(2)} \,, \ I_\text{TBub2}^{(2)} \,, \\ 
& I_\text{TT0}^{(2)} \,, \ I_\text{TT3a}^{(2)} \,, \ I_\text{TT3b}^{(2)}\,, \ I_\text{TBox1}^{(2)} \,, \ I_\text{TBox2a}^{(2)} \,, \ I_\text{TBox2b}^{(2)} \,. \nonumber
\end{align} 
Their coefficients are related to the coefficients of other integrals by the previous bootstrap constraints. We find that the unitarity-cut results are fully consistent with the bootstrap computation. This provides a strong consistency check for our results.

%%%%%%%%%%%%%%%%%%%%%%%%%%%%%%%
\subsection{Equivalence between $\mathcal{N}=4$ SYM and QCD}
\label{sec:MTPtrF3}

In this subsection, we will first consider the difference of the form factors between different gauge theories. We find that for the maximally transcendental parts of the form factors, the difference  between any gauge theory and the pure YM theory will only depend on two free parameters, where the key idea is to apply some universal unitarity cuts. 
Next, we obtain the form factor result in $\mathcal{N}=4$ SYM by computing the difference with a simple four-dimensional unitarity cut. We will show that the difference comes from only the contribution that contains a fermion loop.
Finally, we compare the form factors between $\mathcal{N}=4$ SYM and QCD and show that the maximal transcendentality principle still holds.

%%%%%%%%%%%%%%%%%%%%%%%%%%%%%
\subsubsection{Difference between different theories}

As already mentioned at the end of Section~\ref{sec:bootstrapFF4pt2loop}, different gauge theories can only be different by the value for the remaining 20 parameters. Let us denote a general gauge theory that contains a YM sector as ``Theory-X''. We define the difference between the form factor results of Theory-X and pure YM theory as $\Delta^{(2), \text{Theory-X}}_{\text{M.T.}}$:
\begin{equation}
	\Delta^{(2), \text{Theory-X}}_{\text{M.T.}} = \left( \mathcal{I}_{\mathrm{tr}(F^3),4}^{(2), \text{M.T.}} \right) \Big|_{\text{Theory-X}} - \left( \mathcal{I}_{\mathrm{tr}(F^3),4}^{(2), \text{M.T.}} \right) \Big|_{\text{pure YM}} \, .
\end{equation}
It should be clear that $\Delta^{(2), \text{Theory-X}}_{\text{M.T.}}$ can only come from the contribution of particles other than gluons.

\begin{figure}[t]
	\centering
	\subfloat[]{\includegraphics[scale=0.45]{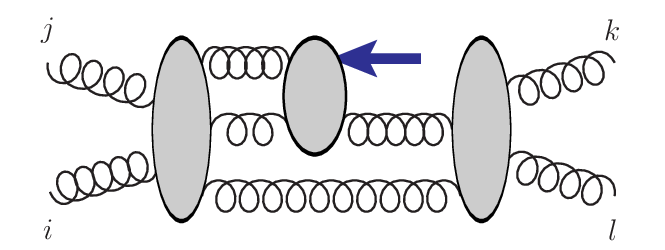}}
	\subfloat[]{\includegraphics[scale=0.45]{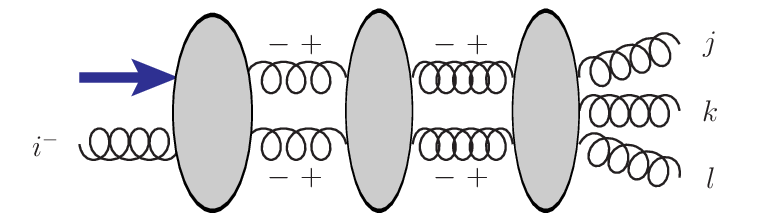}}
	
	\subfloat[]{\includegraphics[scale=0.45]{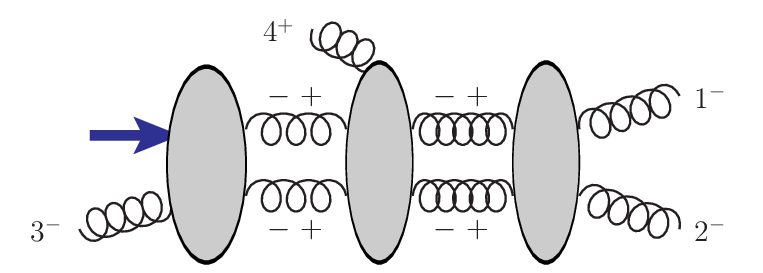}}
	\subfloat[]{\includegraphics[scale=0.45]{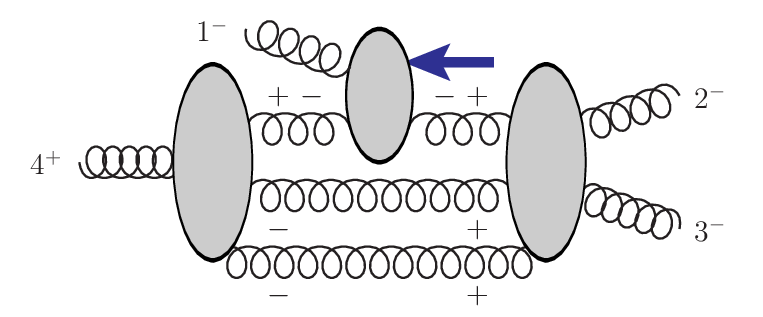}}
	\caption{The unitarity cuts which are same for general gauge theories.}
	\label{fig:samecuts2loop4point}
\end{figure}

Rather than naively expressing $\Delta^{(2)}_{\text{M.T.}}$ as a linear combination of 20 building-blocks listed in \eqref{eq:BuildingBlock}, 
one can show that $\Delta^{(2)}_{\text{M.T.}}$ has a much more compact form that has only two free parameters.
The idea is to consider some universal unitarity cuts that are theory-independent. 
To be concrete, we consider four types of unitarity cuts shown in Figure~\ref{fig:samecuts2loop4point}. These cuts are special in the sense that they can only allow internal gluon configuration, therefore, the coefficients of the master integrals which can be detected by these cuts must be the same for any gauge theory that contains a Yang-Mills sector. The master integrals detected by these cuts are listed below:
\begin{align}
\textrm{cut-(a)}: \quad & I_\text{dBox2a}^{(2)},\ I_\text{dBox2b}^{(2)},\ I_\text{dBox2c}^{(2)},\ I_\text{BPa}^{(2)},\ I_\text{BPb}^{(2)}, \ I_\text{TP}^{(2)} \ \ \textrm{for all orderings of external particles.} \nonumber\\
\textrm{cut-(b)}: \quad& I_\text{BubBox0}^{(2)} \ \ \textrm{for all orderings of external particles.} \nonumber\\
\textrm{cut-(c)}: \quad& I_\text{dBub}^{(2)}(1,2;1,2,4),\ I_\text{TT0}^{(2)}(4,1,2),\ I_\text{TBox0}^{(2)}(2,3,4).\nonumber \\
\textrm{cut-(d)}: \quad& I_\text{TT1a}^{(2)}(4,1,2),\ I_\text{dBox1a}^{(2)}(2,3,4),\ I_\text{TBox0}^{(2)}(4,1,2).\nonumber
\end{align}

%%%%%%%%%%%%%%%%%%%%%%%%%%%
%%%%%%%%%%%%%%%%%%%%%%%%%%%

Since these master integrals are determined by the pure YM theory, they cannot occur in $\Delta^{(2)}_{\text{M.T.}}$. By inspecting their relation with the remaining building blocks in \eqref{eq:BuildingBlock}, we find  that $\Delta^{(2)}_{\text{M.T.}}$ only depends on two functions of $\tilde{G}_{3,\gamma}$ and can be given as
\begin{equation}
	\label{eq:difference}
	\Delta^{(2)}_{\text{M.T.}} = B_4 \left( x_1 \tilde{G}_{3,1}^{(2)} + x_2 \tilde{G}_{3,3}^{(2)} \right) + (p_1 \leftrightarrow p_3) \,.
\end{equation}
The two free parameters $x_{1,2}$ can be fixed by the coefficients of $I_{\text{TBox0}}^{(2)}(1, 4, 3)$ and $I_{\text{dBox1a}}^{(2)}(1, 4, 3)$, which have coefficients $B_5( -\frac{3}{4} x_1 + x_2)$ and $\frac{1}{4} B_5 x_1$ respectively in the above formula. 
Since they are free from $\mu$-terms, it is enough to apply a four-dimensional unitarity cut to fix them.

%%%%%%%%%%%%%%%%%%%%%%%%%%%%%%%%%%%%%%%%%%%%%
\subsubsection{Two-loop result of $\mathcal{N}=4$ SYM}

%%%%%%%%%%%%%%%%%%%%%%%%%%%
\begin{figure}[t]
	\centering
	\includegraphics[scale=0.5]{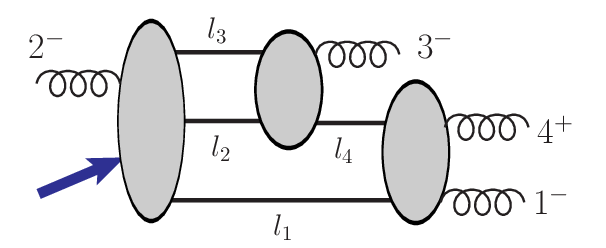}
	\caption{The unitarity cut to determine  $\Delta^{(2), \mathcal{N}=4}_{\text{M.T.}}$.}
	\label{fig:unitaritycut2}
\end{figure}
%%%%%%%%%%%%%%%%%%%%%%%%%%%

Now we can consider the form factor result in $\mathcal{N}=4$ super Yang-Mills. 
Since we have already obtained the pure YM result, it will be enough to compute $\Delta^{(2), \mathcal{N}=4}_{\text{M.T.}}$ given in \eqref{eq:difference}. 
As we discussed before, we will consider a four-dimensional cut as in Figure~\ref{fig:unitaritycut} with the ordering of external particles chosen as $(ijkl)=(2341)$,  which is given in Figure~\ref{fig:unitaritycut2}. This cut can fix the coefficients of $I_{\text{TBox0}}^{(2)}(1,4,3)$ and $I_{\text{dBox1a}}^{(2)}(1,4,3)$.

The cut integrand corresponds to the tree product of a four-point tree form factor and two four-point tree amplitudes:
\begin{align}
	\int \prod_{i=1}^4 d\eta_{l_i}^4 \mathcal{F}_{\mathrm{tr}(F^3), 4}^{(0), \text{MHV}}(2^-, l_3, l_2, l_1) \mathcal{A}_4^{(0), \text{MHV}}(3^-, l_4, -l_2, -l_3) \mathcal{A}_4^{(0), \text{MHV}} (4^+, 1^-, -l_1, -l_4) \,.
\end{align}
To determine $\Delta^{(2), \mathcal{N}=4}_{\text{M.T.}}$,  it is enough to consider configurations that are different from pure YM theory.
This means that we only need to consider the internal particles that contain scalars or fermions, which are
\begin{align}
	& \mathcal{F}_{\mathrm{tr}(F^3), 4}^{(0)}(2^-, l_3^{\psi/\bar{\psi} /\phi}, l_2^{\bar{\psi}/\psi /\bar{\phi}}, l_1^{-}) \mathcal{A}_4^{(0)}(3^-, l_4^{+}, -l_2^{\psi/\bar{\psi} /\phi}, -l_3^{\bar{\psi}/\psi/\bar{\phi}}) \mathcal{A}_4^{(0)}(4^+,1^-,-l_1^{+},-l_4^{-}) \,, \\
	& \mathcal{F}_{\mathrm{tr}(F^3), 4}^{(0)}(2^-, l_3^-, l_2^{\psi/\bar{\psi} /\phi}, l_1^{\bar{\psi}/\psi/\bar{\phi}}) \mathcal{A}_4^{(0)}(3^-, l_4^{\psi/\bar{\psi}/\phi}, -l_2^{\bar{\psi}/\psi/\bar{\phi}}, -l_3^+) \mathcal{A}_4^{(0)}(4^+,1^-,-l_1^{\psi/ \bar{\psi}/\phi},-l_4^{\bar{\psi}/\psi/\bar{\phi}})\,,\nonumber
\end{align}

After performing a unitarity-IBP computation similar to Section~\ref{sec:unitarity4pt}, we find that only the following configuration contributes to the maximally transcendental part:
\begin{align}
\label{eq:fermioncut}
	\mathcal{F}_{\mathrm{tr}(F^3), 4}^{(0)}(2^-, l_3^-, l_2^{\psi/\bar{\psi}}, l_1^{\bar{\psi}/\psi}) \mathcal{A}_4^{(0)}(3^-, l_4^{\psi/\bar{\psi}}, -l_2^{\bar{\psi}/\psi}, -l_3^+) \mathcal{A}_4^{(0)}(4^+, 1^-, -l_1^{\psi/\bar{\psi}}, -l_4^{\bar{\psi}/\psi}) \,.
\end{align}
In particular, all configurations that involve scalars contribute only to lower transcendental parts. 
This is a very important fact as we will discuss shortly: it implies that the maximal transcendentality principle still holds between ${\cal N}$=4 SYM and QCD.

%%%%%%%%%%%%%%%%%%%%%%%%%%%
\begin{table}[t]
	\centering
	\vskip .1 cm 
	\begin{tabular}{|m{2.5cm}<{\centering} |m{2.5cm}<{\centering}|m{3.5cm}<{\centering}| m{3.5cm}<{\centering} |} 
		\hline
		Master & Topology  & $\Delta^{(2), \mathcal{N}=4}_{\text{M.T.}}$ & pure YM  \\ \hline %\hline
		
		$I_{\text{TBub2}}^{(2)}(3, 4, 1)$ & \includegraphics[scale=0.3]{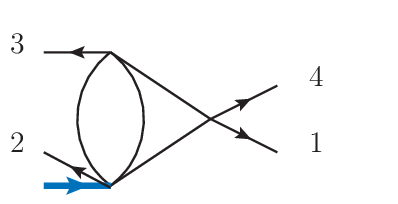} & $\left( 0, 0, 0, 0, -\frac{9}{2} \right)$ & $\left( \frac{11}{8}, \frac{7}{4}, -\frac{3}{8}, 0, \frac{9}{2} \right)$ \\ \hline
		
		$I_{\text{BubBox}}^{(2)}(1, 4, 3)$ & \includegraphics[scale=0.3]{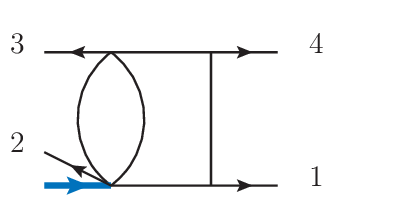} & $\left( 0, 0, 0, 0, 7 \right)$ & $\left( -3, -3, 1, 0, -7 \right)$ \\ \hline
		
		$I_{\text{dBox1a}}^{(2)}(1, 4, 3)$  & \includegraphics[scale=0.3]{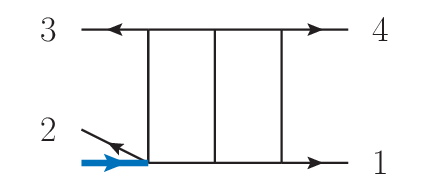} & $\left( 0, 0, 0, 0, -1 \right)$ & $\left( \frac{1}{2}, \frac{1}{2}, 0, 0, 1 \right)$ \\ \hline
		
		$I_{\text{dBox1b}}^{(2)}(1, 4, 3)$ & \includegraphics[scale=0.3]{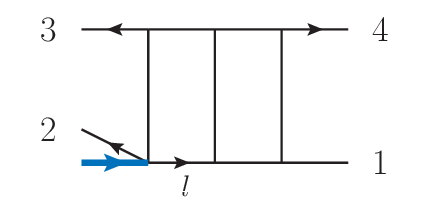} & $\left( 0, 0, 0, 0, -3 \right)$ & $\left( \frac{1}{2}, \frac{1}{2}, 0, 0, 3 \right)$ \\ \hline
		
		$I_{\text{TT0}}^{(2)}(3, 1, 4)$ & \includegraphics[scale=0.3]{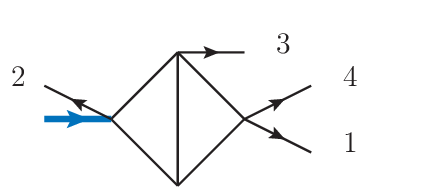} & $\left( 0, 0, 0, 0, -2 \right)$ & $\left( \frac{1}{2}, \frac{1}{2}, 0, 0, 2 \right)$ \\ \hline
		
		$I_{\text{TBox0}}^{(2)}(1, 4, 3)$ & \includegraphics[scale=0.3]{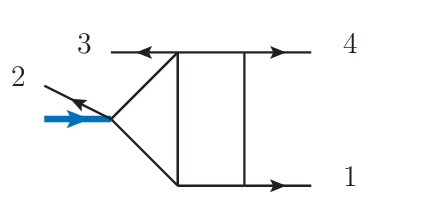} & $\left( 0, 0, 0, 0, 1 \right)$ & $\left( 0, 0, 0, 0, -1 \right)$ \\ \hline
	\end{tabular} 
	\caption{The coefficients of the master integrals detected by the unitarity cut shown in Figure~\ref{fig:unitaritycut} in $\mathcal{N}=4$ super Yang-Mills and pure Yang-Mills, the vector $(c_1,c_2,c_3,c_4,c_5)$ means $\sum_{a} c_a B_a$.
		\label{table:unitarityresult}
	}
\end{table}
%%%%%%%%%%%%%%%%%%%%%%%%%%%

The maximally transcendental contributions detected by the cut of \eqref{eq:fermioncut} are collected in Table~\ref{table:unitarityresult}. 
We also show the corresponding result in pure YM for comparison. 
One note that the difference $\Delta^{(2), \mathcal{N}=4}_{\text{M.T.}}$ are related to $B_5$.
By comparing the coefficients of $I_{\text{TBox0}}^{(2)}(1,4,3)$ and $I_{\text{dBox1a}}^{(2)}(1,4,3)$ with that of $\Delta^{(2), \mathcal{N}=4}_{\text{M.T.}}$ in \eqref{eq:difference}, one can fix the two free parameters $x_i$. One has (note that $B_4$ and $B_5$ are related by symmetry)
\begin{equation}
	B_5 \left( -\frac{3}{4} x_1 + x_2 \right) = -B_5 \,, \qquad \frac{1}{4} B_5 x_1 = B_5 \,, \nonumber
\end{equation}
which gives $x_1=4$ and $x_2 = 2$. Thus we obtain the difference 
\begin{equation}
\Delta^{(2), \mathcal{N}=4}_{\text{M.T.}}= B_4 \Big( 4 \tilde{G}_{3,1}^{(2)} +2 \tilde{G}_{3,3}^{(2)} \Big) + \left( p_1 \leftrightarrow p_3 \right) \,. 
\end{equation}
Together with the pure-YM result, we obtain the full maximally transcendental part for the form factor in $\mathcal{N} = 4$ SYM theory.
The results of the form factors are provided in the auxiliary files.
See also Appendix~\ref{app:fullFF} for some further discussion of the results.

\begin{figure}[t]
	\centering
	\includegraphics[scale=0.65]{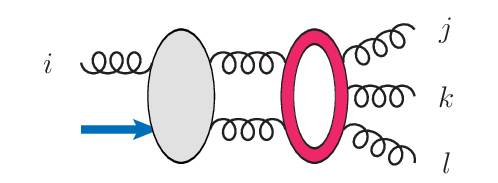}
	\caption{Diagrams that contributing to $\Delta^{(2), \mathcal{N}=4}_{\text{M.T.}}$ must involve a fermions loop, as indicated by the red color.}
	\label{fig:fermionloop}
\end{figure}

Finally, we can understand the result of $\Delta^{(2), \mathcal{\cal N}=4}_{\text{M.T.}}$ from the Feynman diagram point of view. 
The cut configuration \eqref{eq:fermioncut} means that the internal legs $\{ l_1, l_2, l_4\}$ are all fermions. Since the internal fermions must form a fermion loop, this implies that the diagrams that contribute to $\Delta^{(2), \mathcal{\cal N}=4}_{\text{M.T.}}$ must contain a fermion loop as shown in Figure~\ref{fig:fermionloop}. 
This observation will be used in the next subsection.

%%%%%%%%%%%%%%%%%%%%%%%%%%%
%%%%%%%%%%%%%%%%%%%%%%%%%%%

%%%%%%%%%%%%%%%%%%%%%%%%%%%%%%%%%%%%%
\subsubsection{Maximal transcendentality principle and color factors}

The previous computation shows that the difference $\Delta^{(2), {\cal N}=4}_{\text{M.T.}}$ comes from the configurations that involve only fermions but not scalars. Since the theory of QCD consists of the pure YM theory plus fermions, this means that the kinematic parts of the planar form factors between $\mathcal{N}=4$ SYM and QCD should be identical, while only the color factors can be different because of the different color representations of fermions in the two theories.\footnote{Note that both fermions are Dirac fermions in the two theories.}
Therefore, by properly converting the fermion representation from fundamental to adjoint, one will find the maximally transcendental part in QCD and $\mathcal{N}=4$ SYM are equivalent. 
Below we discuss the color factors in more detail.

\begin{figure}
	\centering
	\subfloat[]{\includegraphics[scale=0.45]{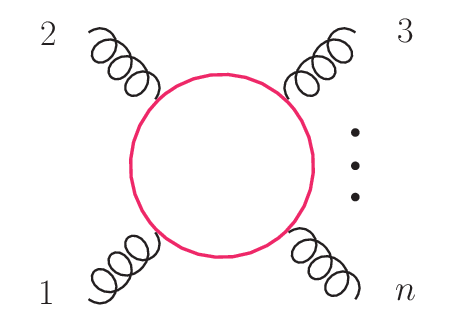}}
	\hskip 2cm
	\subfloat[]{\includegraphics[scale=0.45]{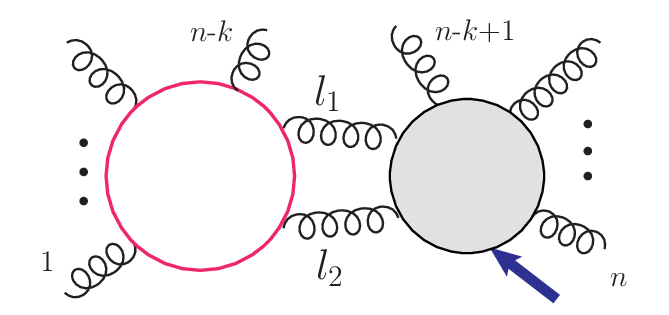}}
	\caption{Color factors of the one- and two-loop cases.\label{fig:colorfactor}}
\end{figure}

It is instructive to start with a one-loop Feynman diagram shown in Figure~\ref{fig:colorfactor}(a), in which there are $n$ external gluons and all propagators are fermions. 
The corresponding color factors for the case of fundamental fermions in QCD and adjoint fermions ${\cal N}=4$ SYM are respectively
\begin{align}
	\text{QCD (fundamental)} : & \qquad n_f \mathbf{C}_{\text{single-trace}} \,, \label{eq:qcdcolor1} \\
	\text{${\cal N}=4$ (adjoint)} : &	\qquad 4 \prod_{k = 1}^n f^{b_{k} a_k b_{k+1}} = 4 N_c \mathbf{C}_{\text{single-trace}} + (\text{double traces}) \,, \label{eq:neq4color1} 
\end{align}
where
\begin{equation}\label{eq:Csingletrace}
	\mathbf{C}_{\text{single-trace}} = \mathrm{tr}(T^{a_1} T^{a_2} \ldots T^{a_n}) + (-1)^n \mathrm{tr}(T^{a_n} T^{a_{n-1}} \ldots T^{a_1}) \,,
\end{equation}
and we have included $n_f$ in \eqref{eq:qcdcolor1} as the flavor number of quarks in QCD and similarly, the factor $4$ in \eqref{eq:neq4color1} is because there are four gluinos in the ${\cal N}=4$ supermultiplet. 
We can see that the two color factors can be identified if we apply the following rule: 
\begin{equation}
\label{eq:colorRule4pt}
n_f \rightarrow 4N_c \,,  \qquad \textrm{and} \qquad N_c \rightarrow \infty \,,
\end{equation} 
where the large $N_c$ limit is to keep only the single-trace terms.

%%%%%%%%%%%%%%%%%%%%%
%%%%%%%%%%%%%%%%%%%%%

Now we consider the two-loop $n$-gluon form factor. 
For the form factor problem under consideration, a related two-loop diagram is shown in Figure~\ref{fig:colorfactor}(b).
The part on the LHS is a one-loop $(n-k+2)$-gluon amplitude with a fermion loop, which can be regarded as Figure~\ref{fig:colorfactor}(a), and we denote its color factor as $\mathbf{C}^{(1)}_{\cal A}$. The part on the RHS is a tree-level $(k+2)$-gluon form factor, and its color factor has only single-trace terms which we denote as $\mathbf{C}^{(0)}_{\cal F}$. The color factor of the two-loop diagram is given by the product as
\begin{equation}
	\sum_{a_{l_1}, a_{l_2}} \mathbf{C}^{(1)}_{\cal A}(T^{a_1,} , \ldots, T^{a_{n-k}}, T^{a_{l_1}}, T^{a_{l_2}}) \mathbf{C}^{(0)}_{\cal F}(T^{a_{n-k+1}}, \ldots, T^{a_{n}}, T^{a_{l_2}}, T^{a_{l_1}}) \,.
\end{equation}
Note that if $l_1$ and $l_2$ are not adjacent in the fermion loop, the color factor will be sub-leading in the large $N_c$ expansion.
Keeping only the leading-$N_c$  color factors, one has
\begin{equation}
	\begin{aligned}
	\text{QCD} : & \qquad n_f N_c \mathbf{C}_{\text{single-trace}} \,,\\
	\text{${\cal N}=4$} : &	\qquad 4 N_c^2 \mathbf{C}_{\text{single-trace}} \,,
	\end{aligned}
\end{equation}
where $\mathbf{C}_{\text{single-trace}}$ is given in \eqref{eq:Csingletrace}.
Therefore, the two-loop color factors are also identified with each other by using the rule \eqref{eq:colorRule4pt}.
The same argument holds for other two-loop diagrams that involve a fermion loop. 

Thus we can conclude that
\begin{equation}
\Delta^{(2), {\rm QCD}}_{\text{M.T.}} \Big|_{n_f \rightarrow 4N_c, N_c \rightarrow \infty} = \Delta^{(2), \mathcal{N}=4}_{\text{M.T.}} \,.
\end{equation}
In other words, we find the following correspondence for the four-point form factor up to two-loop order in the limit of $N_c\rightarrow \infty$:
\begin{align}
&	\mathcal{F}_{\mathrm{tr}(F^3)}^{(1),{\cal N}=4}(1^-,2^-,3^-,4^+) = \mathcal{F}_{\mathrm{tr}(F^3)}^{(1), \text{QCD}}(1^-,2^-,3^-,4^+)  \,, \\
&	\mathcal{F}_{\mathrm{tr}(F^3)}^{(2),{\cal N}=4}(1^-,2^-,3^-,4^+) = \mathcal{F}_{\mathrm{tr}(F^3)}^{(2), \text{QCD}}(1^-,2^-,3^-,4^+) \Big|_{n_f \rightarrow 4N_c} \,, \nonumber
\end{align}
which mean that the principle of maximally transcendental principle still holds for the four-point form factor of $\mathrm{tr}(F^3)$.

To the best of the authors' knowledge, this form factor provides the first example of the MTP where the fermion-loop diagrams contribute to the maximally transcendental part and thus the changing of color factors involves the $n_f$ factor.

%%%%%%%%%%%%%%%%%%%%%%%%%
\subsection{A relation to the form factor of $\mathrm{tr}(\phi^3)$}
\label{sec:phi3andF3}

There is another interesting correspondence between two different operators $\mathrm{tr}(F^3)$ and $\mathrm{tr}(\phi^3)$ in $\mathcal{N}=4$ super Yang-Mills theory. The four-point form factor $\mathcal{F}_{\mathrm{tr}(\phi^3), 4}^{(L)}$ is half-BPS and  has only the maximally transcendentality part. The results up to two loops was obtained in \cite{Guo:2021bym} and take the following form:
\begin{equation}\label{eq:F4phi3structure}
	\mathcal{F}_{\mathrm{tr}(\phi^3), 4}^{(L)}(1^\phi, 2^\phi, 3^\phi, 4^+) = \mathcal{F}_{\mathrm{tr}(\phi^3), 4}^{(0)}(1^\phi, 2^\phi, 3^\phi, 4^+) \left( B_1 \mathcal{G}_1^{(L)} + B_2 \mathcal{G}_2^{(L)} \right) \,,
\end{equation}
which has two spinor factors $B_1$ and $B_2$, while the form factor of $\mathrm{tr}(F^3)$ in \eqref{eq:ansatzFullF3} contains five $B_i$'s.
Interestingly, we find that the loop correction functions $\mathcal{G}_1^{(L)}$ and $\mathcal{G}_2^{(L)}$ in \eqref{eq:F4phi3structure} are precisely the same as that of $\mathcal{F}_{\mathrm{tr}(F^3), 4}^{(L), \text{M.T.}}$ in \eqref{eq:ansatzFullF3}. 
In other words, we have the relation
\begin{equation}
	\mathcal{I}_{\mathrm{tr}(F^3), 4}^{(L), \text{M.T.}}(1^-, 2^-, 3^-, 4^+) = \mathcal{I}_{\mathrm{tr}(\phi^3), 4}^{(L)}(1^\phi, 2^\phi, 3^\phi, 4^+) + \sum_{a = 3}^5 B_a \mathcal{G}_{a}^{(L)} \,,
\end{equation}
at least up to two loops.

To understand this relation, we briefly review the computation for the form factor of ${\mathrm{tr}(\phi^3)}$.
The tree-level four-point form factor of ${\mathrm{tr}(\phi^3)}$ is
\begin{equation}
	\mathcal{F}_{\mathrm{tr}(\phi^3), 4}^{(0)}(1^\phi, 2^\phi, 3^\phi, 4^+) = \frac{\langle 13 \rangle}{\langle 34 \rangle \langle 14 \rangle} \,.
\end{equation}
The pole structure implies that only $B_1$ and $B_2$ should occur in the coefficients of the loop correction, as given in \eqref{eq:F4phi3structure}.
Via the bootstrap method, the form factor is first constrained by infrared structure, and then the constraints of the collinear limit at one and two loops respectively are
\begin{align}
	& \mathcal{I}_{\mathrm{tr}(\phi^3), 4}^{(1)} \, \xlongrightarrow{\mbox{$ p_3 \parallel p_4 $}} \mathcal{I}_{\mathrm{tr}(\phi^3), 3}^{(1)} + r_1^{[1], \text{MT}}(s_{34}, z) \,, \\
	& \mathcal{R}_{\mathrm{tr}(\phi^3), 4}^{(2)} \, \xlongrightarrow{\mbox{$ p_3 \parallel p_4 $}} \mathcal{R}_{\mathrm{tr}(\phi^3), 3}^{(2)} \,,
\end{align}
and the cancellation of the spurious pole $\langle 24 \rangle$ provides the constraint
\begin{equation}
	\mathcal{G}_1^{(L)}-\mathcal{G}_2^{(L)} \, \xlongrightarrow{\mbox{$ \langle 24 \rangle \rightarrow 0$}} 0 \, .
\end{equation}

At one loop, the form factor is fixed uniquely by these constraints, and since the same constraints apply also to the one-loop form factor of ${\rm tr}(F^3)$, the one-loop correction functions the $\mathcal{G}_{1,2}^{(1)}$ must be the same for both form factors. 
The case is more non-trivial at two loops.
At the two-loop level, there are still some remaining degrees of freedom that are free from the above constraints. 
These can be organized as eight building blocks which are the same as $\tilde{G}_{1,\alpha}^{(2)}$ listed in \eqref{eq:BuildingBlock}.\footnote{In \cite{Guo:2021bym}, the number of remaining parameters are 10. Here we reduce the number to 8, which is equal to the number of $\tilde{G}_{1,\alpha}^{(2)}$. The difference is because here we apply further the constraints come from the $\mathcal{O}(\epsilon)$ order for the spurious pole $\langle 24 \rangle$ cancellation (see Appendix~\ref{app:HigherOrder}), which was not used in \cite{Guo:2021bym}.}
These building blocks can be fixed by using simple quadruple cuts. 
Since the unitarity cuts are apparently quite different for the form factors of $\mathrm{tr}(F^3)$ and $\mathrm{tr}(\phi^3)$, it is a prior not obvious at all that the coefficients of $\tilde{G}_{1,\alpha}^{(2)}$ are the same for the two form factors. 
Interestingly, the explicit computations show that they are identical for the two form factors. 
We do not yet have a physical explanation for this, and it would be interesting to check if this is true for more general cases.
It is worth pointing out that $\tilde{G}_{1,\alpha}^{(2)}$ contain all the $\mu$-term master integrals and $D$-dimensional cuts are needed to determine them.
Since the unitarity computation for the half-BPS form factor of ${\rm tr}(\phi^3)$ is much simpler, this relation (if true generally) may be used to simplify the computation for the ${\rm tr}(F^3)$ form factor.

%%%%%%%%%%%%%%%%%%%%%
\section{Summary and discussion}
\label{sec:discussion}

In this paper, we study the principle of maximal transcendentality for a class of form factors using the bootstrap method. 
For the minimal form factors up to two loops, we show that the IR divergences plus some symmetry arguments are sufficient to determine the maximally transcendental part. 
Some non-trivial constraints on the lower transcendental parts are also discussed in a similar way.
For the two-loop three-gluon form factors of ${\rm tr}(F^2)$, we show that the IR divergences together with collinear limit constraints are able to fix the maximal transcendental part of form factors. 
While the IR and collinear constraints may not fix all parameters, an important further insight is to apply some unitarity cuts that only depend on gluon states and thus are universal for general gauge theories.

As another important part of the paper, we obtain the maximal transcendental parts of the two-loop four-point planar form factor with $\mathrm{tr}(F^3)$ operator in both $\mathcal{N}$=4 SYM and pure YM theories.
The bootstrap computation using various physical constraints and certain simple $D$-dimensional unitarity cuts are given in detail.
We find that the form factor results are different between the $\mathcal{N}$=4 SYM and pure YM theories. The difference is due to the contribution from the gluino-loop diagrams in $\mathcal{N}=4$ SYM. Importantly, the scalar-loop diagrams in $\mathcal{N}=4$ SYM do not contribute to the maximally transcendental part.
This implies that the maximally transcendental part of the $\mathcal{N}=4$ SYM result is equivalent to the corresponding form factor result in QCD, up to a proper change of color factors associated with the quark loops.
Thus the maximal transcendentality principle still holds for this form factor.

It is instructive to compare with this four-point case of the simpler three-point form factor of ${\rm tr}(F^2)$. In the latter case, the maximally transcendental parts are all the same for the $\mathcal{N}=4$ SYM, QCD, and pure YM theories,
which can be understood as neither fermion- nor scalar-loops contribute to the maximal transcendentality part. 
In contrast, the MTP property for the four-point form factors appears to be more non-trivial, due to the new contribution of fermion loops. 
This seems to be the first such example for the maximal transcendentality principle and suggests the MTP may hold for more general form factors, or Higgs-plue-multi-gluon amplitudes.
%It would be highly interesting to explore and understand it in more general cases.
Similar to the four-point form factor we consider, a key step to understanding or proving MTP would be to show that the scalar-loop diagrams would have no maximally transcendental contributions in more general cases.

We should also point out there are counterexamples of MTP for other observables. For example, for the gluon amplitudes, the MTP is violated even for the simple one-loop four-gluon amplitudes. In Appendix~\ref{app:A4noMT}, we discuss this in detail and show that while the MTP applies to $A_4^{(1)}(1^-,2^-,3^+,4^+)$, it is explicitly violated by $A_4^{(1)}(1^-,2^+,3^-,4^+)$.
Some counterexamples were found in the study of high energy limit of amplitudes \cite{DelDuca:2017peo}.
Through unitarity cuts, the amplitudes can enter as building blocks in high-loop form factors, and it would be very interesting to understand further the different properties between form factors and amplitudes.

%%%%%%%%%%%%%%%%%%%%%%%%%%
%%%%%%%%%%%%%%%%%%%%%%%%%%
\section*{Acknowledgements}

We thank Guanda Lin for the discussion. 
This work is supported in part by the National Natural Science Foundation of China (Grants No.~12175291, 11935013, 11822508, 12047503) %, 12047502, 11947301), 
and by the Key Research Program of the Chinese Academy of Sciences, Grant NO. XDPB15.
We also thank the support of the HPC Cluster of ITP-CAS.

%%%%%%%%%%%%%%%%%%%%
%%%%%%%%%%%
\appendix

%%%%%%%%%%%%%%%%%%%%%%%%%
\section{UT master integrals}
\label{app:UT}

In this appendix, we list all uniform-transcendentality (UT) integrals we have used in the paper.
We will use the notation that the figures represent the propagators of the integrals (the propagator with a dot is quadratic), and the numerators are given by the coefficients in front of the figures.
We use the convention of an $L$-loop integral as
\begin{equation}
I^{(L)}[N(l_i, p_j)]= e^{L\epsilon\gamma_\mathrm{E}} \int\frac{d^D l_1}{i\pi^{\frac{D}{2}}}\dots\frac{d^Dl_L}{i\pi^{\frac{D}{2}}}
\frac{N(l_i, p_j)}{\prod_j D_j} \,.
\end{equation}

The one-loop master integrals are $I_{\text{Bub}}^{(1)}(1,\ldots,n)$ and $I^{(1)}_{\text{Box}}(i,j,k)$
\begin{align}
& I_{\text{Bub}}^{(1)}(1,\ldots,n) = \frac{1-2\epsilon}{\epsilon} \times
\begin{aligned}
	\includegraphics[scale=0.35]{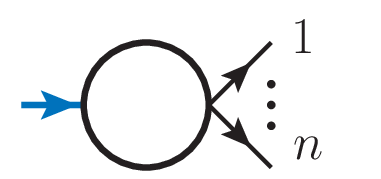}
\end{aligned} \\	
& I_{\text{Box}}^{(1)}(i,j,k) =s_{jk} s_{ij} \times
\begin{aligned}
	\includegraphics[scale=0.3]{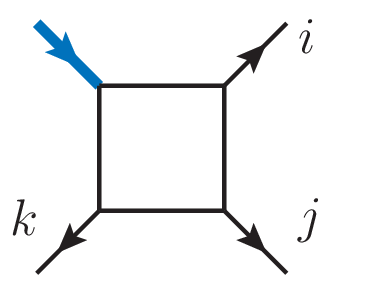}
\end{aligned}
\end{align}
In our convention, the one-loop bubble basis is evaluated as:
\begin{equation}
I_{\text{Bub}}^{(1)}(1,..,n) = (-s_{1..n}^2)^{-\epsilon} \Big[ {1\over\epsilon^{2}} - {\pi^2 \over 12} - {7 \zeta_3 \over3} \epsilon - {47 \pi^4 \over 1440} \epsilon^2  + {\cal O}(\epsilon^3) \Big]
 \,.
 \label{eq:bubbleintegral}
\end{equation}

The planar master integrals for two-loop four-point form factors with $\mathrm{tr}(F^3)$ as follows \cite{Gehrmann:2000zt, Gehrmann:2001ck, Abreu:2020jxa}
\begin{align}
& I_{\text{dBub}}^{(2)}(p_1,\cdots,p_n;p_{m},\cdots,p_{m+k}) = 
\left(\frac{1-2\epsilon}{\epsilon}\right)^2 \times
\begin{aligned}
	\includegraphics[scale=0.5]{figs/dBub.eps}
\end{aligned} \\
& I_{\text{Sun}}^{(2)}(p_1,\cdots,p_n) = \frac{(1-2\epsilon) (1-3\epsilon) (2-3\epsilon)}{\epsilon^3 s_{1{\cdots}n}}\times
\begin{aligned}
	\includegraphics[scale=0.5]{figs/Sun.eps}
\end{aligned} \\
&	I_{\text{TBub0}}^{(2)}(p_1,p_2) = \frac{(1-2\epsilon)(1-3\epsilon)}{\epsilon^2} \times
\begin{aligned}
	\includegraphics[scale=0.4]{figs/TBub0.eps}
\end{aligned} \\
&	I_{\text{TBub1}}^{(2)}(p_1,p_2,p_3) = \frac{(1-2\epsilon)(1-3\epsilon)}{\epsilon^2}\times
\begin{aligned}
	\includegraphics[scale=0.4]{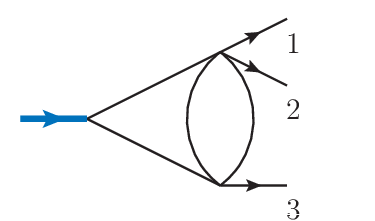}
\end{aligned} \\
&	I_{\text{TBub2}}^{(2)}(p_1,p_2,p_3) = \frac{(1-2\epsilon)(1-3\epsilon)}{\epsilon^2}\times
\begin{aligned}
	\includegraphics[scale=0.4]{figs/TBub2.eps}
\end{aligned} \\
&	I_{\text{TBub2}}^{(2)}(p_1,p_2,p_3,p_4) = \frac{(1-2\epsilon)(1-3\epsilon)}{\epsilon^2}\times
\begin{aligned}
	\includegraphics[scale=0.4]{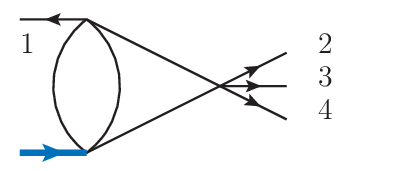}
\end{aligned} \\
&	I_{\text{TBub3a}}^{(2)}(p_1,p_2,p_3,p_4) = \frac{2(1-2\epsilon)(1-3\epsilon)l^2-\epsilon(s_{1234}+s_{12}-s_{34})}{\epsilon^2}\times
\begin{aligned}
	\includegraphics[scale=0.4]{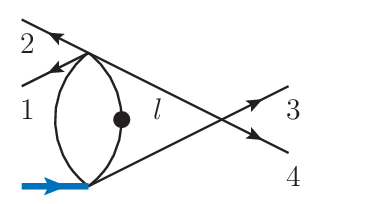}
\end{aligned} \\
&	I_{\text{TBub3b}}^{(2)}(p_1,p_2,p_3,p_4) = \frac{\sqrt{\Delta_3}}{\epsilon} \times
\begin{aligned}
	\includegraphics[scale=0.4]{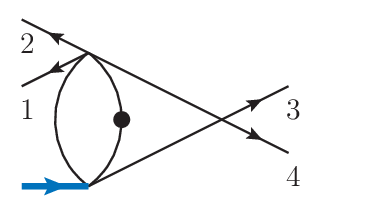}
\end{aligned} \\
&	I_{\text{BubBox}0}^{(2)}(p_1,p_2,p_3) = \frac{(1-2\epsilon)s_{12}s_{23}}{\epsilon}\times
\begin{aligned}
	\includegraphics[scale=0.4]{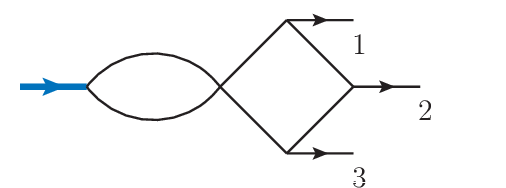}
\end{aligned} \\
&	I_{\text{BoxBub}}^{(2)}(p_1,p_2,p_3) = \frac{(1-2\epsilon)(s_{13}+s_{23})}{\epsilon} \times
\begin{aligned}
	\includegraphics[scale=0.4]{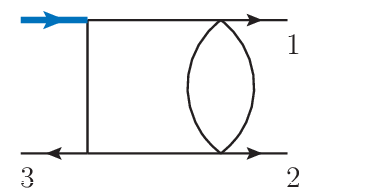}
\end{aligned} \\
&	I_{\text{BubBox}}^{(2)}(p_1,p_2,p_3) = \frac{(1-2\epsilon)s_{12}}{\epsilon}\times
\begin{aligned}
	\includegraphics[scale=0.4]{figs/BubBox.eps}
\end{aligned} \\
&	I_{\text{BubBox1}}^{(2)}(p_1,p_2,p_3,p_4) = \frac{(1-2\epsilon)(s_{13}+s_{23})}{\epsilon}\times
\begin{aligned}
	\includegraphics[scale=0.4]{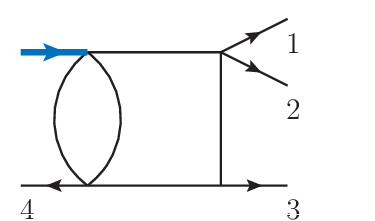}
\end{aligned} \\
&	I_{\text{BubBox2}}^{(2)}(p_1,p_2,p_3,p_4) = -\frac{(1-2\epsilon)(s_{1234}-s_{123})}{\epsilon}\times
\begin{aligned}
	\includegraphics[scale=0.4]{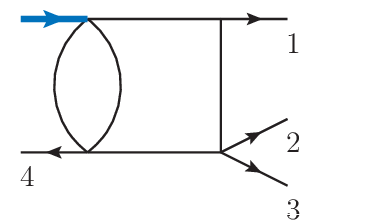}
\end{aligned} \\
&	I_{\text{BubBox3}}^{(2)}(p_1,p_2,p_3,p_4) = \frac{(1-2\epsilon)s_{12}}{\epsilon}\times
\begin{aligned}
	\includegraphics[scale=0.4]{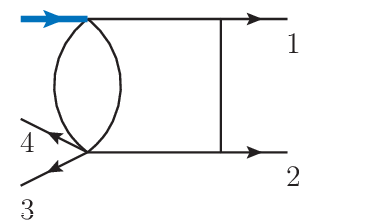}
\end{aligned} \\
&	I_{\text{TT0}}^{(2)}(p_1,p_2,p_3) = (s_{13}+s_{23}) \times
\begin{aligned}
	\includegraphics[scale=0.4]{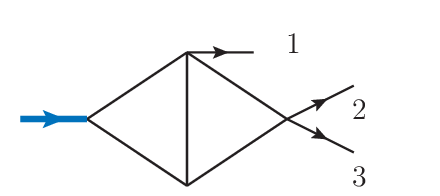}
\end{aligned} \\
&	I_{\text{TT1}}^{(2)}(p_1,p_2,p_3) = \frac{s_{12}s_{23}}{\epsilon}\times
\begin{aligned}
	\includegraphics[scale=0.4]{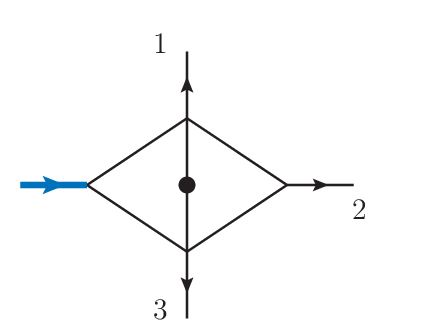}
\end{aligned} \\
&	I_{\text{TT1a}}^{(2)}(p_1,p_2,p_3) = (s_{12}+s_{23}) \times
\begin{aligned}
	\includegraphics[scale=0.4]{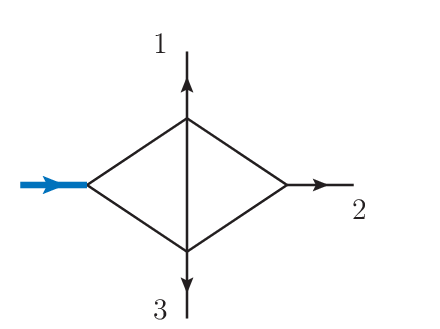}
\end{aligned} \\
&	I_{\text{TT2}}^{(2)}(p_1,p_2,p_3) = s_{13}\times
\begin{aligned}
	\includegraphics[scale=0.4]{figs/TT20.eps}
\end{aligned} \\
&	I_{\text{TT2}}^{(2)}(p_1,p_2,p_3,p_4) = \frac{(s_{12}+s_{13})(s_{24}+s_{34})-s_{14}s_{23}}{\epsilon}\times
\begin{aligned}
	\includegraphics[scale=0.4]{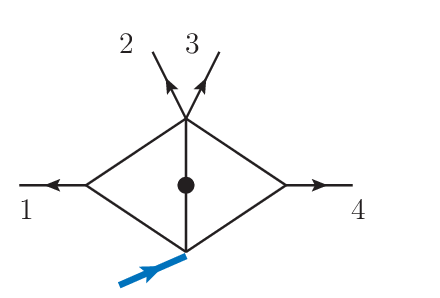}
\end{aligned} \\
&	I_{\text{TT3a}}^{(2)}(p_1,p_2,p_3,p_4) = -(s_{13}+s_{14})\times
\begin{aligned}
	\includegraphics[scale=0.4]{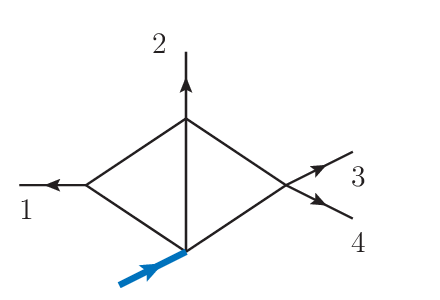}
\end{aligned} \\
&	I_{\text{TT3b}}^{(2)}(p_1,p_2,p_3,p_4) = \frac{s_{12}s_{234}}{\epsilon}\times
\begin{aligned}
	\includegraphics[scale=0.4]{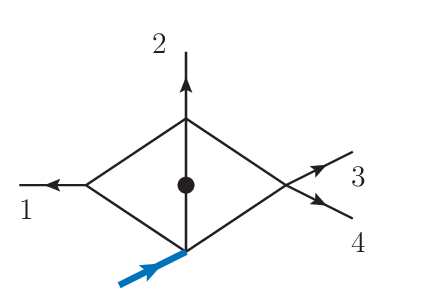}
\end{aligned} \\
&	I_{\text{TT4}}^{(2)}(p_1,p_2,p_3,p_4) = \frac{\sqrt{\Delta_3}}{\epsilon} \times
\begin{aligned}
	\includegraphics[scale=0.4]{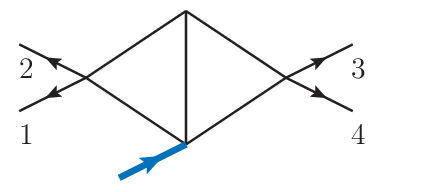}
\end{aligned} \\
&	I_{\text{TBox0}}^{(2)}(p_1,p_2,p_3) = (s_{13}+s_{12})s_{12} \times
\begin{aligned}
	\includegraphics[scale=0.4]{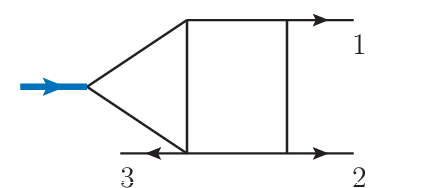}
\end{aligned} \\
&	I_{\text{TBox1}}^{(2)}(p_1,p_2,p_3,p_4) = -(s_{13}+s_{23})s_{34}\times
\begin{aligned}
	\includegraphics[scale=0.4]{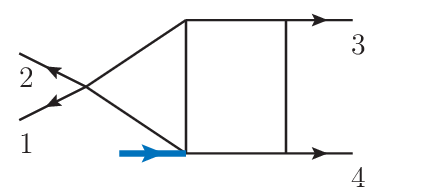}
\end{aligned} \\
&	I_{\text{TBox2a}}^{(2)}(p_1,p_2,p_3,p_4) = s_{34}(s_{12}+s_{13})\times
\begin{aligned}
	\includegraphics[scale=0.4]{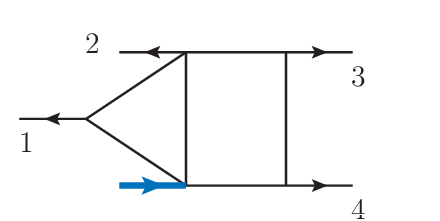}
\end{aligned} \\
&	I_{\text{dBox1a}}^{(2)}(p_1,p_2,p_3) = s_{12}^2s_{23}\times
\begin{aligned}
	\includegraphics[scale=0.4]{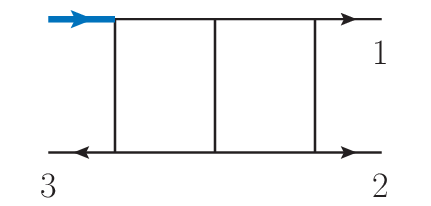}
\end{aligned} \\
&	I_{\text{dBox1b}}^{(2)}(p_1,p_2,p_3) = s_{12}(s_{13}+s_{23})(l-p_1)^2\times
\begin{aligned}
	\includegraphics[scale=0.4]{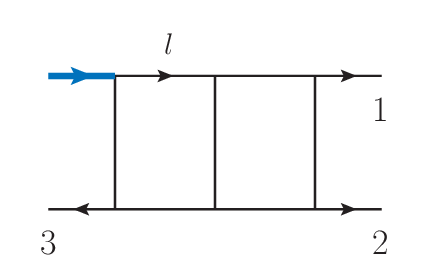}
\end{aligned} \\
&	I_{\text{dBox2a}}^{(2)}(p_1,p_2,p_3,p_4) = s_{12}s_{34}s_{23}\times
\begin{aligned}
	\includegraphics[scale=0.4]{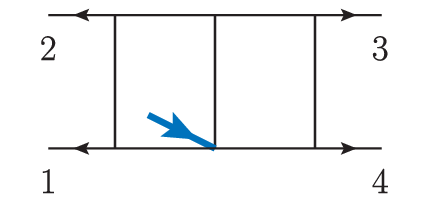}
\end{aligned} \\
&	I_{\text{dBox2b}}^{(2)}(p_1,p_2,p_3,p_4) = s_{12}s_{34}(l-p_1-p_2-p_3)^2\times
\begin{aligned}
	\includegraphics[scale=0.4]{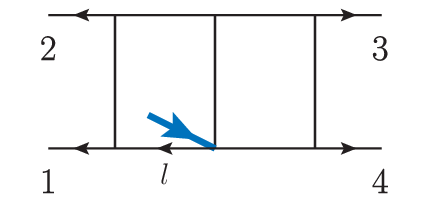}
\end{aligned} \\
&	I_{\text{dBox2c}}^{(2)}(p_1,p_2,p_3,p_4) = \mathrm{tr}_5 \mu_{12} \times
\begin{aligned}
	\includegraphics[scale=0.4]{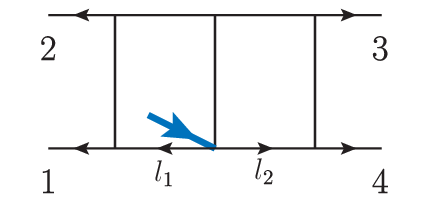}
\end{aligned} \\
&	I_{\text{BPa}}^{(2)}(p_1,p_2,p_3,p_4) = \frac{(1-2\epsilon)s_{23}s_{34}}{\epsilon} \times
\begin{aligned}
	\includegraphics[scale=0.4]{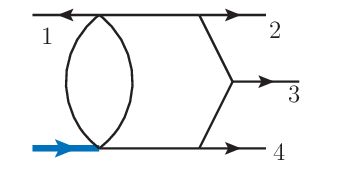}
\end{aligned} \\
&	I_{\text{BPb}}^{(2)}(p_1,p_2,p_3,p_4) = \, \frac{\mathrm{tr}_5\mu_{11}}{\epsilon} \times
\begin{aligned}
	\includegraphics[scale=0.4]{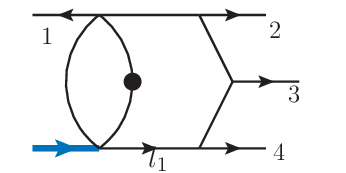}
\end{aligned} \\
&	I_{\text{TP}}^{(2)}(p_1,p_2,p_3,p_4) = \mathrm{tr}_5\mu_{11} \times
\begin{aligned}
	\includegraphics[scale=0.4]{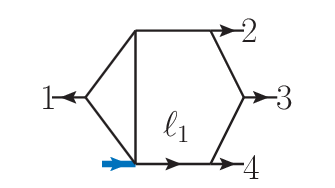}
\end{aligned} 
\end{align}
We point out that, as in \cite{Guo:2021bym}, we have introduced the master integral $I_{\rm TP}^{(2)}$ which is a linear combination of two masters used in \cite{Abreu:2020jxa, Canko:2020ylt} as
\begin{equation}
\hskip -.47cm \begin{tabular}{c}{\includegraphics[scale=0.34]{figs/TP.eps} } \end{tabular} 
\hskip -.31cm {\rm tr}_5 \mu_{11}= 
\hskip -.2cm \begin{tabular}{c}{\includegraphics[scale=0.3]{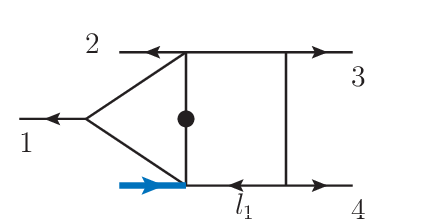} } \end{tabular} 
\hskip -.31cm 
{{\rm tr}_5 \mu_{11} \over 2\epsilon}-
 \hskip -.16cm \begin{tabular}{c}{\includegraphics[scale=0.3]{figs/BPb.eps} }  \end{tabular} 
\hskip -.31cm 
{{\rm tr}_5 \mu_{11} \over \epsilon} \ ,
\label{eq:intRelation}
\end{equation}
and we use $I_{\text{TP}}^{(2)}$ to replace the first integral on the RHS of \eqref{eq:intRelation} as a master integral in the set of 221 master basis. We comment here that $I_{\text{TBub2}}^{(2)}$, $I_{\text{TT4}}^{(2)}$, $I_{\text{TBub3b}}^{(2)}$ and $I_{\text{BPb}}^{(2)}$ do not appear in the maximally transcendental part of the two-loop four-point form factors with $\mathrm{tr}(F^3)$.

There are non-planar integrals for two-loop three-point form factors with the length-$2$ operators \cite{Gehrmann:2001ck} which we list below:
\begin{align}
&	I_{\text{NTBox1}}^{(2)}(p_1,p_2) = s_{12}^2 \times
\begin{aligned}
	\includegraphics[scale=0.4]{figs/TBoxNP1.eps}
\end{aligned} \\
&	I_{\text{NTBox2}}^{(2)}(p_1,p_2,p_3) = (s_{13}+s_{23})^2 \times
\begin{aligned}
	\includegraphics[scale=0.4]{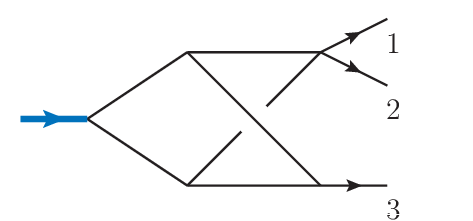}
\end{aligned} \\
&	I_{\text{NTBox3a}}^{(2)}(p_1,p_2,p_3) = s_{12} s_{123} \times
\begin{aligned}
	\includegraphics[scale=0.4]{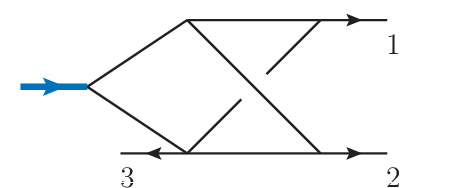}
\end{aligned} \\
&	I_{\text{NTBox3b}}^{(2)}(p_1,p_2,p_3) = 2\left[(s_{12}+s_{13}) l \cdot p_2-(s_{12}+s_{23}) l \cdot p_1\right] \times
\begin{aligned}
	\includegraphics[scale=0.4]{figs/TBoxNP3.eps}
\end{aligned} \\
&	I_{\text{NdBox1a}}^{(2)}(p_1,p_2,p_3) =  s_{12}\left[ s_{23}(l+p_3)^2 + s_{13}(p_{123}-l)^2 \right] \times
\begin{aligned}
	\includegraphics[scale=0.4]{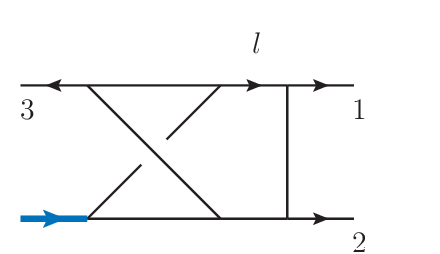}
\end{aligned} \\
&	I_{\text{NdBox1b}}^{(2)}(p_1,p_2,p_3) = s_{12}\left[ s_{23}(l+p_3)^2 - s_{13}(p_{123}-l)^2 \right] \times
\begin{aligned}
	\includegraphics[scale=0.4]{figs/dBoxNP1a.eps}
\end{aligned} \\
&	I_{\text{NdBox2a}}^{(2)}(p_1,p_2,p_3) =  s_{12}\left[s_{13}(l-p_1)^2+s_{23}(l-p_2)^2\right] \times
\begin{aligned}
	\includegraphics[scale=0.4]{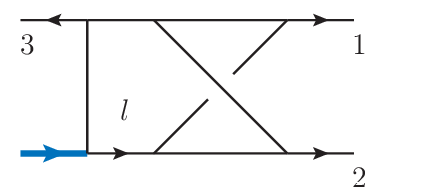}
\end{aligned} \\
&	I_{\text{NdBox2b}}^{(2)}(p_1,p_2,p_3) = s_{12}\left[s_{13}(l-p_1)^2-s_{23}(l-p_2)^2\right] \times
\begin{aligned}
	\includegraphics[scale=0.4]{figs/dBoxNP2a.eps}
\end{aligned}
\end{align}
%

%%%%%%%%%%%%%%%%%%%%%%
\section{Catani IR subtraction formula}
\label{app:catani}

In our discussion, we mainly consider BDS ansatz to subtract the IR divergences, which applies to $\mathcal{N}=4$ SYM. This is enough for our consideration of the maximally transcendental parts. In this appendix, we briefly compare the Catani subtraction and the BDS subtraction.

For general more realistic theories, such as QCD, one needs to use the Catani subtraction formula \cite{Catani:1998bh}
\begin{align}
	\mathcal{F}_n^{(1)} & = I_n^{(1)} \mathcal{F}_n^{(0)} + \mathcal{F}_n^{(1),\text{fin}} + \mathcal{O}(\epsilon) \,, \\
	\mathcal{F}_n^{(2)} & = I_n^{(2)} \mathcal{F}_n^{(0)} + I_n^{(1)} \mathcal{F}_n^{(1)} + \mathcal{F}_n^{(2),\text{fin}} + \mathcal{O}(\epsilon) \,, \nonumber
\end{align}
where the concrete expressions of $I_n^{(1)}$ and $I_n^{(2)}$ depend on the specific physical process, for example, for the form factor with $n$ external gluons,
\begin{align}
	\label{eq:catani}
	I_{n, g}^{(1)}(\epsilon) = & -\frac{e^{\gamma_\text{E} \epsilon}}{\Gamma(1-\epsilon)} \left( \frac{C_A}{\epsilon^2} + \frac{\beta_0}{\epsilon} \right) \sum_{i=1}^n \left( - s_{i,i+1} \right)^{-\epsilon} \,, \\
	I_{n, g}^{(2)}(\epsilon) = & -\frac{1}{2} \left( I_n^{(1)} \right)^2 - \frac{\beta_0}{\epsilon} I_n^{(1)}(\epsilon) + \frac{e^{-\gamma_\text{E} \epsilon} \Gamma(1-2\epsilon)}{\Gamma(1-\epsilon)} \left( \frac{\beta_0}{\epsilon} + \mathcal{K} \right) I_n^{(1)}(2\epsilon) + n \frac{e^{\gamma_\text{E} \epsilon}}{\epsilon \Gamma(1-\epsilon)} \mathcal{H}_g^{(2)} \,, \nonumber
\end{align}
where
\begin{align}
	\beta_0 = & \frac{11}{3}C_A - \frac{2}{3} n_f \, , \\
	\mathcal{K} = & \left( \frac{67}{9} - 2\zeta_2 \right) C_A -\frac{10}{9} n_f \, , \nonumber \\
	\mathcal{H}_g^{(2)} = & \left( \frac{1}{2} \zeta_3 + \frac{11}{24} \zeta_2 +\frac{5}{12} \right) C_A^2 + \frac{5}{27} n_f^2 - \left( \frac{1}{12} \zeta_2 + \frac{89}{108} \right) C_A n_f - \frac{1}{4} \frac{n_f}{C_A} \,. \nonumber
\end{align}

For simplicity, below we only focus on the leading color part of two-loop $n$-point minimal form factor, for which the Catani subtraction is given as
\begin{align}
	\mathcal{F}_{\mathrm{tr}(F^n), n}^{(2), \text{Can}} = \sum_{a=2L}^{-\infty} \epsilon^{2L-a} \mathcal{F}_{\mathrm{tr}(F^n), n}^{(2), \text{Can}, \text{deg-a}} = - I_{n,g}^{(2)} \mathcal{F}_{\mathrm{tr}(F^n), n}^{(0)} - I_{n,g}^{(1)} \mathcal{F}_{\mathrm{tr}(F^n), n}^{(1)} \,.
\end{align}
The maximal transcendental part is not affected by ultraviolet divergence, and the formula $\mathcal{I}_{\text{min}, n}^{(2), \text{Can}}$ and $\mathcal{I}_{\text{min}, n}^{(2), \text{BDS}}$ provide exactly the same subtraction. They differ only in the finite part as
\begin{align}
	& \mathcal{I}_{\mathrm{tr}(F^n), n}^{(L), \text{Can}, \text{deg-4}} - \mathcal{I}_{\mathrm{tr}(F^n), n}^{(L), \text{BDS}, \text{deg-4}} \\
	= & - I_{n,g}^{(2), \text{deg-4}} \mathcal{I}_{\mathrm{tr}(F^n), n}^{(0)} - I_{n,g}^{(1), \text{deg-4}} \mathcal{I}_{\mathrm{tr}(F^3), n}^{(1), \text{deg-4}} \nonumber \\
	= & -\frac{1}{2}\left( \mathcal{I}_{\mathrm{tr}(F^n), n}^{(1),\text{fin}, \text{deg-2}} \right)^2 +  2\zeta_2 \mathcal{I}_{\mathrm{tr}(F^n), n}^{(1),\text{fin}, \text{deg-2}} - \zeta_3 \sum_{i=1}^n \log \left( - s_{i,i+1}\right) - \frac{11}{8} n \zeta_4 + \mathcal{O}(\epsilon) \,, \nonumber
\end{align}
where we factorize out the tree-level form factor from the maximal transcendental part of the loop-level form factor, and we give $I_{n,g}^{(1), \text{deg-4}}$ and $I_{n,g}^{(2), \text{deg-4}}$ as follows
\begin{align}
	I_{n, g}^{(1), \text{deg-4}}(\epsilon) = & -\frac{e^{\gamma_\text{E} \epsilon}}{\epsilon^2 \Gamma(1-\epsilon)} \sum_{i=1}^n \left( - s_{i,i+1} \right)^{-\epsilon} \,, \\
	I_{n, g}^{(2), \text{deg-4}}(\epsilon) = & -\frac{1}{2} \left( I_{n,g}^{(1), \text{deg-2}} \right)^2 - 2 \zeta_2 \frac{e^{\gamma_\text{E}} \Gamma(1-2\epsilon)}{\Gamma(1-\epsilon)} I_{n,g}^{(1), \text{deg-2}}(2\epsilon) + \frac{n \zeta_3}{2}  \frac{e^{-\gamma_\text{E}\epsilon}}{\epsilon \Gamma(1-\epsilon)} \,. \nonumber
\end{align}
%

%%%%%%%%%%%%%%%%%%%%%%%%%%%%%
\section{Two-loop symbol letters and collinear limit}
\label{app:letterandcollinear}

In this appendix, we present the symbol letters that appeared in the two-loop four-point form factor and also give their collinear limits.

%%%%%%%%%%%%%%%
\subsection{Symbol letters}
\label{app:letters}

The letters that appear in the symbol of the master integrals are given in \cite{Abreu:2020jxa}.
We can separate the letters that appear in the remainder into two parts. The first part are simple ratio variables defined in \eqref{eq:defuij} and linear combinations of them:
\begin{align}
	\label{eq:letters1}
	& u_{12} \,, \ u_{13} \,, \ u_{14} \,, \ u_{23} \,, \ u_{24} \,, \ u_{34} \,, \\ 
	& u_{123} \,, \ u_{124} \,, \ u_{134} \,, \ u_{234} \,, \nonumber \\
	& u_{123}-u_{12} \,, \ u_{123}-u_{23} \,, \ u_{124}-u_{12} \,, \ u_{124}-u_{14} \,, \nonumber \\
	& u_{134}-u_{13} \,, \ u_{134}-u_{34} \,, \ u_{234}-u_{23} \,, \ u_{234}-u_{34} \,, \nonumber \\
	& 1-u_{123} \,, \ 1-u_{124} \,, \ 1-u_{134} \,, \ 1-u_{234} \,. \nonumber
\end{align}
There are from $W_1 \sim W_{21}$ in the notation of  \cite{Abreu:2020jxa}.

The second part are the letters that contain square roots. 
Following the notation of Section~\ref{sec:collinearFF}, we define the variables
\begin{align}
	x_{ijkl}^{\pm} & = \frac{q^2+s_{ij}-s_{kl}\pm\sqrt{\Delta_{3,ijkl}}}{2 s_{ij}} \,, \\
	y_{ijkl}^{\pm} & = \frac{\mathrm{tr}_{\pm}(ijkl)}{2 s_{ij} s_{il}} = \frac{s_{ij}  s_{kl} -s_{ik} s_{jl} +s_{il} s_{jk} \pm P(ijkl)\sqrt{\Delta_5}}{2 s_{ij} s_{il}} \,, \nonumber \\
	z_{ijkl}^{\pm\pm} & = 1+y_{ijkl}^{\pm}-x_{lijk}^{\pm}
	=  \frac{s_{il}s_{jk} -(s_{ij}+s_{ik})(s_{ij}+s_{jl}) \mp s_{ij} \sqrt{\Delta_{3,ijkl}} \pm P(ijkl) \sqrt{\Delta_5}}{2 s_{ij} s_{il}} \,, \nonumber
\end{align}
where
$\Delta_3$ appears in 3-massive triangle integral
\begin{equation}
	\Delta_{3,ijkl} = (q^2-s_{ij}-s_{kl})^2 -4s_{ij} s_{kl} \,,
\end{equation}
$P(ijkl)=\pm1$ is the number of inversions, and $\Delta_5$ is the Gram determinant
\begin{equation}
	\Delta_5 = \mathrm{tr}_5^2 = (s_{12} s_{34} -s_{13} s_{24} +s_{14} s_{23})^2 -4 s_{12} s_{23} s_{34} s_{14} \,.
\end{equation}
With these definition, other letters occur in remainder can be expressed in the following form:
\begin{align}
	&	X_1(p_i+p_j,p_k,p_l) = \frac{u_{ij} x_{ijkl}^{+}-u_{ijl}}{u_{ij} x_{ijkl}^{-}-u_{ijl}} \,, \qquad 
	X_2(p_i+p_j,p_k+p_l) = \frac{x_{ijkl}^{+}}{x_{ijkl}^{-}} \,, \\
	&	Y_0(p_i+p_j,p_k+p_l) = u_{ikl}u_{jkl}-u_{kl} \,, \qquad
	Y_1(p_i,p_j,p_k,p_l) = \frac{\mathrm{tr}_{+}(ijkl)}{\mathrm{tr}_{-}(ijkl)}=\frac{y_{ijkl}^{+}}{y_{ijkl}^{-}} \,, \nonumber \\
	&	Y_2(p_i,p_j,p_k,p_l)=\frac{y_{ijkl}^{+}+1}{y_{ijkl}^{-}+1} \,, \qquad
	\qquad\quad\ \  Z(p_i,p_j,p_k,p_l)=\frac{z_{ijkl}^{++}z_{ijkl}^{--}}{z_{ijkl}^{+-}z_{ijkl}^{-+}} \,. \nonumber
\end{align}
More explicitly, all the letters are
\begin{align}
	\label{eq:letters2}
	& X_1(p_1 + p_2, p_3, p_4) \,, \ X_1(p_2 + p_3, p_1, p_4) \,, \ X_1(p_1+ p_4, p_2, p_3) \,, \ X_1(p_3 + p_4, p_1, p_2) \,, \nonumber \\
	& X_2(p_1 + p_2, p_3 + p_4) \,, \ X_2(p_2 + p_3, p_1 + p_4) \,, \ X_2(p_1 + p_4, p_2 + p_3) \,, \ X_2(p_3 + p_4, p_1 + p_2) \,, \nonumber \\
	& Y_0(p_1 + p_2, p_3 + p_4) \,, \ Y_0(p_2 + p_3, p_1 + p_4) \,, \ Y_0(p_1 + p_4, p_2 + p_3) \,, \ Y_0(p_3 + p_4, p_1 + p_2) \,, \nonumber \\
	& Y_1(p_1, p_2, p_3, p_4) \,, \ Y_1(p_1, p_3, p_2, p_4) \nonumber\\
	& Y_2(p_1, p_3, p_2, p_4) \,, \ Y_2(p_3, p_1, p_2, p_4) \,, \ Y_2(p_1, p_3, p_4, p_2) \,, \ Y_2(p_3, p_1, p_4, p_2) \,, \nonumber \\
	& Z(p_1, p_2, p_3, p_4) \,, \ Z(p_3, p_2, p_1, p_4) \,.
\end{align}
To make connection to the notation of  \cite{Abreu:2020jxa}: $X_1$ comes from $W_{37} \sim W_{39}, W_{54}$, $X_2$ comes from $W_{33} \sim W_{36}$, $Y_0$ comes from $W_{22} \sim W_{24}, W_{51}$, $Y_1$ comes from $W_{40}$, $Y_2$ comes from $W_{42} \sim W_{45}$. Here we use the relation of $Y_2$ to simplify all $Y_2$ by introduce $Y_1(p_1, p_3, p_2, p_4)$.

Now all letters can be expressed by the variables:
\begin{align}
	& x_{1234}^\pm \,, \ x_{2314}^\pm \,, \ x_{1423}^\pm \,, \ x_{3412}^\pm \,, \\
	& y_{1234}^{\pm} \,, \ y_{1324}^{\pm} \,, \ y_{3124}^{\pm} \,, \ y_{1342}^{\pm} \,, \ y_{3142}^{\pm} \,.
\end{align}
All letters appear in pairs with exchanging $p_1$ and $p_3$ except $y^{\pm}_{1234}$, which appear only in $Y_1(p_1, p_2, p_3, p_4)$, $Z(p_1, p_2, p_3, p_4)$ and $Z(p_3, p_2, p_1, p_4)$.

The letters that appear in the master integrals but not in the remainder function are: $q^2$, $\sqrt{\Delta_{3,1234}}$, $\sqrt{\Delta_{3,1423}}$, and $\sqrt{\Delta_5}$.

The full symbol of the remainder can be expressed in terms of the letters \eqref{eq:letters1} and \eqref{eq:letters2}:
\begin{equation}
	\mathcal{S} \left( R_{\mathrm{tr}(\mathcal{O}), n}^{(2)} \right) =  \left( A \otimes B \otimes C \otimes D \right) + \ldots \,.
\end{equation}
Let us discuss a few properties of the Symbol. 
Integrability of the symbol is the following requirement
\begin{equation}
	\sum d W_i \wedge d W_{i+1} \left( W_1 \otimes \cdots \otimes W_{i-1} \otimes W_{i+2} \otimes \cdots W_n \right) = 0 \,,
\end{equation}
where $W_i$ is the letter of $i$-th-entry.
The entry conditions can be summarized as follows:
\begin{itemize}
	\item[1.] The first-entry condition is required by physics, which is means the first letters of the symbol must be physical poles $u_{12},u_{14},u_{23},u_{34},u_{123},u_{124},u_{134},u_{234}$.
	\item[2.] The second-entry and last-entry letters can appear at one loop, actually the second-entry is free from $y_{ijkl}^{\pm}$ and $z_{ijkl}^{\pm}$, and the last-entry is free from $z_{ijkl}^{\pm}$.
	\item[3.] For the third-entry, we find that all letters appear.
\end{itemize}

%%%%%%%%%%%%%%%%%%%%%%%
\subsection{Collinear limit of the letters}
\label{app:letterCL}

Below we give the collinear limit for various letters following  Section~\ref{sec:collinearFF}. 

For convenience, we introduce a new variable $t$ as:
\begin{equation}
	\tau = \frac{t-1}{t}\frac{s_{12}+s_{13}}{s_{12}+s_{23}} \,.
\end{equation}
One can check that $\langle 34 \rangle \propto \delta$, $[34] \propto \frac{\eta}{\delta}$. Keep the leading term in the collinear limit, 
the behavior of letters at the collinear limit as follows
\begin{align}
	& u_{12} \rightarrow u'_{12} \,, & u_{23} \rightarrow (1-t)u'_{23} \,, \\
	& u_{34} \rightarrow -\eta u'_{13}u'_{23} \,, & u_{14} \rightarrow t u'_{13} \,, \nonumber \\
	& u_{123} \rightarrow 1-t(u'_{13}+u'_{23}) \,, & u_{234} \rightarrow u'_{23} \,,  \nonumber \\
	& u_{341} \rightarrow u'_{13} \,, & u_{412} \rightarrow u'_{12}+t(u'_{13}+u'_{23}) \,, \nonumber \\
	& u_{13} \rightarrow (1-t)u'_{13}, & u_{24} \rightarrow t u'_{23} \,. \nonumber
\end{align}

The collinear limits of $x_{ijkl}^\pm$ and $y_{ijkl}^\pm$ are more subtle. First, $\sqrt{\Delta_{3,ijkl}}$ is hard to be expressed rationally with momentum twistor. The limit behaviors of $\Delta_{3,ijkl}$ are
\begin{align}
	\Delta_{3,1234} \rightarrow & (1-u'_{12})^2 \,, \\
	\Delta_{3,1423} \rightarrow & (1+\frac{u'_{13}}{1+t}-u'_{23})^2-\frac{4u'_{13}}{1+t} \,, \nonumber
\end{align}
where only $\sqrt{\Delta_{3,1234}}$ can turn to a rational function when $p_3 \parallel p_4$, which makes $x_{1234}^{\pm}$ and $x_{3412}^{\pm}$ free from square root. Additionally, the limit behavior of $x_{3412}^{\pm}$ is
\begin{equation}
	x_{3412}^{-} \rightarrow \frac{1}{1-u'_{12}}+\mathcal{O}(\eta) \,,
\end{equation}
where $u'_{12}<1$. And the following non-trivial relation is necessary
\begin{equation}
	\frac{X_1(p_i+p_j,p_k,p_l)X_1(p_k+p_l,p_i,p_j)}{X_2(p_i+p_j,p_k+p_l)X_2(p_k+p_l,p_i+p_j)}  \xlongrightarrow[\mbox{}]{\mbox{$p_j \parallel p_k$}}  1 \,.
\end{equation}
Some $y_{ijkl}^\pm$ will vanish in the limit:
\begin{align}
	& y_{1234}^+ \rightarrow \frac{(1-t)\delta}{t}\frac{(1-u'_{23})u'_{23}}{u'_{12}} \,, & y_{1234}^- \rightarrow -\frac{\eta}{\delta}\frac{u'_{23}}{1-u'_{23}} \,, \\
	& y_{1324}^+ \rightarrow \frac{u'_{23}}{u'_{13}} \,, & y_{1324}^- \rightarrow \frac{u'_{23}}{u'_{13}} \,, \nonumber \\
	& y_{3124}^+ \rightarrow -\frac{t}{(1-t)\delta}\frac{u'_{12}}{u'_{13}(1-u'_{23})} \,, & y_{3124}^- \rightarrow \frac{\delta}{\eta}\frac{1-u'_{23}}{u'_{13}} \,, \nonumber \\
	& y_{1342}^+ \rightarrow \frac{t\eta}{(1-t)\delta}\frac{u'_{23}}{1-u'_{23}} \,, & y_{1342}^- \rightarrow -\delta\frac{(1-u'_{23})u'_{23}}{u'_{12}} \,, \nonumber \\
	& y_{3142}^+ \rightarrow \frac{t}{1-t} \,, & y_{3142}^- \rightarrow \frac{t}{1-t} \,. \nonumber
\end{align}

%%%%%%%%%%%%%%%%%%%%%%%%%%
\section{Constraints from higher order of $\epsilon$-expansion}
\label{app:HigherOrder}

In Section~\ref{sec:bootstrapFF4pt2loop}, we point out that the spurious poles should cancel at the high $\mathcal{O}(\epsilon)$ orders of the form factor results, and in particular, the $\epsilon^1$-order can provide useful new constraints for the two-loop four-point form factor.
In this appendix, we show that once the cancellation of spurious poles is satisfied at $\epsilon^a$-order where $a$ is a finite number, it will hold for the higher order of $\epsilon$. 

Let's review our problem first, for example, the cancellation of the spurious pole $\langle 24 \rangle$ (see \eqref{eq:4pt2loopSP}) requires the following equation holds for any order of $\epsilon$-expansion at the symbol level 
\begin{equation}
	\label{eq:collG12}
	\mathcal{S}\left( \mathcal{G}_{1}^{(2)} - \mathcal{G}_{2}^{(2)} \right) \xrightarrow{ \langle 24 \rangle \rightarrow 0 } 0 \,,
\end{equation}
where the LHS can be expressed in terms of mater integrals as $\mathcal{G}_{1}^{(2)} - \mathcal{G}_{2}^{(2)} = \mathbf{c} \cdot \mathbf{I}^{(2)}$, in the formula $\mathbf{c}$ and $\mathbf{I}^{(2)}$ are the vectors of the coefficients and 221 master integrals. Then our goal is to solve the parameters in $\mathbf{c}$ or verify whether a given $\mathbf{c}$ satisfies \eqref{eq:collG12}. This can be done by using the iterative properties of master integrals as follows.

The canonical differential equation of the master integrals we used has been given in \cite{Abreu:2020jxa}
\begin{equation}
	d\mathbf{I}^{(2)} = \epsilon \sum_{i = 1}^{50} \left( \mathbf{A}_{i} \cdot \mathbf{I}^{(2)} \right) d\log(W_i) \,,
\end{equation}
where $W_i$ are the symbol letters, $\{\mathbf{A}_i\}$ is a set of $221 \times 221$ rational matrices, which gives the symbol of the master integrals as the following iterative formula by the definition of symbol
\begin{equation}
	\label{eq:iteration}
	\mathcal{S}(\mathbf{I}^{(2)}) = \epsilon \sum_{i = 1}^{50} \mathbf{A}_{i} \cdot \left( \mathcal{S}(\mathbf{I}^{(2)}) \otimes W_i \right) \,.
\end{equation}
Most notably, $\mathbf{I}^{(2)}$ starts from of the $\epsilon^{-4}$-order
\begin{equation}
	\mathbf{I}^{(2)} = \mathbf{I}_0^{(2)} \epsilon^{-4} + \sum_{k=-3}^\infty \left( \mathbf{I}^{(2)} \big|_{\text{$\epsilon^{k}$-order}} \right) \epsilon^{k} \,,
\end{equation}
where $\mathbf{I}_0^{(2)}$ contains only rational number. By using \eqref{eq:iteration}, each order of $\mathbf{I}^{(2)}$ can be obtained iteratively. 

Consider first the $\epsilon^{-3}$-order of $\mathbf{I}^{(2)}$, one has
\begin{equation}
	\mathcal{S}(\mathbf{I}^{(2)}) \big|_{\text{$\epsilon^{-3}$-order}} = \sum_{i = 1}^{50} \mathbf{A}_{i} \cdot \mathbf{I}_0^{(2)} \left( W_i \right) \,.
\end{equation}
The cancellation of spurious pole \eqref{eq:collG12} at this order requires
\begin{equation}
	\mathcal{S}(\mathcal{G}_1^{(2)}-\mathcal{G}_2^{(2)}) \big|_{\text{$\epsilon^{-3}$-order}} \xrightarrow{ \langle 24 \rangle \rightarrow 0 } \sum_{i=1}^{50} \mathbf{c} \cdot \mathbf{A}_{i} \cdot \mathbf{I}_0^{(2)} \left(  W_i \big|_{\langle 24 \rangle \rightarrow 0} \right) = \sum_{i=1}^{n'} \mathbf{c} \cdot \mathbf{A}_{i}' \cdot \mathbf{I}_0^{(2)} \left(  W_i' \right) = 0 \,.
\end{equation}
Here it should be noticed that $\{W_i\}$ are no longer linearly independent when $\langle 24 \rangle \rightarrow 0$, thus they need to be expressed as another set of linearly independent basis $\{W_i'\}$ with $n' < 50$, meanwhile, the matrices $\{\mathbf{A}_{i}\}$ also transform linearly to $\{\mathbf{A}_{i}'\}$ correspondingly.
Then, it requires the $n'$ equations $\mathbf{c} \cdot \mathbf{A}_{i}' \cdot \mathbf{I}_0^{(2)} = 0$ where $i = 1,\ldots,n'$ at $\epsilon^{-3}$-order. 

Similarly, the $\epsilon^{-2}$-order of $\mathcal{S}(\mathcal{G}_1^{(2)}-\mathcal{G}_2^{(2)})=0$ is
\begin{equation}
	\mathcal{S}(\mathcal{G}_1^{(2)}-\mathcal{G}_2^{(2)}) \big|_{\text{$\epsilon^{-2}$-order}} \xrightarrow{ \langle 24 \rangle \rightarrow 0 } \sum_{i,j = 1}^{n'} \mathbf{c} \cdot \mathbf{A}_{i}' \cdot \mathbf{A}_{j}' \cdot \mathbf{I}_0^{(2)} \left( W_j' \otimes W_i' \right) \,,
\end{equation}
or equivalently the more compact $(n')^2$ equations $\mathbf{c} \cdot \mathbf{A}_{j}' \cdot \mathbf{A}_{i}' \cdot \mathbf{I}_0^{(2)} = 0$ where $i,j = 1,\ldots,n'$.

More generally, we can abbreviate the equations of the $\epsilon^{m-4}$-order ($m$ can be any positive integer) as
\begin{equation}
	\label{eq:InfiniteEquations}
	\mathbf{C}_{i_1,\cdots,i_m} \cdot \mathbf{I}_0^{(2)} = 0 \,,
\end{equation}
where $i_x=1,..,n'$, and $\mathbf{C}_{i_1,\cdots,i_m}$ can be given iteratively as
\begin{equation}
	\mathbf{C}_{i} = \mathbf{c} \cdot \mathbf{A}_i'\,, \qquad \mathbf{C}_{i_1,\cdots,i_{m-1},i_{m}} = \mathbf{C}_{i_1,\cdots,i_{m-1}} \cdot \mathbf{A}_{i_m}' \,.
\end{equation}

Obviously, we do not need to solve the equations \eqref{eq:InfiniteEquations} for all $m$, because the vectors live in $221$-dimensional linear space. This means that there are no more than 221 linearly independent $\{\mathbf{C}_{i_1,\cdots,i_m}\}$ (or independent equations), although the index $m$ can turn to infinite.
Therefore, for a given $\mathbf{c}$, we can generate the finite dimensional linear space of $\{\mathbf{C}_{i_1,\cdots,i_m}\}$ using a finite number of iterations. For example, five steps of iterations are sufficient for the problem of the cancellation of the spurious pole $\langle 24 \rangle$. More specifically, if we find a certain $\mathbf{c}$ makes the equation \eqref{eq:InfiniteEquations} holds for $m \leqslant 5$, then any $\mathbf{C}_{i_1,\cdots,i_m}$ generated by such $\mathbf{c}$ for $m>5$ always can be expressed with $\mathbf{C}_{i_1}$, $\mathbf{C}_{i_1,i_2}$, $\mathbf{C}_{i_1,i_2,i_3}$, $\mathbf{C}_{i_1,i_2,i_3,i_4}$ and $\mathbf{C}_{i_1,i_2,i_3,i_4,i_5}$, thus will not provide new independent equations. In other words, if the cancellation is satisfied at $\epsilon^1$-order, it will also hold at the higher order of $\epsilon$ at symbol level. We also comment here that the solution family of $\mathbf{c}$ require $\{\mathbf{C}_{i_1,\cdots,i_m}\}$ to belong to the orthogonal complement space of $\mathbf{I}_0^{(2)}$.

%%%%%%%%%%%%%%%%%%%%%%%%%%
\section{Building-blocks for 2-loop four-point form factor of ${\rm tr}(F^3)$}
\label{app:BuildingBlocks}

In this appendix, we give the expressions of the building blocks in Section~\ref{sec:buildingblocksFF4pt} as follows.

First, for the building blocks $\tilde{G}_{2,\beta}$, where $\beta = 1, \ldots, 7$ in \eqref{eq:BuildingBlock}:
\begin{align*}
	\tilde{G}_{2,1} = & -\frac{1}{4} I_{\text{dBub}}(p_1,p_2;p_1,p_2) -\frac{1}{4} I_{\text{dBub}}(p_1,p_2;p_1,p_2,p_4) -\frac{1}{4} I_{\text{dBub}}(p_2,p_3;p_2,p_3) \\
	& -\frac{1}{4} I_{\text{dBub}}(p_2,p_3;p_2,p_3,p_4) + I_{\text{BoxBub}}(p_2,p_1,p_4) + I_{\text{BoxBub}}(p_2,p_3,p_4) \\
	& +\frac{1}{4} I_{\text{dBox1a}}(p_2,p_1,p_4) + \frac{1}{4} I_{\text{dBox1a}}(p_2,p_3,p_4) +\frac{1}{4} I_{\text{dBox1b}}(p_2,p_1,p_4) +\frac{1}{4} I_{\text{dBox1b}}(p_2,p_3,p_4) \\
	& -\frac{5}{2} I_{\text{Sun}}(p_1,p_2) +\frac{1}{4} I_{\text{Sun}}(p_1,p_4) -\frac{5}{2} I_{\text{Sun}}(p_2,p_3) +\frac{1}{4} I_{\text{Sun}}(p_3,p_4) -\frac{1}{4} I_{\text{Sun}}(p_1,p_2,p_4) \\
	& -\frac{1}{4} I_{\text{Sun}}(p_2,p_3,p_4) +\frac{1}{2} I_{\text{TBox0}}(p_2,p_1,p_4)+\frac{1}{2} I_{\text{TBox0}}(p_2,p_3,p_4)+\frac{5}{8} I_{\text{TBub0}}(p_1,p_2) \\
	& +\frac{5}{8} I_{\text{TBub0}}(p_2,p_3) +\frac{3}{16} I_{\text{TBub1}}(p_1,p_2,p_4) -\frac{21}{16} I_{\text{TBub1}}(p_1,p_4,p_2) +\frac{3}{16} I_{\text{TBub1}}(p_2,p_3,p_4) \\
	& -\frac{21}{16} I_{\text{TBub1}}(p_3,p_4,p_2)-\frac{1}{8} I_{\text{TBub2}}(p_4,p_1,p_2)-\frac{1}{8} I_{\text{TBub2}}(p_4,p_2,p_3)-\frac{1}{4} I_{\text{TT0}}(p_4,p_1,p_2) \\
	& -\frac{1}{4} I_{\text{TT0}}(p_4,p_2,p_3) +\frac{1}{16} I_{\text{TT1}}(p_2,p_1,p_4) +\frac{1}{16} I_{\text{TT1}}(p_2,p_3,p_4) -\frac{1}{8} I_{\text{TT1a}}(p_2,p_1,p_4) \\
	& -\frac{1}{8} I_{\text{TT1a}}(p_2,p_3,p_4)+\frac{3}{4} I_{\text{TT2}}(p_2,p_1,p_4)+\frac{3}{4} I_{\text{TT2}}(p_2,p_3,p_4) \, , \\
	%%%%%%%%
	\tilde{G}_{2,2} = & \tilde{G}_{9} \big|_{p_1 \leftrightarrow p_3} \, , \\
	%%%%%%%%
	\tilde{G}_{2,3} = & -I_{\text{dBub}}(p_1,p_2;p_1,p_2,p_4) -I_{\text{dBub}}(p_2,p_3;p_2,p_3,p_4) +I_{\text{BubBox}}(p_2,p_1,p_4) \\
	& +I_{\text{BubBox}}(p_2,p_3,p_4) -I_{\text{Sun}}(p_1,p_2) -I_{\text{Sun}}(p_2,p_3) -I_{\text{Sun}}(p_1,p_2,p_4) -I_{\text{Sun}}(p_2,p_3,p_4) \\
	& +I_{\text{TBox0}}(p_2,p_1,p_4) +I_{\text{TBox0}}(p_2,p_3,p_4) +\frac{3}{4} I_{\text{TBub1}}(p_1,p_2,p_4) -\frac{5}{4} I_{\text{TBub1}}(p_1,p_4,p_2) \\
	& +\frac{3}{4} I_{\text{TBub1}}(p_2,p_3,p_4) -\frac{5}{4} I_{\text{TBub1}}(p_3,p_4,p_2) -I_{\text{TT0}}(p_4,p_1,p_2) -I_{\text{TT0}}(p_4,p_2,p_3) \\
	& +\frac{1}{4} I_{\text{TT1}}(p_2,p_1,p_4)+\frac{1}{4} I_{\text{TT1}}(p_2,p_3,p_4)-\frac{1}{2} I_{\text{TT1a}}(p_2,p_1,p_4)-\frac{1}{2} I_{\text{TT1a}}(p_2,p_3,p_4) \, , \\
	%%%%%%%%
	\tilde{G}_{2,4} = & \tilde{G}_{11} \big|_{p_1 \leftrightarrow p_3} \, , \\
	%%%%%%%%
	\tilde{G}_{2,5} = & -4 I_{\text{dBub}}(p_1,p_2;p_1,p_2,p_4) -4 I_{\text{dBub}}(p_1,p_4;p_1,p_2,p_4) -4 I_{\text{dBub}}(p_2,p_3;p_2,p_3,p_4) \\
	& -4 I_{\text{dBub}}(p_3,p_4;p_2,p_3,p_4) +2 I_{\text{dBub}}(p_1,p_2,p_4;p_1,p_2,p_4) +2 I_{\text{dBub}}(p_2,p_3,p_4;p_2,p_3,p_4) \\
	& +I_{\text{BubBox0}}(p_2,p_1,p_4) +I_{\text{BubBox0}}(p_2,p_3,p_4) -4 I_{\text{Sun}}(p_1,p_2) -4 I_{\text{Sun}}(p_1,p_4) -4 I_{\text{Sun}}(p_2,p_3) \\
	& -4 I_{\text{Sun}}(p_3,p_4) -4 I_{\text{Sun}}(p_1,p_2,p_4)-4 I_{\text{Sun}}(p_2,p_3,p_4) +2 I_{\text{TBox0}}(p_2,p_1,p_4)  \\
	& +2 I_{\text{TBox0}}(p_2,p_3,p_4) +2 I_{\text{TBox0}}(p_4,p_1,p_2) +2 I_{\text{TBox0}}(p_4,p_3,p_2) -\frac{1}{2} I_{\text{TBub1}}(p_1,p_2,p_4) \\
	& -\frac{1}{2} I_{\text{TBub1}}(p_1,p_4,p_2) -\frac{1}{2} I_{\text{TBub1}}(p_2,p_3,p_4)-\frac{1}{2} I_{\text{TBub1}}(p_3,p_4,p_2)+2 I_{\text{TBub2}}(p_2,p_1,p_4) \\
	& +2 I_{\text{TBub2}}(p_2,p_3,p_4) +2 I_{\text{TBub2}}(p_4,p_1,p_2) +2 I_{\text{TBub2}}(p_4,p_2,p_3) -2 I_{\text{TT0}}(p_2,p_1,p_4) \\
	& -2 I_{\text{TT0}}(p_2,p_3,p_4) -2 I_{\text{TT0}}(p_4,p_1,p_2) -2 I_{\text{TT0}}(p_4,p_2,p_3) +\frac{1}{2} I_{\text{TT1}}(p_2,p_1,p_4) \\
	& +\frac{1}{2} I_{\text{TT1}}(p_2,p_3,p_4) -3 I_{\text{TT1a}}(p_2,p_1,p_4) -3 I_{\text{TT1a}}(p_2,p_3,p_4) \,, \\
	%%%%%%%%
	\tilde{G}_{2,6} = & 2 I_{\text{BubBox}}(p_1,p_2,p_3) +2 I_{\text{BubBox}}(p_3,p_2,p_1) +I_{\text{BubBox3}}(p_1,p_2,p_3,p_4) +I_{\text{BubBox3}}(p_3,p_2,p_1,p_4) \\
	& -\frac{3}{2} I_{\text{BubBox3}}(p_4,p_1,p_2,p_3) -\frac{3}{2} I_{\text{BubBox3}}(p_4,p_3,p_1,p_2) -\frac{1}{2} I_{\text{dBox2a}}(p_1,p_2,p_3,p_4) \\
	& -\frac{1}{2} I_{\text{dBox2a}}(p_3,p_2,p_1,p_4) +\frac{1}{2} I_{\text{dBox2b}}(p_1,p_2,p_3,p_4) +\frac{1}{2} I_{\text{dBox2b}}(p_3,p_2,p_1,p_4) \\
	& +I_{\text{dBox2c}}(p_1,p_2,p_3,p_4)+I_{\text{dBox2c}}(p_3,p_2,p_1,p_4)+2 I_{\text{dBub}}(p_1,p_4;p_1,p_2,p_4) \\
	& +2 I_{\text{dBub}}(p_3,p_4;p_2,p_3,p_4)+\frac{11}{2} I_{\text{Sun}}(p_1,p_2)+3 I_{\text{Sun}}(p_1,p_4)+\frac{11}{2} I_{\text{Sun}}(p_2,p_3) \\
	& +3 I_{\text{Sun}}(p_3,p_4) +I_{\text{Sun}}(p_1,p_2,p_3) +\frac{5}{2} I_{\text{Sun}}(p_1,p_2,p_4) +\frac{5}{2} I_{\text{Sun}}(p_2,p_3,p_4) -I_{\text{Sun}}(p_1,p_2,p_3,p_4) \\
	& -2 I_{\text{TBox0}}(p_4,p_1,p_2) -2 I_{\text{TBox0}}(p_4,p_3,p_2) -\frac{1}{2} I_{\text{TBox1}}(p_1,p_2,p_3,p_4) +I_{\text{TBox1}}(p_1,p_4,p_2,p_3) \\
	& -\frac{1}{2} I_{\text{TBox1}}(p_2,p_3,p_1,p_4) +I_{\text{TBox1}}(p_3,p_4,p_2,p_1) -\frac{1}{2} I_{\text{TBox2b}}(p_1,p_2,p_3,p_4) \\
	& -\frac{1}{2} I_{\text{TBox2b}}(p_3,p_2,p_1,p_4) +\frac{1}{2} I_{\text{TBox2b}}(p_4,p_1,p_2,p_3) +\frac{1}{2} I_{\text{TBox2b}}(p_4,p_3,p_2,p_1) \\
	& +\frac{5}{2} I_{\text{TBub1}}(p_1,p_2,p_4) -\frac{3}{2} I_{\text{TBub1}}(p_1,p_4,p_2) +\frac{5}{2} I_{\text{TBub1}}(p_2,p_3,p_4) -\frac{3}{2} I_{\text{TBub1}}(p_3,p_4,p_2) \\
	& -\frac{3}{4} I_{\text{TBub2}}(p_1,p_2,p_3) -2 I_{\text{TBub2}}(p_2,p_1,p_4) -2 I_{\text{TBub2}}(p_2,p_3,p_4) -\frac{3}{4} I_{\text{TBub2}}(p_3,p_1,p_2) \\
	& +\frac{1}{2} I_{\text{TBub3}}(p_1,p_2,p_3,p_4)-\frac{1}{2} I_{\text{TBub3}}(p_1,p_4,p_2,p_3)+\frac{1}{2} I_{\text{TBub3}}(p_2,p_3,p_1,p_4) \\
	& -\frac{1}{2} I_{\text{TBub3}}(p_3,p_4,p_1,p_2) +2 I_{\text{TT0}}(p_2,p_1,p_4) +2 I_{\text{TT0}}(p_2,p_3,p_4) -\frac{1}{2} I_{\text{TT1}}(p_2,p_1,p_4) \\
	& -\frac{1}{2} I_{\text{TT1}}(p_2,p_3,p_4)+I_{\text{TT1a}}(p_2,p_1,p_4)+I_{\text{TT1a}}(p_2,p_3,p_4)-3 I_{\text{TT2}}(p_1,p_2,p_3) \\
	& -\frac{3}{2} I_{\text{TT2}}(p_2,p_1,p_4) -\frac{3}{2} I_{\text{TT2}}(p_2,p_3,p_4) -\frac{1}{12} I_{\text{TT2}}(p_1,p_2,p_3,p_4) -\frac{1}{12} I_{\text{TT2}}(p_3,p_1,p_2,p_4) \\
	& +\frac{1}{2} I_{\text{TT3a}}(p_4,p_1,p_2,p_3) +\frac{1}{2} I_{\text{TT3a}}(p_4,p_3,p_1,p_2) +\frac{1}{4} I_{\text{TT3b}}(p_1,p_2,p_3,p_4) \\
	& +\frac{1}{4} I_{\text{TT3b}}(p_3,p_2,p_1,p_4)-\frac{1}{4} I_{\text{TT3b}}(p_4,p_1,p_2,p_3)-\frac{1}{4} I_{\text{TT3b}}(p_4,p_3,p_1,p_2) \, , \\
	%%%%%%%%
	\tilde{G}_{2,7} = & 2 I_{\text{dBub}}(p_1,p_2;p_1,p_2,p_4)+2 I_{\text{dBub}}(p_2,p_3;p_2,p_3,p_4) +2 I_{\text{BubBox}}(p_1,p_4,p_3) \\
	& +2 I_{\text{BubBox}}(p_3,p_4,p_1) -\frac{1}{2} I_{\text{BubBox3}}(p_1,p_4,p_2,p_3) -\frac{1}{2} I_{\text{BubBox3}}(p_3,p_4,p_1,p_2)\\
	& -\frac{1}{2} I_{\text{dBox2a}}(p_1,p_4,p_3,p_2)-\frac{1}{2} I_{\text{dBox2a}}(p_2,p_1,p_4,p_3) +\frac{1}{2} I_{\text{dBox2b}}(p_2,p_1,p_4,p_3) \\
	& +\frac{1}{2} I_{\text{dBox2b}}(p_2,p_3,p_4,p_1) +I_{\text{dBox2c}}(p_1,p_4,p_3,p_2) +I_{\text{dBox2c}}(p_2,p_1,p_4,p_3) +2 I_{\text{Sun}}(p_1,p_2) \\
	& +\frac{13}{2} I_{\text{Sun}}(p_1,p_4)+2 I_{\text{Sun}}(p_2,p_3)+\frac{13}{2} I_{\text{Sun}}(p_3,p_4)+\frac{5}{2} I_{\text{Sun}}(p_1,p_2,p_4)+I_{\text{Sun}}(p_1,p_3,p_4) \\
	& +\frac{5}{2} I_{\text{Sun}}(p_2,p_3,p_4) -I_{\text{Sun}}(p_1,p_2,p_3,p_4)-2 I_{\text{TBox0}}(p_2,p_1,p_4) -2 I_{\text{TBox0}}(p_2,p_3,p_4) \\
	& +\frac{1}{2} I_{\text{TBox1}}(p_1,p_2,p_4,p_3) +\frac{1}{2} I_{\text{TBox1}}(p_2,p_3,p_4,p_1) -\frac{1}{2} I_{\text{TBox2b}}(p_1,p_4,p_3,p_2) \\
	& +\frac{1}{2} I_{\text{TBox2b}}(p_2,p_1,p_4,p_3) +\frac{1}{2} I_{\text{TBox2b}}(p_2,p_3,p_4,p_1) -\frac{1}{2} I_{\text{TBox2b}}(p_3,p_4,p_1,p_2) \\
	& -\frac{3}{2} I_{\text{TBub1}}(p_1,p_2,p_4) +\frac{5}{2} I_{\text{TBub1}}(p_1,p_4,p_2) -\frac{3}{2} I_{\text{TBub1}}(p_2,p_3,p_4) +\frac{5}{2} I_{\text{TBub1}}(p_3,p_4,p_2) \\
	& -\frac{3}{4} I_{\text{TBub2}}(p_1,p_3,p_4) -\frac{3}{4} I_{\text{TBub2}}(p_3,p_1,p_4) -2 I_{\text{TBub2}}(p_4,p_1,p_2) -2 I_{\text{TBub2}}(p_4,p_2,p_3) \\
	& +2 I_{\text{TT0}}(p_4,p_1,p_2) +2 I_{\text{TT0}}(p_4,p_2,p_3) -\frac{1}{2} I_{\text{TT1}}(p_2,p_1,p_4) -\frac{1}{2} I_{\text{TT1}}(p_2,p_3,p_4) \\
	& +I_{\text{TT1a}}(p_2,p_1,p_4)+I_{\text{TT1a}}(p_2,p_3,p_4) -3 I_{\text{TT2}}(p_1,p_4,p_3)-\frac{3}{2} I_{\text{TT2}}(p_2,p_1,p_4) \\
	& -\frac{3}{2} I_{\text{TT2}}(p_2,p_3,p_4) -\frac{1}{12} I_{\text{TT2}}(p_1,p_3,p_4,p_2) -\frac{1}{12} I_{\text{TT2}}(p_2,p_1,p_4,p_3)+\frac{1}{2} I_{\text{TT3a}}(p_2,p_1,p_3,p_4) \\
	& +\frac{1}{2} I_{\text{TT3a}}(p_2,p_3,p_1,p_4) \,.
\end{align*}

For the building blocks $\tilde{G}_{3,\gamma}$, where $\gamma = 1, \ldots, 5$ in \eqref{eq:BuildingBlock}:
\begin{align*}
	\tilde{G}_{3,1} = & \frac{3}{4} I_{\text{dBub}}(p_3,p_4;p_3,p_4) +I_{\text{BoxBub}}(p_3,p_4,p_1) -\frac{9}{4} I_{\text{BubBox}}(p_3,p_4,p_1) -\frac{1}{2} I_{\text{Sun}}(p_1,p_3,p_4) \\
	& -\frac{3}{4} I_{\text{Sun}}(p_1,p_4) +\frac{11}{4} I_{\text{Sun}}(p_3,p_4)  +\frac{13}{8} I_{\text{TBub0}}(p_3,p_4) -\frac{1}{2} I_{\text{TBub1}}(p_1,p_4,p_3)  \\
	& +\frac{9}{8} I_{\text{TBub2}}(p_1,p_3,p_4) -\frac{3}{4} I_{\text{TBox0}}(p_3,p_4,p_1) -\frac{1}{2} I_{\text{TT1}}(p_1,p_4,p_3) +\frac{7}{4} I_{\text{TT2}}(p_1,p_4,p_3) \\
	& +\frac{1}{4} I_{\text{dBox1a}}(p_3,p_4,p_1) +\frac{3}{4} I_{\text{dBox1b}}(p_3,p_4,p_1)  \, , \\
	%%%%%%%%
	\tilde{G}_{3,2} = & \frac{3}{4} I_{\text{dBub}}(p_1,p_2;p_1,p_2) -\frac{9}{4} I_{\text{dBub}}(p_1,p_2;p_1,p_2,p_4) +I_{\text{BoxBub}}(p_2,p_1,p_4)  \\
	& +\frac{1}{4} I_{\text{dBox1a}}(p_2,p_1,p_4) +\frac{3}{4} I_{\text{dBox1b}}(p_2,p_1,p_4) +\frac{1}{2} I_{\text{Sun}}(p_1,p_2) -\frac{3}{4} I_{\text{Sun}}(p_1,p_4) \\
	& -\frac{11}{4} I_{\text{Sun}}(p_1,p_2,p_4) +\frac{3}{2} I_{\text{TBox0}}(p_2,p_1,p_4) +\frac{13}{8} I_{\text{TBub0}}(p_1,p_2) +\frac{27}{16} I_{\text{TBub1}}(p_1,p_2,p_4) \\
	& -\frac{53}{16} I_{\text{TBub1}}(p_1,p_4,p_2) +\frac{9}{8} I_{\text{TBub2}}(p_4,p_1,p_2) +\frac{9}{4} I_{\text{TT0}}(p_4,p_1,p_2) +\frac{1}{16} I_{\text{TT1}}(p_2,p_1,p_4) \\
	& -\frac{9}{8} I_{\text{TT1a}}(p_2,p_1,p_4) +\frac{7}{4} I_{\text{TT2}}(p_2,p_1,p_4) \, , \\
	%%%%%%%%
	\tilde{G}_{3,3} = & -I_{\text{dBub}}(p_3,p_4;p_1,p_3,p_4) +I_{\text{BubBox}}(p_3,p_4,p_1) -I_{\text{Sun}}(p_3,p_4) -I_{\text{Sun}}(p_1,p_3,p_4) \\
	& +I_{\text{TBox0}}(p_3,p_4,p_1) -\frac{5}{4} I_{\text{TBub1}}(p_1,p_4,p_3) +\frac{3}{4} I_{\text{TBub1}}(p_3,p_4,p_1) +I_{\text{TT0}}(p_1,p_3,p_4) \\
	& +\frac{1}{4} I_{\text{TT1}}(p_1,p_4,p_3) -\frac{1}{2} I_{\text{TT1a}}(p_1,p_4,p_3) \, , \\
	%%%%%%%%
	\tilde{G}_{3,4} = & -I_{\text{dBub}}(p_1,p_2;p_1,p_2,p_4) +I_{\text{BubBox}}(p_2,p_1,p_4) -I_{\text{Sun}}(p_1,p_2) -I_{\text{Sun}}(p_1,p_2,p_4) \\
	& +I_{\text{TBox0}}(p_2,p_1,p_4)+\frac{3}{4} I_{\text{TBub1}}(p_1,p_2,p_4)-\frac{5}{4} I_{\text{TBub1}}(p_1,p_4,p_2)+I_{\text{TT0}}(p_4,p_1,p_2) \\
	& +\frac{1}{4} I_{\text{TT1}}(p_2,p_1,p_4)-\frac{1}{2} I_{\text{TT1a}}(p_2,p_1,p_4) \, , \\
	%%%%%%%%
	\tilde{G}_{3,5} = & 2 I_{\text{dBub}}(p_1,p_2;p_3,p_4) +I_{\text{BubBox}}(p_1,p_2,p_3) +I_{\text{BubBox}}(p_4,p_3,p_2) -4 I_{\text{BubBox3}}(p_1,p_2,p_3,p_4) \\
	& +\frac{1}{2} I_{\text{BubBox3}}(p_4,p_3,p_1,p_2) +\frac{1}{2} I_{\text{dBox2a}}(p_1,p_2,p_3,p_4) -\frac{3}{2} I_{\text{dBox2b}}(p_1,p_2,p_3,p_4) \\
	& -I_{\text{dBox2c}}(p_1,p_2,p_3,p_4) +2 I_{\text{Sun}}(p_1,p_2)-3 I_{\text{Sun}}(p_2,p_3)-I_{\text{Sun}}(p_3,p_4)+\frac{3}{2} I_{\text{Sun}}(p_1,p_2,p_3) \\
	& +\frac{3}{2} I_{\text{Sun}}(p_2,p_3,p_4) -2 I_{\text{Sun}}(p_1,p_2,p_3,p_4) +\frac{1}{2} I_{\text{TBox1}}(p_1,p_2,p_3,p_4) -I_{\text{TBox1}}(p_3,p_4,p_2,p_1) \\
	& +I_{\text{TBox2a}}(p_1,p_2,p_3,p_4) +I_{\text{TBox2a}}(p_4,p_3,p_2,p_1) +\frac{3}{4} I_{\text{TBub2}}(p_2,p_3,p_4) +\frac{3}{4} I_{\text{TBub2}}(p_3,p_1,p_2) \\
	& -I_{\text{TBub3}}(p_1,p_2,p_3,p_4) +\frac{1}{2} I_{\text{TBub3}}(p_3,p_4,p_1,p_2) +\frac{1}{2} I_{\text{TT2}}(p_1,p_2,p_3) +\frac{1}{2} I_{\text{TT2}}(p_2,p_3,p_4) \\
	& -\frac{1}{4} I_{\text{TT2}}(p_1,p_2,p_3,p_4) +\frac{1}{2} I_{\text{TT3a}}(p_1,p_2,p_3,p_4) +\frac{1}{2} I_{\text{TT3a}}(p_4,p_3,p_1,p_2) \\
	& -\frac{1}{2} I_{\text{TT3b}}(p_1,p_2,p_3,p_4) +\frac{1}{4} I_{\text{TT3b}}(p_4,p_3,p_1,p_2) \, .
\end{align*}

In addition, the concrete expression of $\Delta^{(2), \mathcal{N}=4}_{\text{M.T.}}$ is
\begin{align}
	& \Delta^{(2), \mathcal{N}=4}_{\text{M.T.}} = B_4 \left( 4 \tilde{G}_{3,1}^{(2)} +2 \tilde{G}_{3,3}^{(2)} \right) + \left( p_1 \leftrightarrow p_3 \right) \\
	= & B_4 \left( 3 I_{\text{dBub}}^{(2)}(p_3,p_4;p_3,p_4) -2 I_{\text{dBub}}^{(2)}(p_3,p_4; p_1,p_3,p_4) -3 I_{\text{Sun}}^{(2)}(p_1,p_4) +9 I_{\text{Sun}}^{(2)}(p_3,p_4) \right. \nonumber \\
	& -4 I_{\text{Sun}}^{(2)}(p_1,p_3,p_4) +4 I_{\text{BoxBub}}^{(2)}(p_3,p_4,p_1) -7 I_{\text{BubBox}}^{(2)}(p_3,p_4,p_1) +\frac{13}{2} I_{\text{TBub0}}^{(2)}(p_3,p_4) \nonumber \\
	& -\frac{9}{2} I_{\text{TBub1}}^{(2)}(p_1,p_4,p_3) +\frac{3}{2} I_{\text{TBub1}}^{(2)}(p_3,p_4,p_1) +\frac{9}{2} I_{\text{TBub2}}^{(2)}(p_1,p_3,p_4) +2 I_{\text{TT0}}^{(2)}(p_1,p_3,p_4) \nonumber \\
	& -\frac{3}{2} I_{\text{TT1}}^{(2)}(p_1,p_4,p_3) - I_{\text{TT1a}}^{(2)}(p_1,p_4,p_3) +7 I_{\text{TT2}}^{(2)}(p_1,p_4,p_3) - I_{\text{TBox0}}^{(2)}(p_3,p_4,p_1) \nonumber \\
	& \left. +I_{\text{dBox1a}}^{(2)}(p_3,p_4,p_1) +3 I_{\text{dBox1b}}^{(2)}(p_3,p_4,p_1) \right) + \left( p_1 \leftrightarrow p_3 \right) \, . \nonumber
\end{align}

%%%%%%%%%%%%%%%%%%%%%%%%%%%%%%%%%%%%%
\section{Remainders and numerical results for the four-point form factor}
\label{app:fullFF}

The final form factor $\mathcal{F}_{\mathrm{tr}(F^3), 4}^{(2), \text{M.T.}}$ are obtained in terms of 221 master integrals, and their coefficients are given in the ancillary files.

To give some details about letters in each entry: (1) the first-entry contains 8 letters, corresponding to physical poles $u_{i,i+1}$ and $u_{i,i+1,i+2}$; (2) the second entry is free from $\{ X_1, Y_1, Y_2, Z, u_{13}, u_{24}\}$, and there are 28 letters;  (3) third entry contains all 42 letters except $u_{123}$; (4) the last-entry is free from $\{X_1, X_2, Z, u_{ijk}, 1- u_{ijk}, u_{12}-u_{123}, u_{23}-u_{123} \}$, and there are 22 letters. Explicit definitions of $X_i, Y_i, Z$ can be found in the supplemental material.
We provide the full expression of the remainder in the auxiliary file. 

The analytic expression can be obtained in terms of multiple polylogarithm functions with proper analytic continuation. 

%%%%%%%%%%%%%%%%  TABLE   %%%%%%%%%%%%%%%%%%%%%
\begin{table}[t]
	\centering
	\vskip .1 cm 
	\begin{tabular}{| c | c | c |} 
		\hline
		& \multicolumn{2}{|c}{$\mathcal{F}_{\mathrm{tr}(F^3), 4}^{(2), \text{M.T.}} / \mathcal{F}_{\mathrm{tr}(F^3), 4}^{(0)}$}  \\ \hline \hline
		$\epsilon^{-4}$ & \multicolumn{2}{c|}{$8$} \\ \hline
		$\epsilon^{-3}$ & \multicolumn{2}{c|}{$-31.395344718654$} \\ \hline
		$\epsilon^{-2}$ & \multicolumn{2}{c|}{$51.480019340206+0.0023559816692372 i$} \\ \hline
		$\epsilon^{-1}$ & \multicolumn{2}{c|}{$82.059735716289-1.9467291063519 i$}  \\ \hline
		$\epsilon^{ 0}$ & $107.65805575095-3.2229844658046 i$ & $93.198436450457+6.1054492577170 i$\\ \hline \hline
		$\mathcal{R}_{\mathrm{tr}(F^3), 4}^{(2), \text{M.T.}}$ &  $3.7743292318089+3.7743292318089 i$ & $-10.685290068684+12.528949688488 i$ \\ \hline  
		Theory & $\mathcal{N} = 4$ SYM & pure YM \\ \hline
	\end{tabular} 
	\caption{
		The numerical results of the two-loop four-point form factors in the two theories up to finite order,
		and the finite remainders are given in the last line,
		where the kinematics are chosen as:
		\{$q^2 = -13$, $s_{12} = -9/2$, $s_{23} = -67/10$, $s_{34} = -10$, $s_{14} = -17/2$, $s_{13} = 38/5$, $s_{24} = 91/10$ and $\mathrm{tr}_5 = 3 \sqrt{10195351}/100i$\}.
		\label{tab:num2loop}
	}
	
\end{table}
%%%%%%%%%%%%%%%%%%%%%%%%%%%%%%%%%%%%%%%%%%%

In Table~\ref{tab:num2loop}, we provide a high precision numerical data point, which is computed using the C++ library Ginac \cite{Bauer:2000cp}, and the Mathematica package DiffExp \cite{Hidding:2020ytt} can be also used. The method and condition of analytic continuation is discussed in \cite{Gehrmann:2002zr} (see also \cite{Duhr:2012fh}), and we can evaluate the form factor in both the Euclidean and non-Euclidean regions. We find the master integrals which numerator proper to $\sqrt{\Delta_3}$ and $\mathrm{tr}_5$ have the ambiguity of sign, then we add a sign factor to each of these integrals. We have performed cross-checks of the numerical results using FIESTA \cite{Smirnov:2015mct} and pySecDec \cite{Borowka:2017idc}.

%%%%%%%%%%%%%%%%%%%%%%%%%%%
\section{One-loop four-gluon amplitude: a counter-example of MTP}
\label{app:A4noMT}

We consider one-loop four-gluon amplitudes in pure YM theory as counterexamples that break the principle of maximal transcendentality. 
To be concrete, we consider two helicity configurations: $A_{4, \scriptscriptstyle{\rm YM}}^{(1)}(1^-, 2^-, 3^+, 4^+)$ and $A_{4, \scriptscriptstyle{\rm YM}}^{(1)}(1^-, 2^+, 3^-, 4^+)$. For convenience of notation, we define the loop correction function normalized by tree factor as
\begin{equation}
\mathcal{I}_{A_4}^{(1)}  = {A_4^{(1)} \over A_4^{(0)}} \,.
\end{equation}
We also define
\begin{equation}
s=(p_1+p_2)^2\,, \quad t=(p_2+p_3)^2 \,, \quad s_{il}=(p_i+l)^2\,.
\end{equation}

%%%%%%%%%%%%%%%%%%%%%%%%%%%%%
\subsection*{(I) $A_{4, \scriptscriptstyle{\rm YM}}^{(1)}(1^-, 2^-, 3^+, 4^+)$:}

The result of $A_{4, \scriptscriptstyle{\rm YM}}^{(1)}(1^-, 2^-, 3^+, 4^+)$ can be computed using unitarity-cut as\footnote{One can simply use the four-dimensional unitarity cut, which is enough for the consideration of maximal transcendentality part. The missing terms are rational parts.}
\begin{equation}
\mathcal{I}_{A_{4, \scriptscriptstyle{\rm YM}}}^{(1)}(1^-, 2^-, 3^+, 4^+) = - s t\hskip -.2cm  \begin{tabular}{c}{\includegraphics[height=1.2cm]{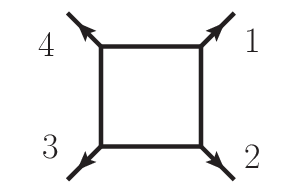} } \end{tabular} 
\hskip -.4cm+ \ c_1 \hskip -.25cm  \begin{tabular}{c}{\includegraphics[height=1.2cm]{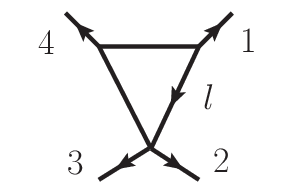} } \end{tabular} 
\hskip -.4cm+ \ c_2 \hskip -.25cm  \begin{tabular}{c}{\includegraphics[height=1.2cm]{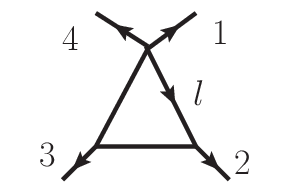} } \end{tabular} 
\hskip -.4cm + \textrm{lower trans.}
\end{equation}
where the diagrams represent box and triangle scalar integrals, the triangle coefficients are
\begin{equation}
c_1 = t \left( 2- 3 {s_{2l} \over s} + 2 {s_{2l}^2 \over s^2} - {s_{2l}^3 \over s^3} \right)  \,, \quad 
c_2 = t \left( 2- 3 {s_{1l} \over s} + 2 {s_{1l}^2 \over s^2} - {s_{1l}^3 \over s^3} \right) \,,
\end{equation}
and `{lower trans.}' represents other lower transcendental contributions from higher $\epsilon$-order terms of box and triangles, as well as the bubble integrals and rational terms.

For the box and triangle integrals, each separately contributes to the maximally transcendental terms of degree-$2$. Interestingly, after integral reduction, one finds that the maximal transcendentality parts of the two triangle integrals cancel with each other such that
\begin{equation}
c_1 \hskip -.2cm  \begin{tabular}{c}{\includegraphics[height=1.2cm]{I3_1} } \end{tabular} 
\hskip -.4cm+ \ c_2 \hskip -.2cm  \begin{tabular}{c}{\includegraphics[height=1.2cm]{I3_2} } \end{tabular} 
\hskip -.4cm  =
 \textrm{lower trans.}
\end{equation}
Thus only the box integral contributions to the maximally transcendental part, which is identical to the well-known $\mathcal{N}=4$ result:
\begin{equation}
\mathcal{I}_{A_{4, \scriptscriptstyle{\rm YM}}}^{(1),{\rm M.T.}}(1^-, 2^-, 3^+, 4^+) = \mathcal{I}_{A_4, \scriptscriptstyle\mathcal{N}=4}^{(1)}  \,.
\end{equation}
This also means that the scalar and fermion loops in ${\cal N}=4$ SYM do not have maximally transcendental contributions.

%%%%%%%%%%%%%%%%%%%%%%%%%%%%%%%%%%%
\subsection*{(II) $A_{4, \scriptscriptstyle{\rm YM}}^{(1)}(1^-, 2^+, 3^-, 4^+)$:}

The result of $A_{4, \scriptscriptstyle{\rm YM}}^{(1)}(1^-, 2^+, 3^-, 4^+)$ takes a different form as:
\begin{equation}
\label{eq:YMmpmp}
\mathcal{I}_{A_{4, \scriptscriptstyle{\rm YM}}}^{(1),{\rm fin}}(1^-, 2^+, 3^-, 4^+) = (- st + {\tilde d}) \hskip -.2cm  \begin{tabular}{c}{\includegraphics[height=1.2cm]{I4} } \end{tabular} 
\hskip -.4cm + \ 
\left( {\tilde c} \hskip -.2cm  \begin{tabular}{c}{\includegraphics[height=1.2cm]{I3_2} } \end{tabular} \hskip -.4cm 
+ \textrm{cyclic perm.} \right)
 + \textrm{lower trans.}
\end{equation}
where
\begin{align}
{\tilde d} = &  {s^2 t^2 ( 2 s^2 + 3 s t + 2 t^2) \over (s+t)^4}  \,, \\ 
{\tilde c} = &  -{s t \over (s+t)^4} \Big[  s_{1l}^3 - 2 s_{1l}^2 (s-t) + 3 s_{1l} (s^2+ t^2) - 2 (s^3- t^3) \Big] \,.
\end{align}

We can see that the coefficients of the box and triangle integrals are much more complicated, in particular, a non-trivial spurious pole $1\over (s+t)^4$ appears, and so do the triangle integrals. Since the IR divergence have a universal maximally transcendental part which is provided by box integral with coefficient $-st$, 
the other IR divergences from the box contribution with the coefficient ${\tilde d}$ and the triangle integrals must cancel with each other, which can be checked.
At finite order, however, the maximally transcendental part receives a new contribution and is therefore different for the $\mathcal{N}=4$ results. Namely
\begin{align}
& \mathcal{I}_{A_{4, \scriptscriptstyle{\rm YM}}}^{(1),{\rm M.T.}}(1^-, 2^+, 3^-, 4^+)|_{\rm I.R.} = \mathcal{I}_{A_4, \scriptscriptstyle\mathcal{N}=4}^{(1)}|_{\rm I.R.} \,, \\
& \mathcal{I}_{A_{4, \scriptscriptstyle{\rm YM}}}^{(1),{\rm M.T.}}(1^-, 2^+, 3^-, 4^+)|_{\rm fin} \neq \mathcal{I}_{A_4, \scriptscriptstyle\mathcal{N}=4}^{(1)}|_{\rm fin} \,.
\end{align}

Let us study the contribution in ${\cal N}=4$ SYM in more detail. The full  ${\cal N}=4$ SYM result can be decomposed as three parts:
\begin{equation}
\mathcal{I}_{A_4, \scriptscriptstyle\mathcal{N}=4}^{(1)} 
= \mathcal{I}_{A_4, \scriptscriptstyle{\rm gluon}}^{(1)} + \mathcal{I}_{A_4, \scriptscriptstyle{\rm fermion}}^{(1)} + \mathcal{I}_{A_4, \scriptscriptstyle{\rm scalar}}^{(1)}   \,,
\end{equation}
where gluon, fermion and scalar indicate the particles that circulate the loop. One can find that both fermion and scalar loops contribute to maximally transcendental part. In particular, the coefficients of scalar box integral are:
\begin{equation}
{\tilde d}_{\scriptscriptstyle{\rm fermion}}  =  -2{s^2 t^2 (s^2+ t^2) \over (s+t)^4} \,, \qquad
{\tilde d}_{\scriptscriptstyle{\rm scalar}}  =  -3 {s^3 t^3 \over (s+t)^4} \,. \\
\end{equation}
They cancel precisely the ${\tilde d}$ in the gluon loop result \eqref{eq:YMmpmp}, which gives the correct ${\cal N}=4$ coefficient.

The contribution from the scalar-loop means that there is no way to match ${\cal N}=4$ SYM and QCD results by simply changing the representation of fermions. Therefore, the principle of maximal transcendentality does not apply for the one-loop four gluon amplitude, or more concretely for the helicity amplitudes $A_{4}^{(1)}(1^-, 2^+, 3^-, 4^+)$. Compared to the form factor cases we consider in the main text, this may be understood that the IR and collinear constraints are not enough to fix the amplitudes uniquely, but allow a larger solution space for the maximal transcendental parts.

%%%%%%%%%%%%%%%%%%%%%%%%%%%%%%%%%%%%%%%%%%
%%%%%%%%%%%%%%%%%%%%%%%%%%%%%%%%%%%%%%%%%%
%\bibliographystyle{JHEP}
%\bibliography{formfactor}

\providecommand{\href}[2]{#2}\begingroup\raggedright\endgroup

\end{document}